%% file: main.tex
\documentclass[12pt,a4paper]{article}
\usepackage[top=1in, bottom=1.25in, left=1in, right=1in]{geometry}
\usepackage[utf8]{inputenc}
\usepackage{graphicx}
\usepackage{color}
\usepackage{subcaption}

\usepackage{caption} %these three command get the figure and table captions automatically small

\title{The role of multi-parton interactions in doubly-heavy hadron production}
\author{U. Egede$^1$, T. Hadavizadeh$^1$, M. Singla$^1$, P. Skands$^1$, \\
M. Vesterinen$^2$\\ 
\small $^1$\emph{School of Physics and Astronomy, Monash University, Melbourne, Australia}\\
\small $^2$\emph{Department of Physics, University of Warwick, Coventry, United Kingdom}}
\date{August 2022}

\usepackage{xspace}
\usepackage{ifthen} % for conditional statements
\newboolean{pdflatex}
\setboolean{pdflatex}{true} % False for eps figures 

\newboolean{articletitles}
\setboolean{articletitles}{true} % False removes titles in references

\newboolean{uprightparticles}
\setboolean{uprightparticles}{false} %True for upright particle symbols

\newboolean{inbibliography}
\setboolean{inbibliography}{false} %True once you enter the bibliography

\def\pythia {\textsc{Pythia}\xspace}

\def\bcvegpy {\textsc{BcVegPy}\xspace}

\def\genxicc {\textsc{GenXicc}\xspace}

\def\vincia {\textsc{Vincia}\xspace}

\def\papercopyright{The authors} % new since 9/Apr/2018
\def\paperlicence{CC-BY-4.0 licence}
\def\paperlicenceurl{https://creativecommons.org/licenses/by/4.0/}

\newcommand\plotwidths{0.45}

\input{lhcb-symbols-def} % Add in the predefined LHCb symbols

\usepackage{hyperxmp}
\usepackage[pdftex,
            pdfauthor={U. Egede, T. Hadavizadeh, M. Singla, P. Skands, M. Vesterinen},
            pdftitle={The role of multi-parton interactions in doubly-heavy hadron production},
            pdfcopyright={Copyright (C) \papercopyright},
            pdflicenseurl={\paperlicenceurl}]{hyperref}
\usepackage[all]{hypcap} % Internal hyperlinks to floats.

\usepackage{cite} % Allows for ranges in citations
\usepackage{mciteplus}

\begin{document}
\maketitle

\begin{abstract}
\noindent Beauty and charm quarks are ideal probes of pertubative Quantum Chromodymanics in proton-proton collisions, owing to their large masses. In this paper the role of multi-parton interactions in the production of doubly-heavy hadrons is studied using simulation samples generated with \pythia, a Monte Carlo event generator. Comparisons are made to the stand-alone generators \bcvegpy and \mbox{\genxicc}. New methods of speeding up \pythia simulations for events containing heavy quarks are described, enabling the production of large samples with multiple heavy-quark pairs.
We show that significantly higher production rates of doubly-heavy hadrons are predicted in models that allow heavy quarks originating from different parton-parton interactions (within the same hadron-hadron collision) to combine to form such hadrons. Quantitative predictions are sensitive to the modelling of colour reconnections. We suggest a set of experimental measurements capable of differentiating these additional contributions.
\end{abstract}

{\footnotesize 
\centerline{\copyright~\papercopyright,  \href{\paperlicenceurl}{\paperlicence}.}}

%%%%%%%%%%%%%%%%%%%%%%%%%%%%%%%%%%%%%%%%%%%%%%%%%%%%%%%%%%%%%%%%%%%%
\section{Introduction}
\label{sec:introduction}
%%%%%%%%%%%%%%%%%%%%%%%%%%%%%%%%%%%%%%%%%%%%%%%%%%%%%%%%%%%%%%%%%%%%

The masses of the charm and beauty quarks are so high that their production at hadron colliders is dominated by perturbative QCD processes making them ideal probes of the partonic interactions that led to their formation. 
Hadrons containing two beauty and/or charm quarks are referred to as doubly-heavy hadrons and, here, include both mesons and baryons with a net heavy flavour such as \Bc mesons and \Xiccpp baryons, and quarkonia with zero net flavour.

Large samples of doubly-heavy hadrons have been collected at the Large Hadron Collider~(LHC) and the study of their properties can provide a unique insight into the role of multi-parton interactions (MPIs) in hadron formation. Because $c$ and $b$ quarks are too heavy to be produced non-perturbatively, doubly-heavy hadrons can only be formed by the effective coalescence of two perturbatively produced heavy quarks. This is fundamentally different from singly-heavy hadrons, in which the heavy quark is confined together with light quark(s) that can be produced non-perturbatively. The fact that heavy hadrons can be produced via MPI has been confirmed not only in  inclusive measurements of heavy-hadron-pair cross sections~\cite{LHCb:2012aiv,D0:2014vql,CMS:2014cmt,LHCb:2015wvu} but also in differential measurements of $\jpsi$ rates vs charged-track multiplicity~\cite{ALICE:2012pet}. 
However, these measurements do not by themselves show conclusively whether two heavy quarks from \emph{different} parton-parton interactions can join to form hadrons.\footnote{Throughout, all parton-parton interactions are within the context of a single proton-proton collision.} Recent measurements of the newly-discovered doubly-charmed tetraquark $T_{cc}^{++}$ indicate the production has similarities with processes involving different parton-parton interactions \cite{LHCb-PAPER-2021-032}. In this paper, we point out that this question can be addressed by considering non-onium doubly-heavy hadrons, such as $B_c^+$ and $\Xi_{cc}^{++}$. Moreover, such measurements will place constraints on models of colour reconnections (CR), which are relevant to a broad range of hadron-collider physics studies~\cite{Argyropoulos:2014zoa,Christiansen:2015yqa,Bierlich:2015rha,Bellm:2019wrh}. 

Samples of doubly-heavy hadrons can be simulated using Monte Carlo event generators such as \pythia~\cite{Sjostrand:2006za,Bierlich:2022pfr}. However, as these hadrons require the chance coalescence of two heavy particles into a bound state, their generation can be prohibitively slow. For this reason some rare doubly-heavy hadrons, for example, \Bcp, $\Xi_{cc}$ and $\Xi_{bc}$, are at the moment usually generated using dedicated generators, such as \bcvegpy~\cite{Chang:2015qea} or \genxicc~\cite{Wang:2012vj}, that perform fixed-order matrix-element calculations. These generators are then interfaced with event generators to simulate the rest of the event evolution and hadronisation.     
However, the fixed-order matrix-element generators assume that both of the heavy quarks are produced in a  \emph{single} parton-parton interaction, for example $gg\to B_{c}^{+} b \bar{c} $ or $q\bar{q}\to B_{c}^{+} b \bar{c} $. This ignores the role that MPIs could play in generating heavy quarks that could contribute to the formation of such hadrons.

In this paper, we develop a method to enhance the rate of hadrons containing one or more heavy quarks in  simulation samples generated with \pythia. This enables us to investigate the role of MPIs in the production of doubly-heavy hadrons in the context of inclusive event samples, in which the formation of such hadrons would normally be exceedingly rare. The generated samples are studied to identify properties that can differentiate between the contributions from single parton scattering (SPS) and double parton scatering (DPS) mechanisms. Simulations are carried out with \pythia version 8.306 using the default Monash tune~\cite{Skands:2014pea} and at a proton-proton centre of mass energy of $13\tev$. The hard interaction is simulated using \pythia~8.306, \bcvegpy~2.2 or \genxicc~2.1 while the fragmentation process is simulated using the default \pythia Simple Shower framework~\cite{Sjostrand:2004ef,Corke:2010yf}. Finally, measurements using LHC data that are able to shed light on the production mechanism are proposed and their feasibility discussed.

Throughout this paper charge conjugation is not implied when describing cross sections. The generator \bcvegpy only produces \Bcp mesons, therefore to avoid ambiguity all cross sections are explicitly only produced for the explicitly specified charge. In general the cross section predictions from \bcvegpy should be doubled when comparing to measurements of $\sigma(\decay{\proton\proton}{B_{c}^{\pm} X})$.  

%%%%%%%%%%%%%%%%%%%%%%%%%%%%%%%%%%%%%%%%%%%%%%%%%%%%%%%%%%%%%%%%%%%%
\section{Sources of heavy quarks}
\label{sec:heavyquarksources}
%%%%%%%%%%%%%%%%%%%%%%%%%%%%%%%%%%%%%%%%%%%%%%%%%%%%%%%%%%%%%%%%%%%%
\begin{figure}[tbp]
    \centering
    \begin{subfigure}{.32\textwidth}
        \includegraphics[width=\linewidth]{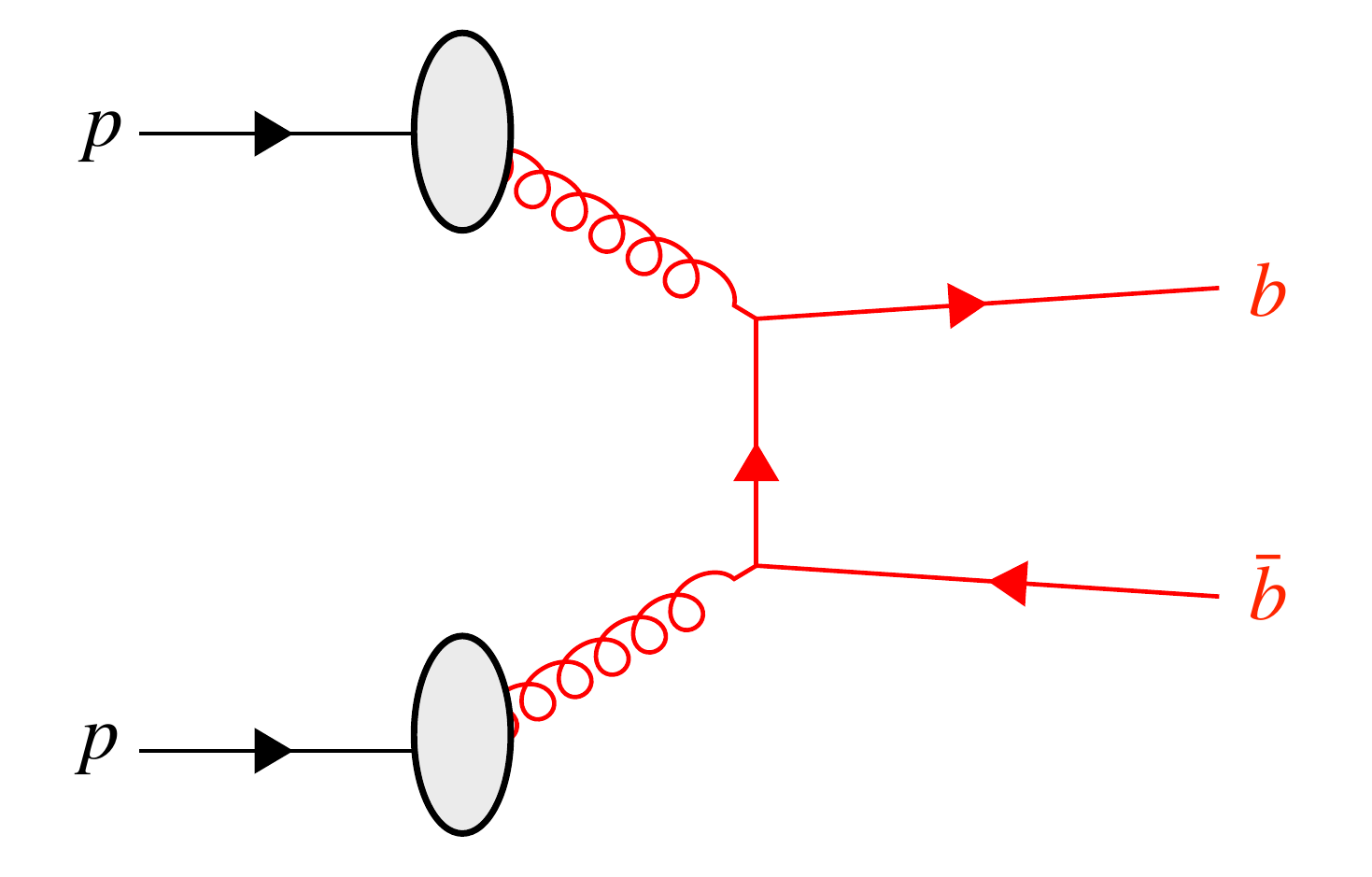}
        \caption{\small  Pair creation.}
        \label{fig:feyn_bb}
    \end{subfigure}%
    \begin{subfigure}{.32\textwidth}
        \includegraphics[width=\linewidth]{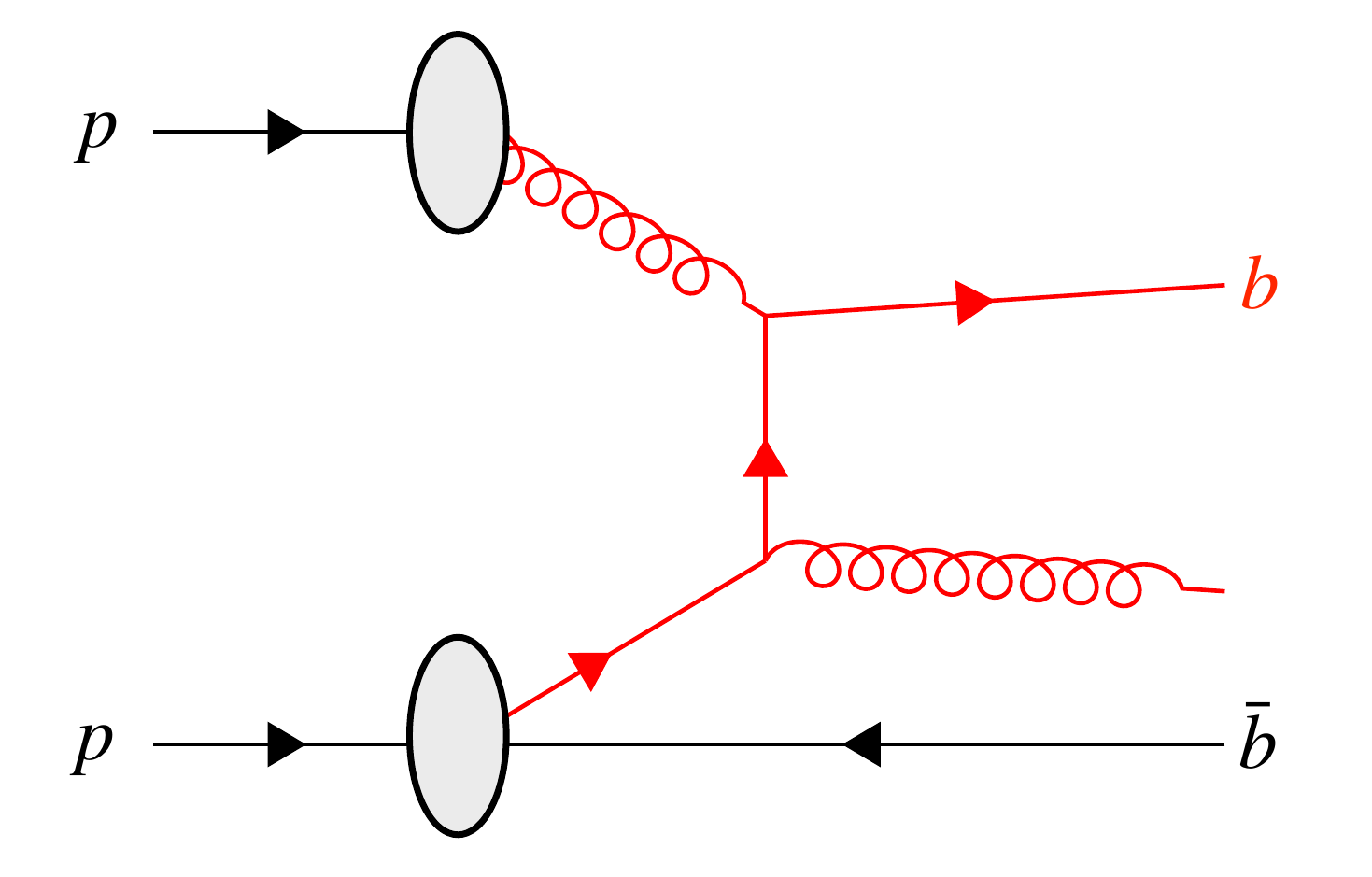}
        \caption{\small  Flavour excitation.}
        \label{fig:feyn_bx}
    \end{subfigure}%
    \begin{subfigure}{.32\textwidth}
        \includegraphics[width=\linewidth]{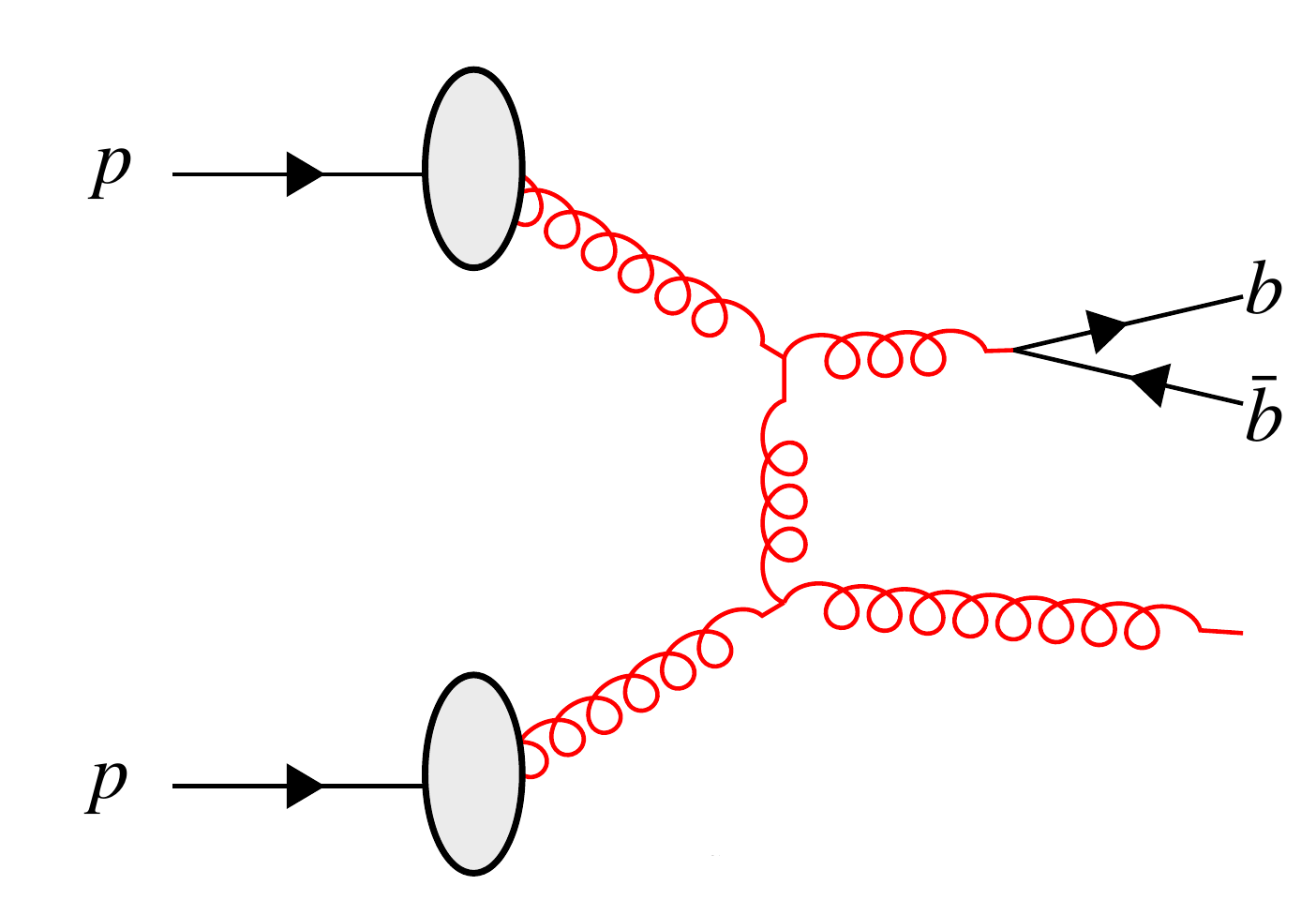}
        \caption{\small  Gluon splitting in parton shower.}
        \label{fig:feyn_shower}
    \end{subfigure}%
    \caption{Examples of production mechanisms for heavy quarks in proton-proton collisions. The incoming, outgoing and intermediate particles of the process considered to be the hardest process are highlighted in red. In the case of flavour excitation, the $\bar{b}$ quark shown at the bottom represents the companion quark produced as a result of the initial-state evolution.}
    \label{fig:singly_heavy_production}
\end{figure}

In proton-proton collisions, the QCD production mechanisms for the heavy $c$ and $b$ quarks can be split into three categories referred to as 
pair creation, flavour excitation and gluon splitting in parton showers~\cite{Norrbin:2000zc}. The processes are classified according to the interaction with the largest momentum transfer, referred to here as the \textit{hard} interaction.

\begin{figure}[tbp]
    \centering
    \includegraphics[width=0.6\linewidth]{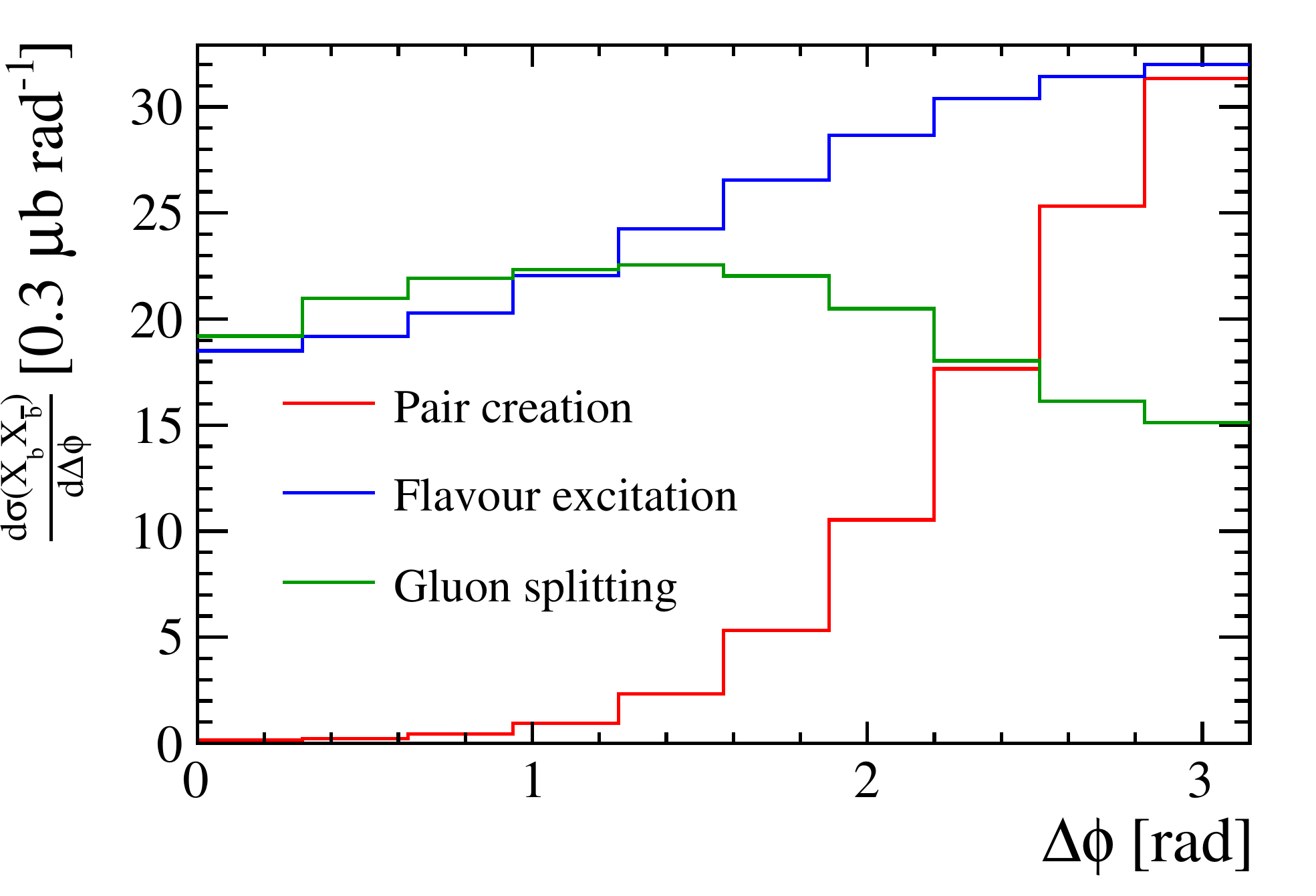}
    \caption{Differential cross-section in the transverse angle between the two $b$-hadrons in \pythia simulations with exactly two $b$-hadrons.}
    \label{fig:delta_phi_bb}
\end{figure}

Pair creation involves a $gg \to Q\bar{Q}$\footnote{$Q$ here represents heavy quarks and $q$ represents light quarks} or $q\bar{q} \to Q\bar{Q}$ hard interaction, as shown in Fig.~\ref{fig:feyn_bb}, that, in the absence of significant initial-state radiation, creates outgoing heavy quarks with equal and opposite transverse momenta. The resulting heavy hadrons formed from the heavy quarks similarly have a strong tendency to be back-to-back in the transverse plane, as shown in Fig.~\ref{fig:delta_phi_bb} for $b\bar{b}$ production. 

Flavour excitation is the process involving one heavy quark: $Qg \to Qg$ or $Qq \to Qq$, represented in Fig.~\ref{fig:feyn_bx}. In this process a virtual $Q\bar{Q}$ pair is produced as part of the initial-state evolution of one of the incoming protons, and one of them, say the $Q$, subsequently interacts with a (non-heavy) parton from the other proton. The $\bar{Q}$ (a.k.a.\ the ``companion'' quark of the scattered heavy quark~\cite{Sjostrand:2004pf}) is ejected as part of the initial-state evolution of the incoming remnant at a lower scale, with less transverse momentum and significantly less correlation with the direction of the $Q$, as shown in Fig.~\ref{fig:delta_phi_bb}.

Heavy quarks can also be produced via gluon splittings during parton showers. A typical example would be a hard $gg \to gg$ interaction followed by a subsequent $g\to Q\bar{Q}$ splitting in the subsequent initial- or final-state shower evolution, as shown in Fig.~\ref{fig:feyn_shower}. Although this figure shows one of the outgoing gluons from the hard interaction directly splitting to heavy quarks, that is just for simplicity; in principle any gluon produced within a shower above the heavy quark-mass threshold could result in heavy quarks. As gluon-gluon interactions have a large cross-section at the LHC, this constitutes a significant contribution to the heavy-quark production mechanisms. For final-state gluon splittings, the resulting $Q\bar{Q}$ pair will be boosted in the direction of the parent gluon. Events in which two singly-heavy hadron are produced by this mechanism tend to have smaller angles between the two heavy hadrons, as shown in Fig.~\ref{fig:delta_phi_bb}.

%%%%%%%%%%%%%%%%%%%%%%%%%%%%%%%%%%%%%%%%%%%%%%%%%%%%%%%%%%%%%%%%%%%%
\section{Sources of doubly-heavy hadrons}
\label{sec:doublyheavyquarksources}
%%%%%%%%%%%%%%%%%%%%%%%%%%%%%%%%%%%%%%%%%%%%%%%%%%%%%%%%%%%%%%%%%%%%

\begin{figure}[h]
    \centering
    \begin{subfigure}{.32\textwidth}
        \includegraphics[width=\linewidth]{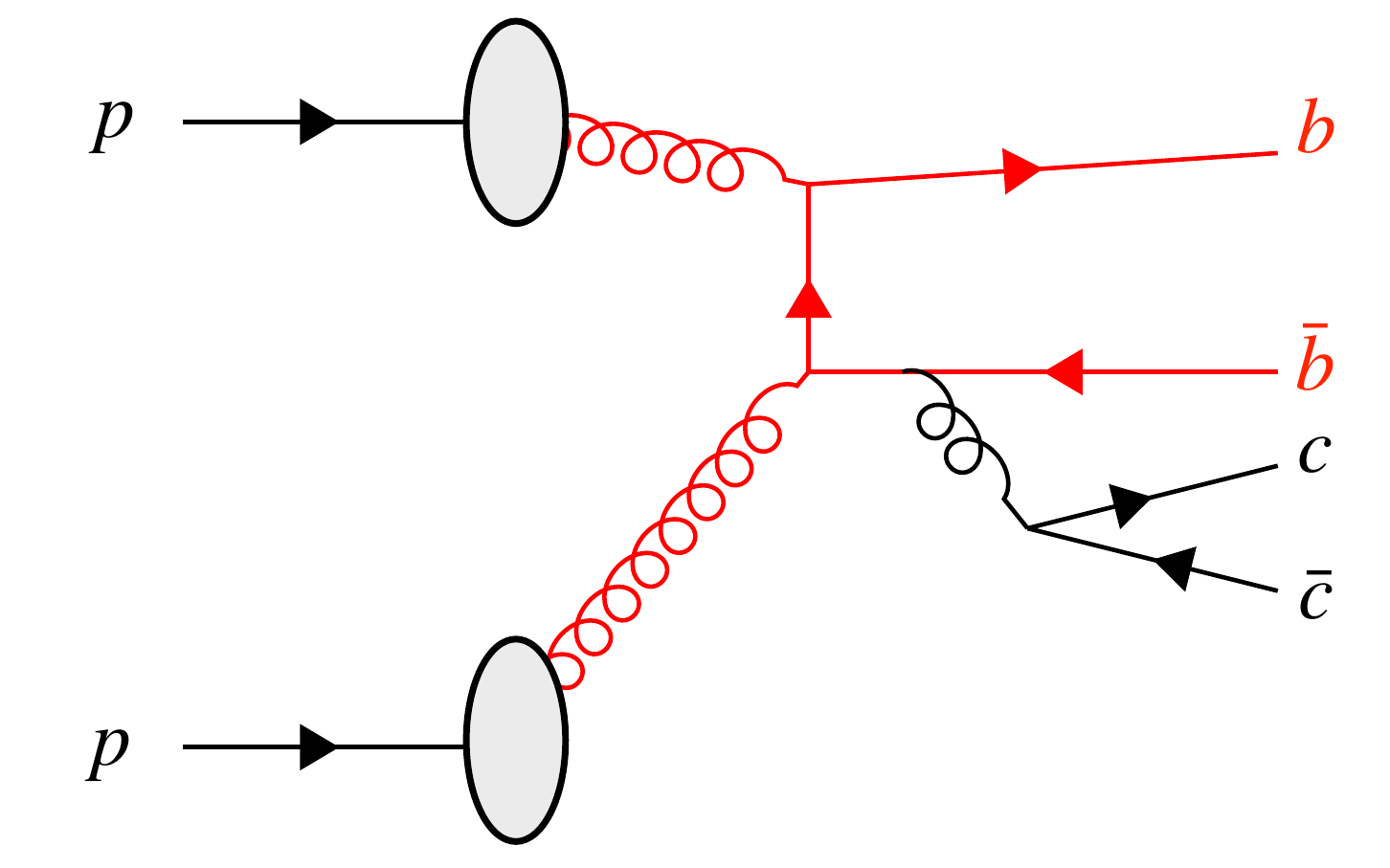}
        \caption{\small  \textbf{Example of SPS:} Pair creation and gluon splitting.}
        \label{fig:feyn_MPI_shower}
    \end{subfigure}%
    \hfill
    \begin{subfigure}{.32\textwidth}
        \includegraphics[width=\linewidth]{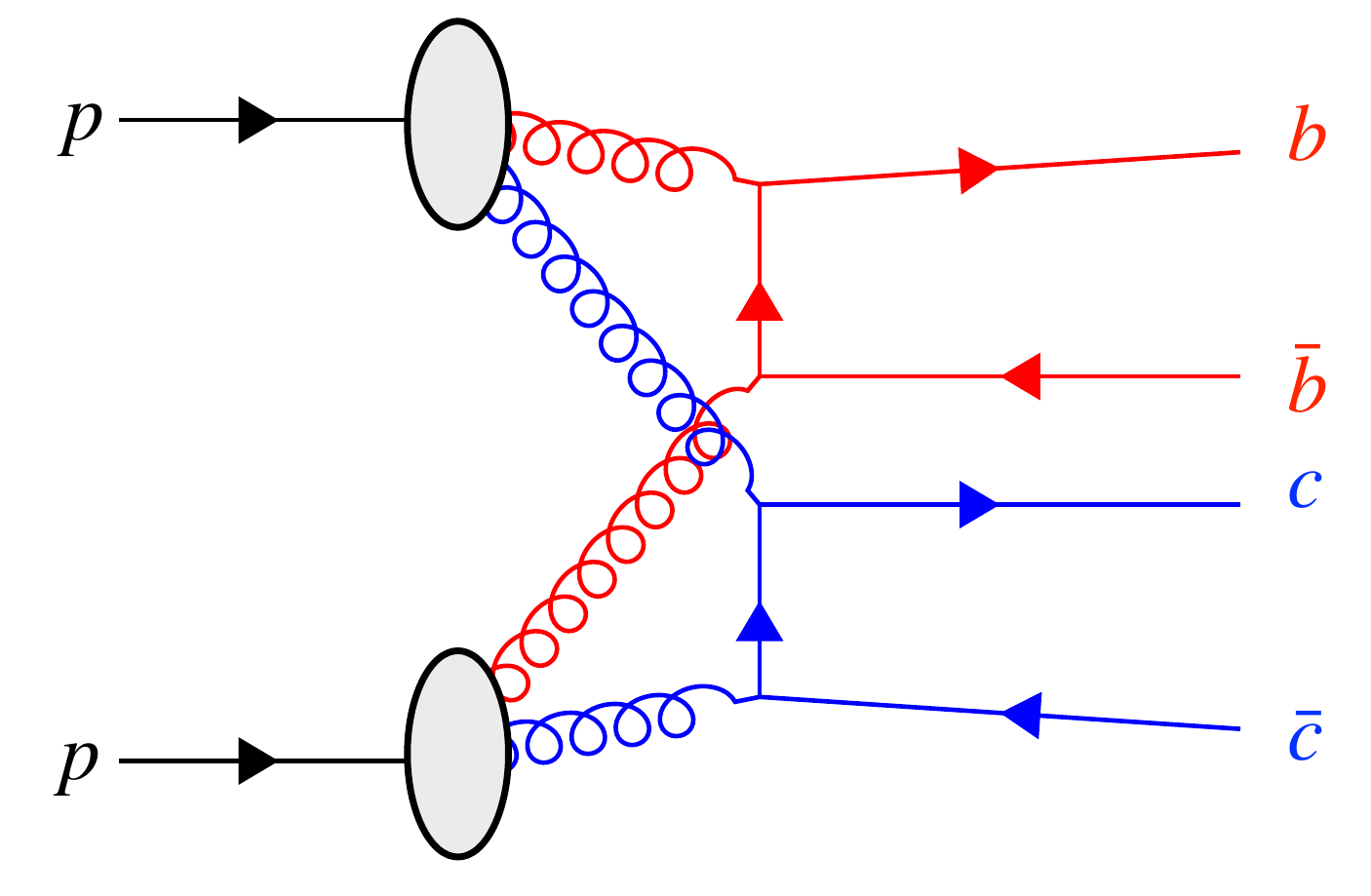}
        \caption{\small  \textbf{Example of DPS:} Double pair creation.}
        \label{fig:feyn_MPI_bb_cc}
    \end{subfigure}%
    \hfill
    \begin{subfigure}{.32\textwidth}
        \includegraphics[width=\linewidth]{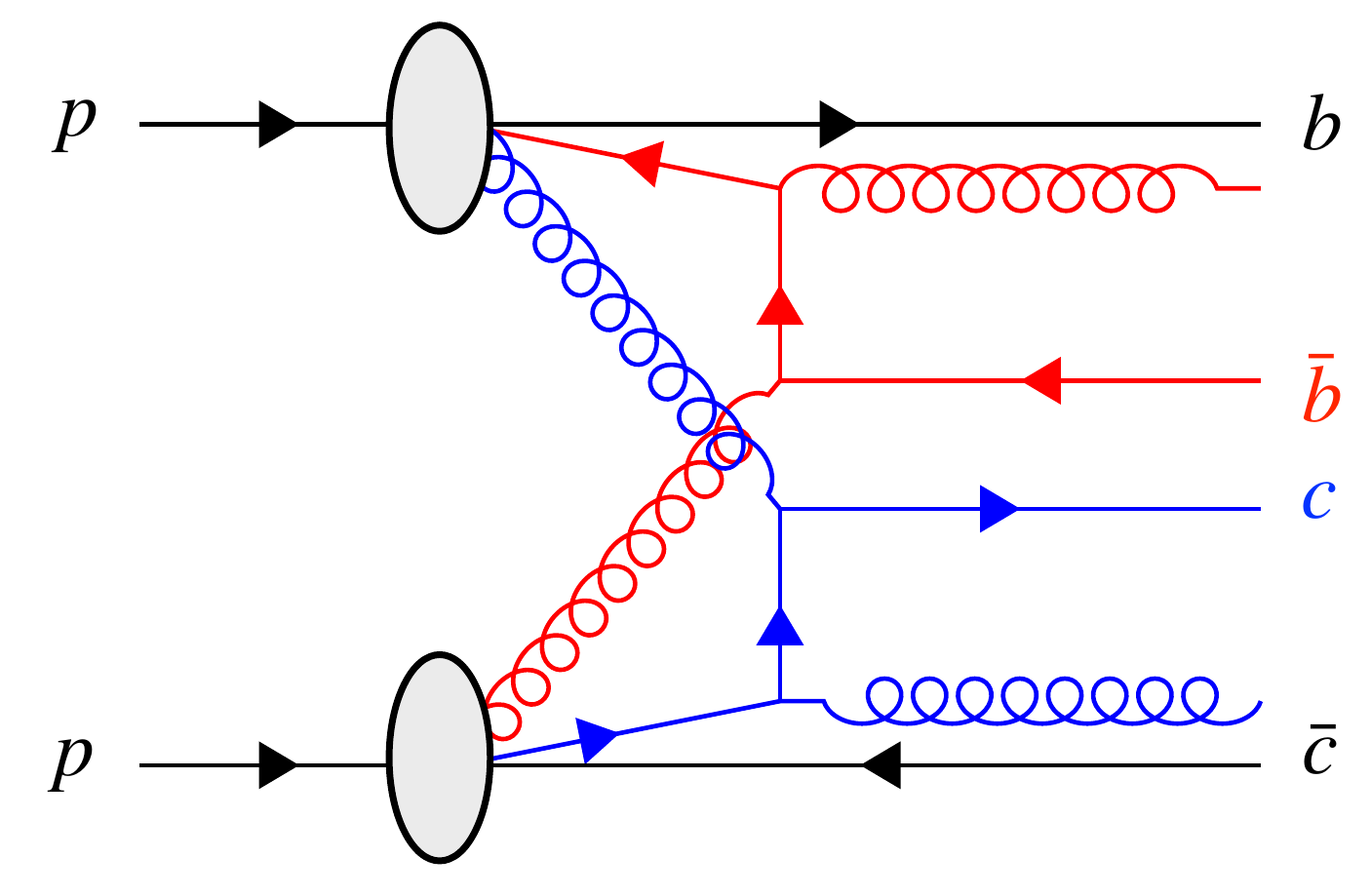}
        \caption{\small  \textbf{Example of DPS:} Double flavour excitation.}
        \label{fig:feyn_MPI_bx_cx}
    \end{subfigure}%
    \caption{\small  Production mechanisms for events with both a $b\bar{b}$ and $c\bar{c}$ pair. The incoming, outgoing and intermediate particles of each parton-parton interaction are shown in red and (where relevant) blue. In the case of double flavour excitation, $b$ and $\bar{c}$ quarks shown at the  top and bottom represents the companion quark produced as a result of the initial-state evolution.}
    \label{fig:doubly_heavy_production}
\end{figure}

To create doubly-heavy hadrons that are not quarkonium states, two  $Q\bar{Q}$ pairs must be produced during the perturbative evolution of the collision. An example of an SPS mechanism contributing to this process is shown in  Fig.~\ref{fig:feyn_MPI_shower}: hard $b\bar{b}$ pair creation followed by a $g\to c\bar{c}$ splitting during the shower evolution. Equivalent processes involving flavour excitation or double gluon splitting within a single SPS are of course also possible. 

When allowing for MPI, the two $Q\bar{Q}$ pairs may also be produced in two \emph{different} parton-parton interactions (still within the context of a single hadron-hadron collision). This is what we label DPS. Two examples, double pair creation and double flavour excitation, are shown in Figs.~\ref{fig:feyn_MPI_bb_cc} and \ref{fig:feyn_MPI_bx_cx} respectively, again with other combinations of pair creation, flavour excitation, and/or gluon splittings obviously also possible. In these diagrams the two parton interactions have been highlighted in different colours to clarify the origin of the partons.

In events with more than two parton-parton interactions, SPS mechanisms could contribute from any one of the \emph{single} parton-parton interactions, whilst DPS mechanisms could contribute from the combination of any two. 

Once the appropriate quarks have been produced in the collision, only pairs that are sufficiently close in phase space and which have a non-zero probability to be in an overall colour-singlet state, have a chance to form an on-shell doubly-heavy hadron.

Measurements of the cross sections of multiple heavy hadrons suggest that MPIs play a significant role in the production of multiple heavy quark pairs at hadron colliders~\cite{ALICE:2012pet,LHCb:2012aiv,D0:2014vql,CMS:2014cmt,LHCb:2015wvu}. 
However, the question of how partons originating from \emph{different} parts of the protons become bound into hadrons is still afflicted with significant uncertainties. In general-purpose event generators like \pythia{}, this is controlled by a combination of perturbative heavy-quark production mechanisms (hard scatterings, MPI, and parton showers) and semi-empirical models of colour reconnections with~\cite{Ferreres-Sole:2018vgo,Bellm:2019wrh} and without~\cite{Argyropoulos:2014zoa,Christiansen:2015yqa} space-time dependence.  The simple diagrams in Fig.~\ref{fig:doubly_heavy_production} demonstrate how $B_c^+$ mesons formed from the $\bar{b}c$ combinations could provide an ideal probe into the hadronisation process. This is unique to doubly-heavy hadrons, since light quarks are mainly created nonperturbatively and hence do not have the same character of being associated with specific short-distance processes in the colliding protons.

%%%%%%%%%%%%%%%%%%%%%%%%%%%%%%%%%%%%%%%%%%%%%%%%%%%%%%%%%%%%%%%%%%%%
\section{Efficient simulation of events with heavy hadrons in Pythia}
\label{sec:enhancements}
%%%%%%%%%%%%%%%%%%%%%%%%%%%%%%%%%%%%%%%%%%%%%%%%%%%%%%%%%%%%%%%%%%%%

Generating unbiased events with multiple pairs of heavy quarks and doubly-heavy hadrons with Monte Carlo event generators can be very time consuming as few events will fulfil the requirements to form the doubly-heavy hadrons. 
A method of enhancing the efficiency to produce events containing heavy quarks in \pythia is outlined here, and can be applied to both singly- and doubly-heavy hadrons.

\pythia provides user-configurable classes called \texttt{UserHooks} aimed at allowing the user to inspect and veto events at different stages during the event evolution. These can be exploited to veto events that do not contain the requisite heavy quarks early on in the generation, removing time spent evolving and hadronising events that will never be accepted. 
\begin{figure}[h]
    \centering
    \begin{subfigure}[b]{.24\textwidth}
        \includegraphics[width=\linewidth]{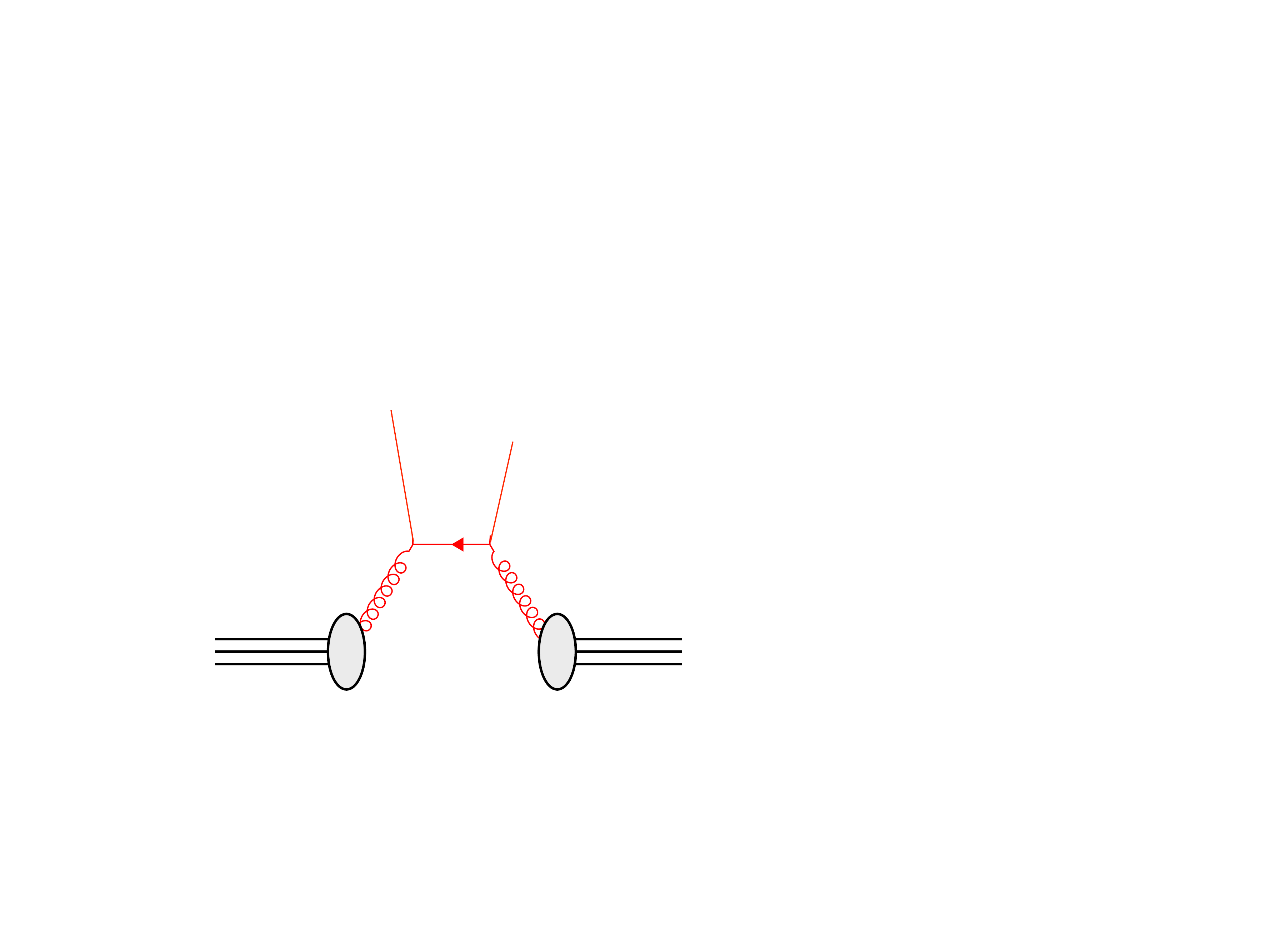}
        \caption{\small  Process Level.}
        \label{fig:uh_process_level}
    \end{subfigure}%
    \hfill
    \begin{subfigure}[b]{.24\textwidth}
        \includegraphics[width=\linewidth]{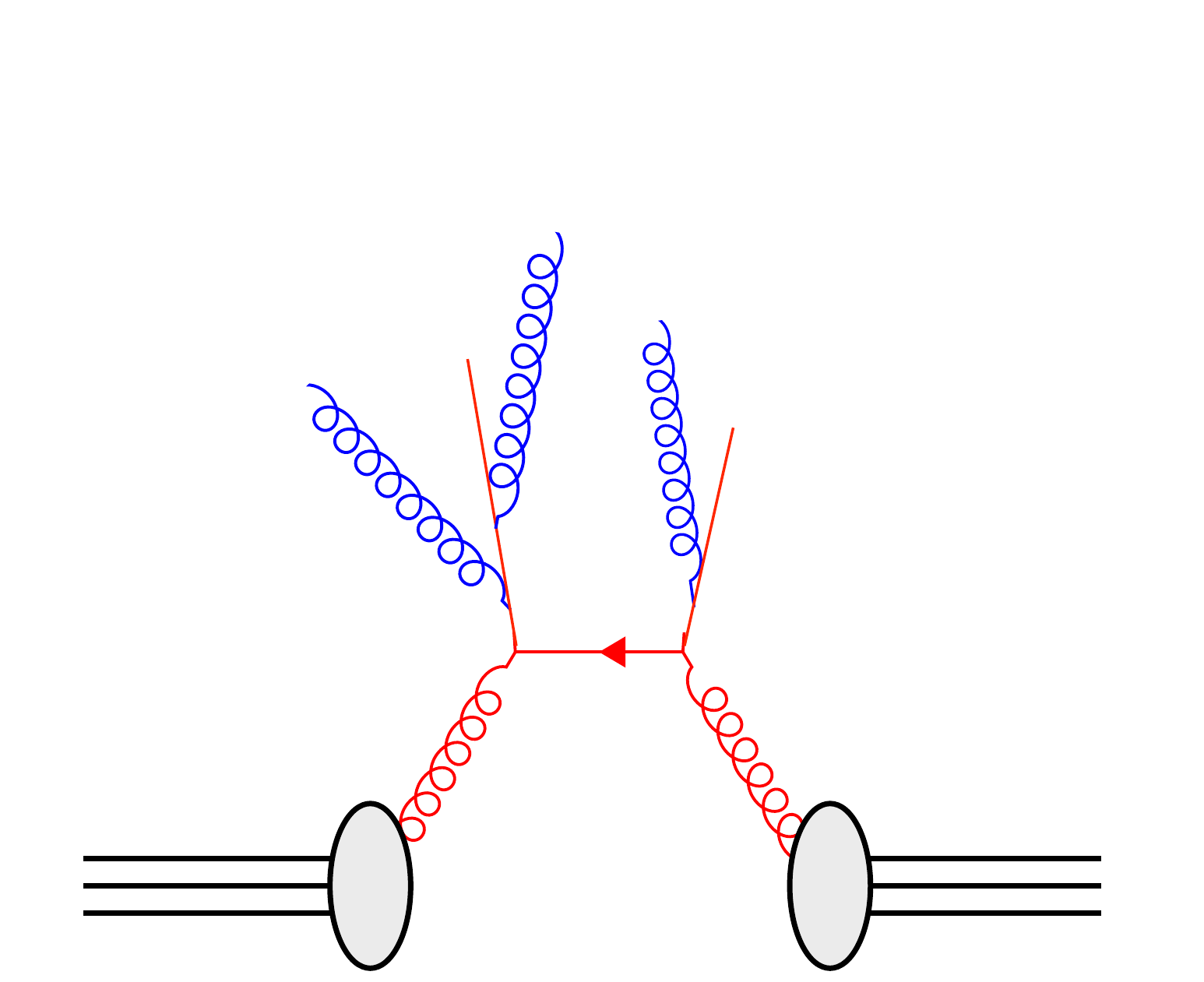}
        \caption{\small  During evolution.}
        \label{fig:uh_veto_pt}
    \end{subfigure}
    \hfill
    \begin{subfigure}[b]{.24\textwidth}
        \includegraphics[width=\linewidth]{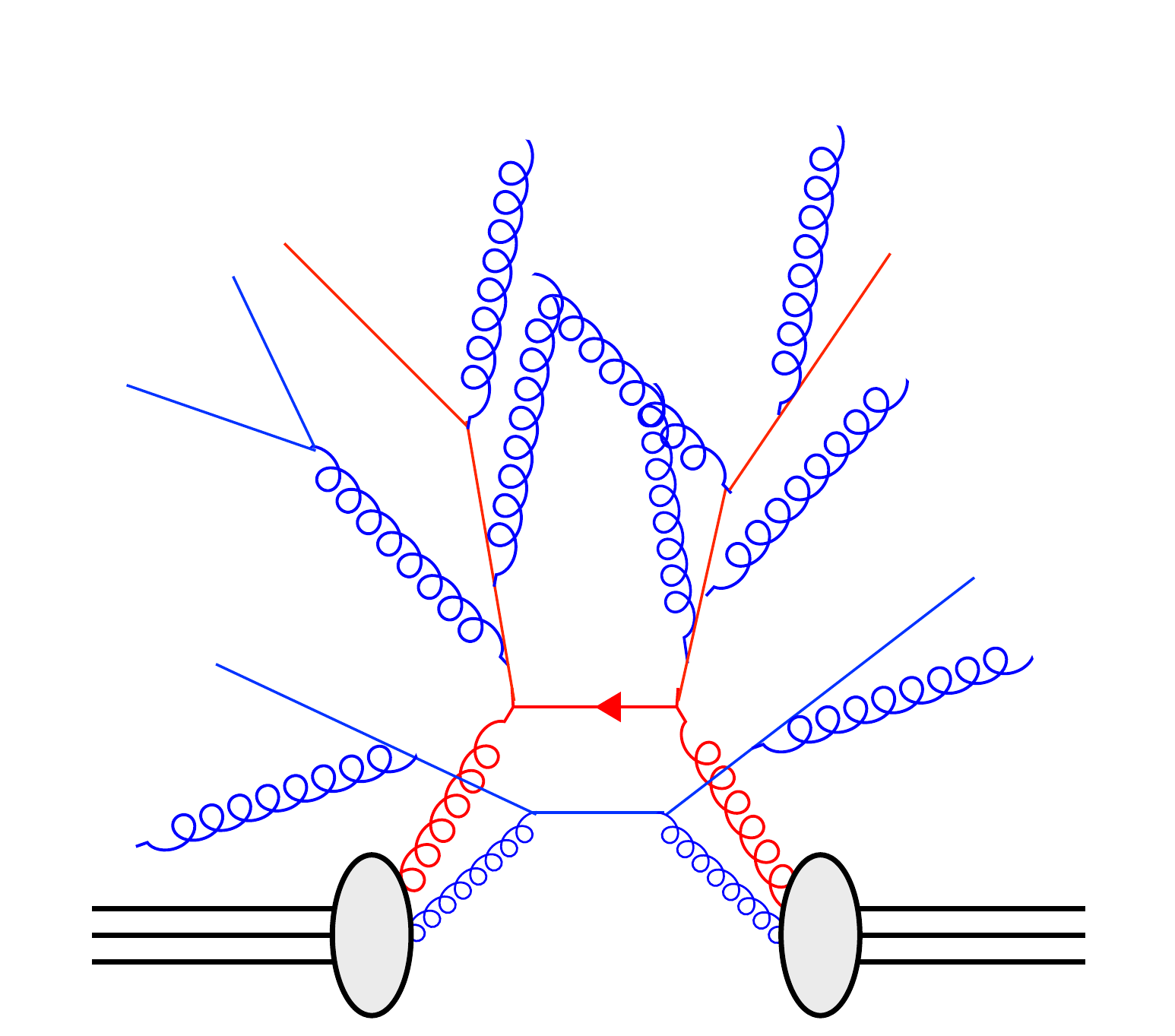}
        \caption{\small  Parton Level.}
        \label{fig:uh_parton_level}
    \end{subfigure}%
    \hfill
    \begin{subfigure}[b]{.24\textwidth}
        \includegraphics[width=\linewidth]{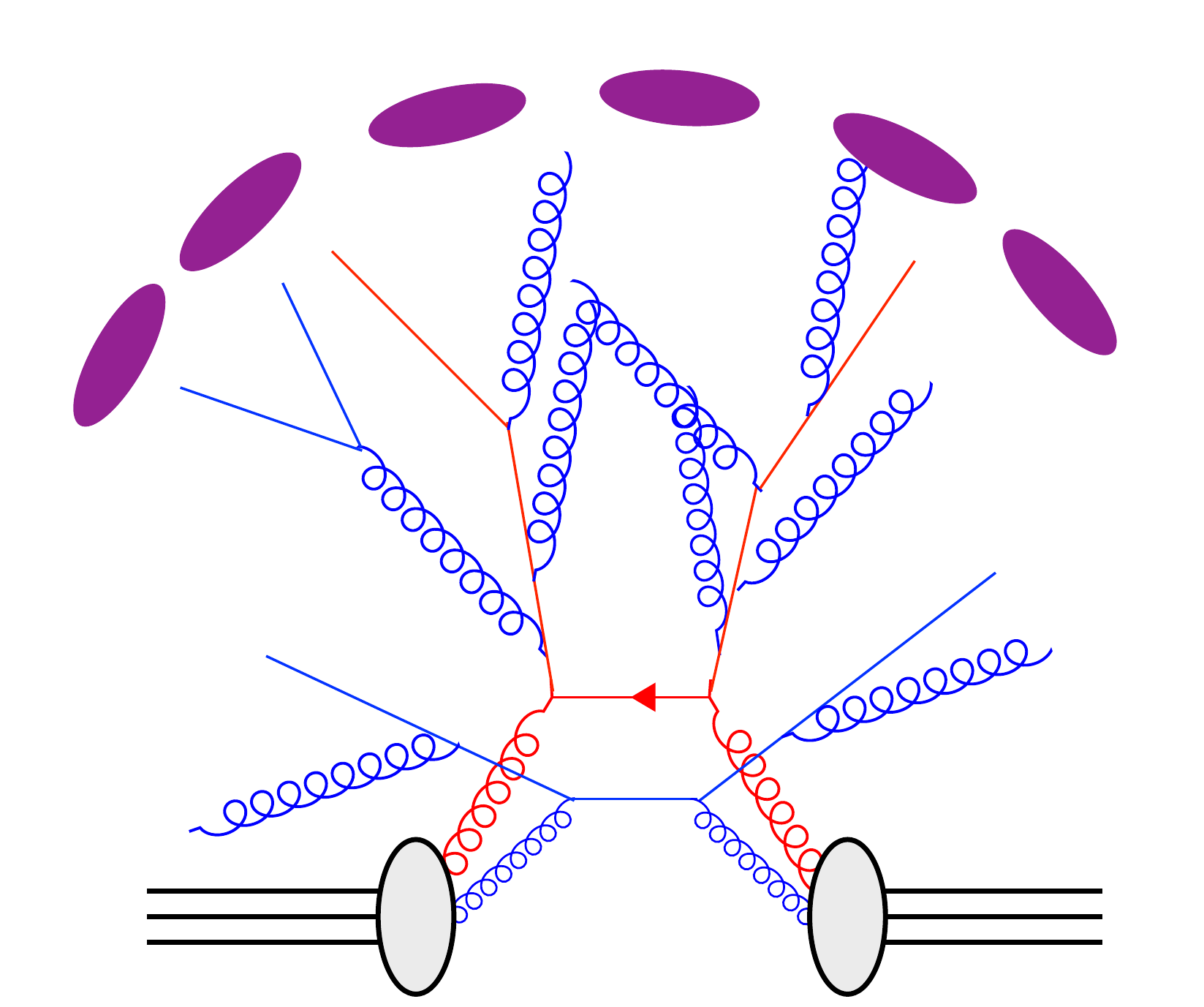}
        \caption{\small  Hadronisation.}
        \label{fig:uh_hadronisation}
    \end{subfigure}%
    \caption{\small  Simple representations of different stages during the event evolution in \pythia.}
    \label{fig:UserHook_stages}
\end{figure}

The \texttt{UserHook} stages that are utilised to improve the efficiency are:
\begin{itemize}
    \item \textbf{Hard-process-level veto:} This veto inspects the event after the most energetic parton interaction has occurred, as represented in Fig~\ref{fig:uh_process_level};
    \item \textbf{Event-evolution veto:} In \pythia the event is evolved from the hard-interaction scale down to the hadronisation scale. During this process, the event can be inspected when the evolution reaches an arbitrary (user-defined) value of the evolution scale, illustrated in Fig~\ref{fig:uh_veto_pt};
    \item \textbf{Parton-level veto:} Once the evolution has reached the hadronisation scale, the event can be inspected before any hadrons are created, leaving the complete partonic event as shown in  Fig~\ref{fig:uh_parton_level};
    \item \textbf{Post-hadronisation veto:} It is possible to inspect the event after the hadronic states have been formed as shown in Fig~\ref{fig:uh_hadronisation}.
\end{itemize}
Obviously, vetoes that can be placed early on can lead to far bigger efficiency gains than ones placed later, and a veto at the post-hadronisation level, after most of the event activity has already been treated (apart from decays and, optionally, hadronic rescatterings) will not make much of a difference at all.

%%%%%%%%%%%%%%%%%%%%%%%%%%%%%%%%%%%%%%%%%%%%%%%%%%%%%%%%%%%%%%%%%%%%%%%%%%%%
\subsection{Simulating final states involving one \boldmath{$Q\bar{Q}$} pair} 
\label{sec:sim_QQ}
%%%%%%%%%%%%%%%%%%%%%%%%%%%%%%%%%%%%%%%%%%%%%%%%%%%%%%%%%%%%%%%%%%%%%%%%%%%%
The creation of a heavy quark, $Q$, involves a physical momentum transfer of at least ${\cal O}(m_Q)$, regardless of whether the production mechanism is a hard process, MPI, or a $\decay{g}{Q\bar{Q}}$ shower branching.
Given an event-generator evolution algorithm that is ordered in a measure of momentum transfer, when the evolution reaches a scale of order $m_Q$ one could thus immediately veto any events that do not contain a specific desired number of heavy quarks\footnote{Including not only quarks of flavour $Q$ but also states that contain them, like $Q$-onia, or which can decay to them, e.g.\ via $b\to c$ and/or $t\to b$ transitions.}. Doing so effectively ``saves'' the time it would otherwise have taken to evolve such events from $m_Q$ to the perturbative cutoff, as well as the time required to hadronise them. 
Since the running value of $\alpha_s$ is largest near the cutoff, where the number of evolving partons is also highest, and hadronisation can consume a further significant amount of time (in particular when colour reconnections are involved), a veto at the scale $m_Q$ should  produce order-of-magnitude efficiency gains, compared with a standard simulation in which events are fully evolved and hadronised before the event is inspected.

\begin{figure}[t]
\centering\includegraphics*[width=0.5\textwidth]{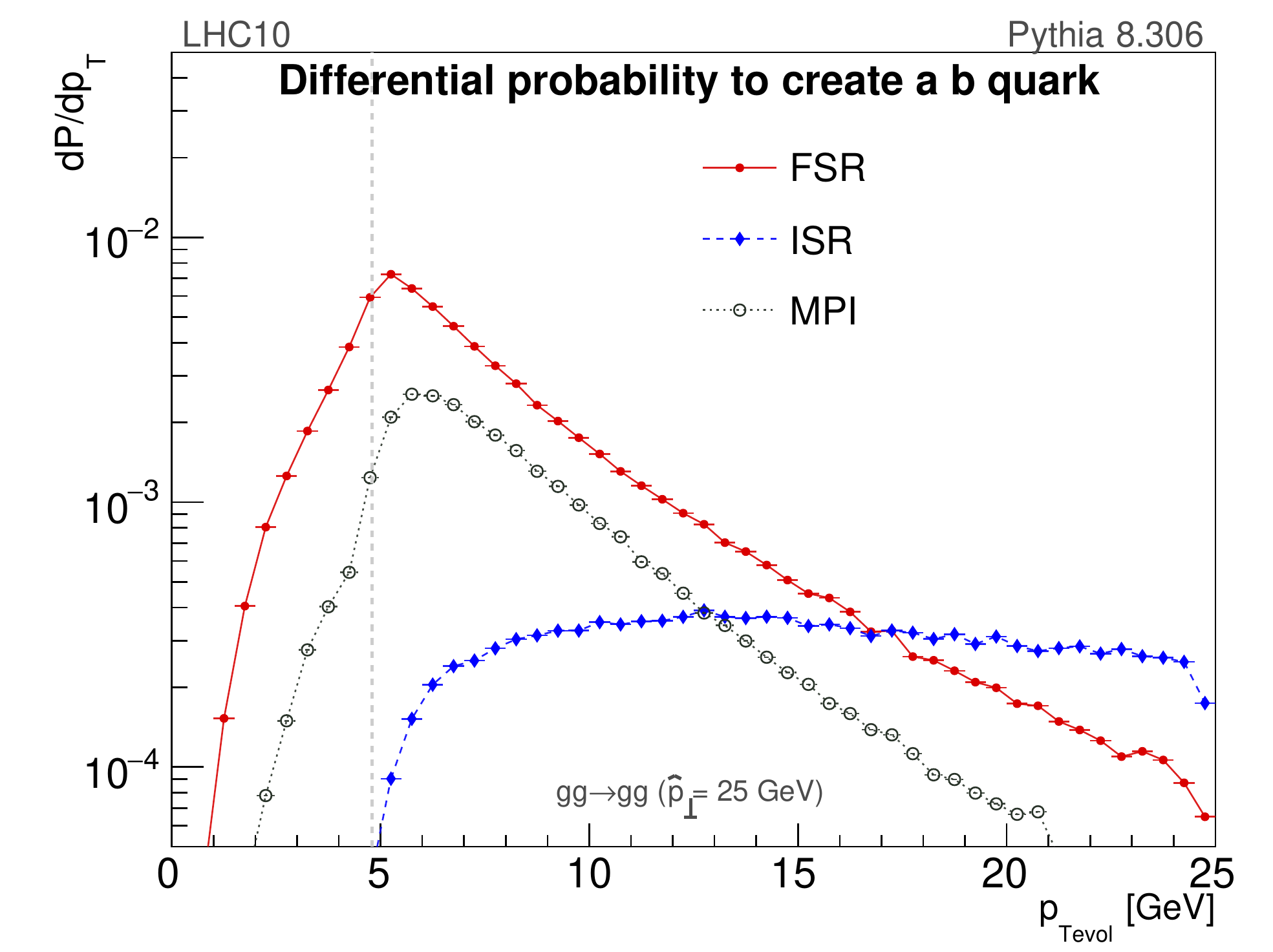}
\caption{Differential creation probabilities for $b$ quarks, as a function of the \pythia $p_\perp$ evolution parameter $p_{\perp\mathrm{evol}}$, for a reference $gg\to gg$ hard process with $\hat{p}_\perp = 25~\mathrm{GeV}$ in proton-proton collisions at $\sqrt{s}=10\,\mathrm{TeV}$. The solid red line shows FSR $g\to b\bar{b}$ branchings, the dashed blue one shows ISR gluons backwards-evolving to $b$ quarks, and the dotted black one shows MPI pair-creation and flavour-excitation processes. The vertical dashed gray line indicates the default value of the $b$ quark mass in \pythia, $m_b = 4.8\,\mathrm{GeV}$.
\label{fig:PTevol}}
\end{figure}
A caveat to implementing such a strategy for \pythia, however, is that its default transverse-momentum evolution scales~\cite{Sjostrand:2004ef,Corke:2010yf} for heavy-quark processes are only guaranteed to be greater than $m_Q$ for hard processes and for ISR, whereas the probability densities for FSR $\decay{g}{Q\bar{Q}}$ splittings and MPI $\decay{gg}{Q\bar{Q}}$ processes are non-zero all the way down to $p_{\perp\mathrm{evol}}\to 0$. 
This is illustrated in Fig.~\ref{fig:PTevol}, for a representative case of $\decay{gg}{gg}$ hard-scattering events with $\hat{p}_\perp = 25\pm0.5\,\mathrm{GeV}$ in proton-proton collisions with $\sqrt{s}=10~\mathrm{TeV}$. 

An interesting follow-up question for future work is thus whether it would be physically justifiable to reformulate these algorithms in terms of a measure that would associate all heavy-quark production processes with scales $\ge~m_Q$. (This is, \eg, the choice made in the \vincia shower~\cite{Brooks:2020upa}, which however still relies on \pythia's MPI evolution.)

For now, we accept that this subtlety will force a tradeoff between realising the full possible efficiency gains and ``missing'' a small fraction of signal events, which we will comment on further below. We therefore phrase our implementation in terms of an arbitrary veto scale, not necessarily equal to the heavy-quark mass, which can be varied to determine if an ``acceptable'' tradeoff can be found, which may vary from application to application. Specifically, we define the following  \texttt{Userhook} vetoes:
\begin{description}
\item[Hard-process-level veto:] At this stage, nothing is known about the subsequent shower or MPI evolution, except for what the starting scale for those evolutions will be.  Our veto function only accepts events that fulfil at least one of the following two conditions: 1) the hard process itself  contains the requisite heavy flavour (by which we include any onium containing it or a heavier quark that can decay to it), in which case a flag may also be set to bypass any downstream vetoes, or 2) the  starting scale for MPI and showers is above our user-defined veto scale, so that we want to give MPI and/or showers a chance to produce the heavy flavour. This essentially means that $\decay{gg}{gg}$ events with $\hat{p}_\perp < {\cal O}(m_Q)$ can be rejected already at this stage, with minimum processing. 
\item[Event-evolution veto:] If the hard-scattering process did not contain the requisite heavy flavour but was allowed a chance to produce it via MPI and/or showers, the event is inspected again when the evolution reaches our veto scale, and is now rejected if the required flavour (again including onia and/or heavier flavours) is still not present in the event. 
\end{description}

\begin{figure}[tbp]
    \centering
    \includegraphics[width=0.49\linewidth]{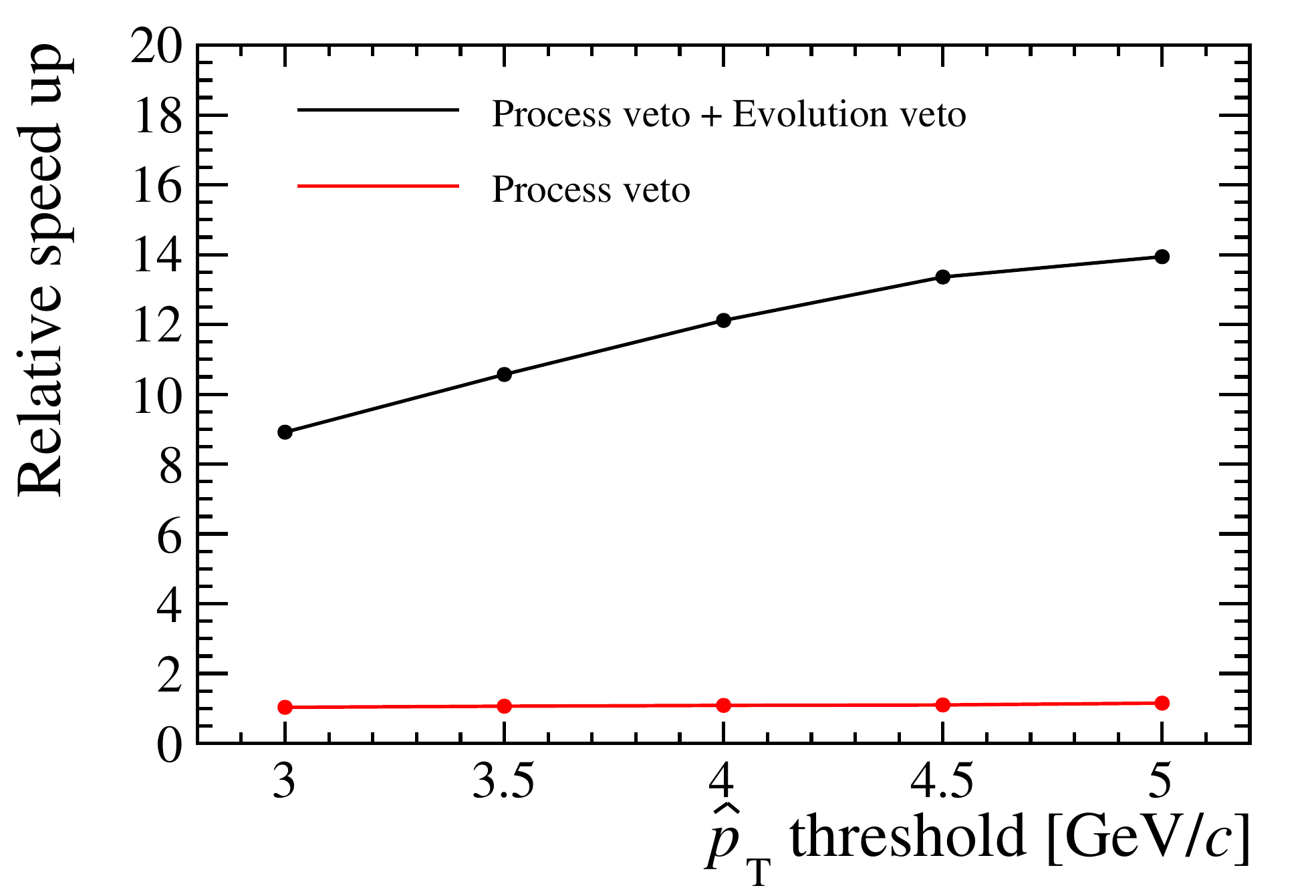}
    \includegraphics[width=0.49\linewidth]{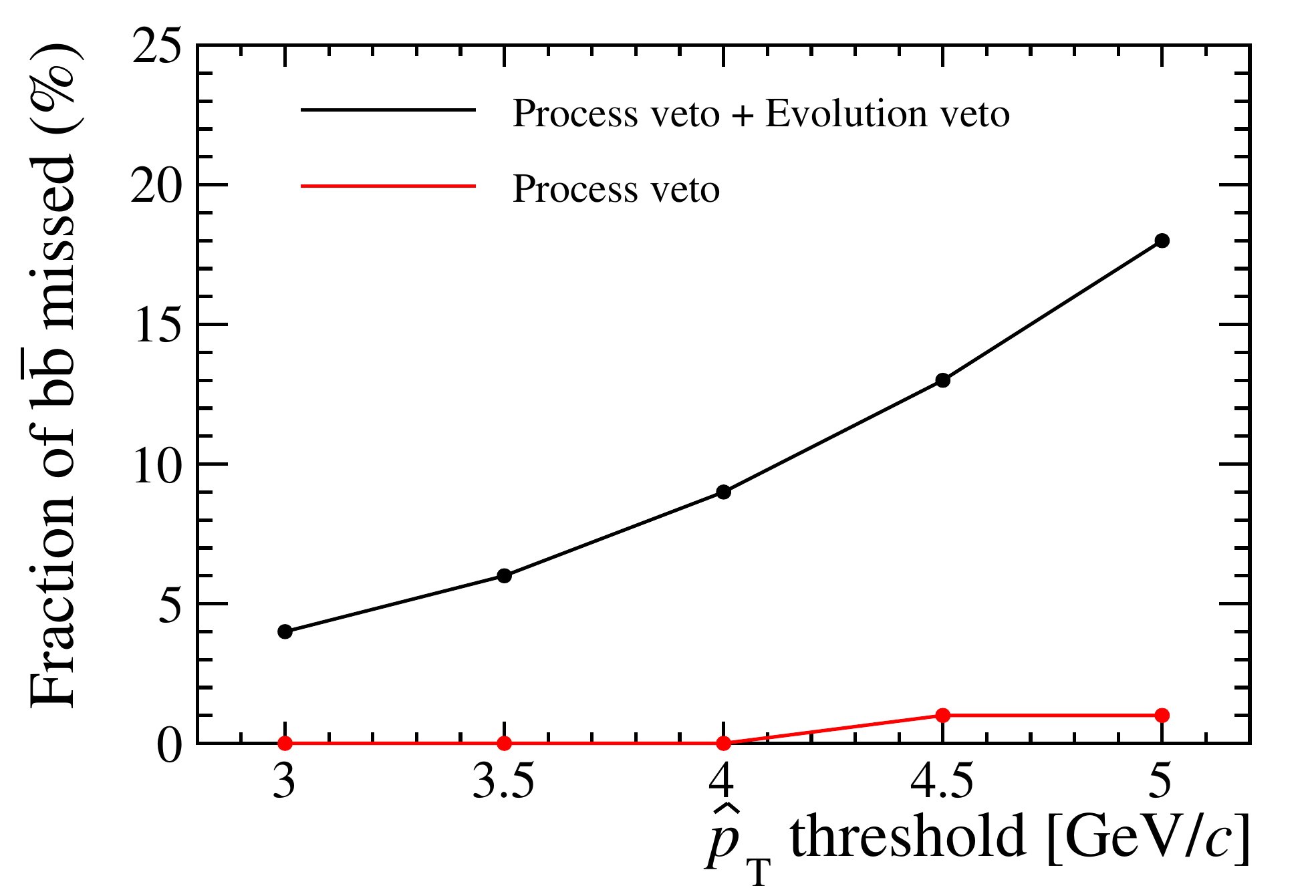}
    \caption{\small (Left) Relative speed enhancements of \pythia when generating $b\bar{b}$ events with process-level and evolution-level \texttt{UserHooks}. (Right) The fractions of $b\bar{b}$ events that are not retained by the \texttt{UserHooks} as a function of $\hat{p}_T$ scale.}
    \label{fig:Enhancements_bb}
\end{figure}

\noindent The improvement in efficiency when generating samples with these two \texttt{UserHooks} is investigated for samples of events containing $b\bar{b}$ or $c\bar{c}$. The time taken to generate the $Q\bar{Q}$ pairs is compared to a baseline without the \texttt{UserHooks} included. All timing tests are performed using an Apple M1 MacBook Pro.\footnote{The timing studies were performed using single-core jobs. Benchmarking tests suggest in this configuration the machine has a CPU power of approximately 44 HS06.} The relative speed-up and fraction of events missed due to the evolution scale definition are shown for $b\bar{b}$ pairs in Fig.~\ref{fig:Enhancements_bb}.
 A significant improvement in efficiency is found when generating $b\bar{b}$ pairs with the \texttt{UserHooks}. The improvement is less significant when generating $c\bar{c}$ pairs because the smaller $c$-quark mass means the event evolution must continue further before the event can be vetoed.  

\begin{figure}[tbp]
    \centering
    \includegraphics[width=0.49\linewidth]{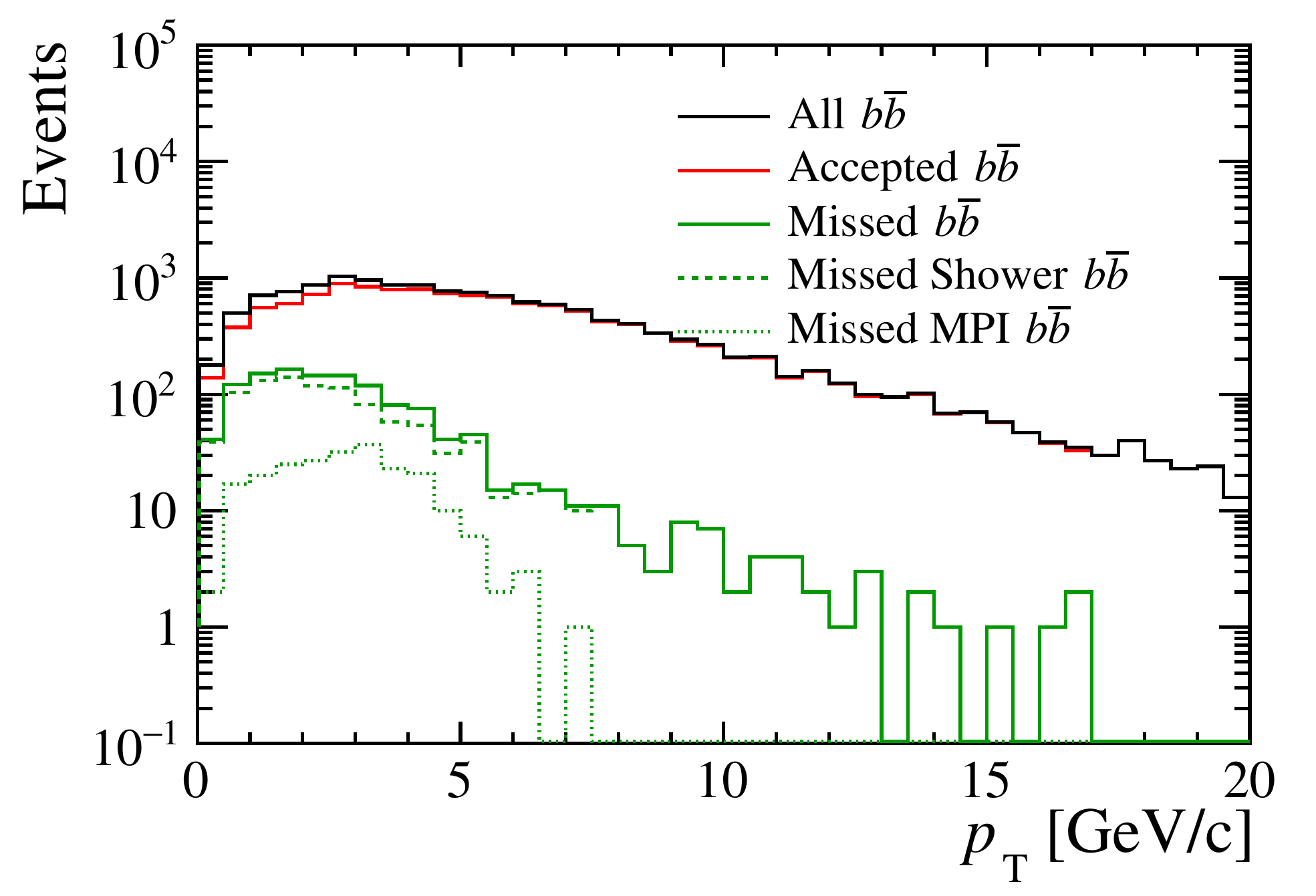}
    \caption{Kinematic distribution of $b$-hadrons in \pythia samples generated with the Simple Shower model. Those that are accepted or missed by the \texttt{UserHooks} discussed in the text with a $\hat{p}_T$ scale of $4\gev$ are highlighted and further split according to whether the missed heavy quarks originated during the parton shower or as an additional MPI process.}
    \label{fig:pt_missing_bb}
\end{figure}

The $p_T$ distribution of $B$ hadrons in events that are not retained by the \texttt{UserHooks} are shown in Fig.~\ref{fig:pt_missing_bb}. This sample, produced with the Simple Shower model misses $\bquark\bquarkbar$ pairs produced in both the parton shower and as additional MPI interactions. Overall, setting a $\hat{p}_T$ scale of $4\gev$ gives a factor 10 improvement in simulation speed, but leads to a small distortion in the $p_T$ spectra of the generated $b$ hadrons.

The impact of these efficiency improvements can further contextualised in terms of the typical time taken to generate specific singly-heavy hadrons with \pythia. The typical times with and without the developed \texttt{UserHooks} are listed in Table~\ref{tab:timing_hadrons_bb}.

\begin{table}[h]
    \centering
    \begin{tabular}[t]{c|c | c}
    \hline \hline 
     Hadron  &  No \texttt{UH} & With \texttt{UH} \\
    \hline     
     $B^{+}$           &  0.15  & 0.013 \\
     $B^{0}$           &  0.15  & 0.013 \\
     $B_{s}^{0}$       &  0.55  & 0.045 \\
     $\Lambda_{b}^{0}$ &  1.5  & 0.13 \\           
     $\Sigma_{b}^{-}$  &  27  & 1.9 \\           
     $\Sigma_{b}^0$    &  26  & 2.3 \\           
     $\Sigma_b^+$      &  26  & 2.3 \\           
     $\Xi_b^-$         &  11  & 0.9 \\          
     $\Xi_b^0$         &  13  & 0.8 \\           
     ${\Xi'}_{b}^{-}$  & 220    & 18 \\           
     ${\Xi'}_b^0$      & 150    & 17 \\       
     $\Omega_b^-$      & 310    & 28 \\
         \hline \hline
    \end{tabular}
    \begin{tabular}[t]{c|c | c}
    \hline \hline 
     Hadron  &  No \texttt{UH} & With \texttt{UH} \\
     \hline 

    $D^0$           &   0.011     &    0.006   \\
    $D^+$           &   0.017     &    0.009   \\
    $D_s^+$         &   0.04      &    0.02    \\
    $\Lambda_c^0$   &   0.11      &    0.06    \\
    $\Sigma_c^{++}$ &   2.2       &    1.2    \\
    $\Sigma_c^{+}$  &   2.0       &    1.0   \\
    &&\\
    $\Xi_c^{+}$     &   0.7       &    0.4 \\
    $\Xi_c^0$       &   0.8       &    0.4 \\
    ${\Xi'}_{c}^{+}$&   11        &      6  \\
    ${\Xi'}_c^0$    &   12        &    9  \\
    $\Omega_c^0$    &   36        &   12   \\
         \hline \hline
    \end{tabular}
    \caption{\small Typical generation times in seconds for single-heavy hadrons with and without the \texttt{Userhooks} (\texttt{UH}) described in the text. The tests are performed using the Simple Shower model. The vetoes are imposed with $\hat{p}_{T}$ cut-off scales of $4.0\gevc$ and $1.5\gevc$ for the $b\bar{b}$ and $c\bar{c}$ samples, respectively. Due to the size of the generated samples the uncertainty in the typical generation time for $\Sigma$, $\Xi$ and $\Omega$ baryons is up to 7\%, 15\% and 30\%, respectively. The corresponding uncertainties for the meson and $\Lambda$ samples are negligible.}
    \label{tab:timing_hadrons_bb}
\end{table}

%%%%%%%%%%%%%%%%%%%%%%%%%%%%%%%%%%%%%%%%%%%%%%%%%%%%%%%%%%%%%%%%%%%%%%%%%%%%
\subsection{Simulating final states involving multiple \boldmath{$Q\bar{Q}$} pairs} 
\label{sec:sim_QQQQ}
%%%%%%%%%%%%%%%%%%%%%%%%%%%%%%%%%%%%%%%%%%%%%%%%%%%%%%%%%%%%%%%%%%%%%%%%%%%%

When simulating events with multiple pairs of heavy quarks, the same \texttt{Userhook} vetoes described previously in Section~\ref{sec:sim_QQ} can be utilised. In principle, for $B_c^+$ production and the like, it would be useful to apply an event-evolution veto first at ${\cal O}(m_b)$ and then again at ${\cal O}(m_c)$. However, with current versions of \pythia, the event can only be inspected at a single value of the evolution scale. Therefore, we set the the event-evolution threshold according to the heaviest quark being simulated, while the end-of-evolution parton-level veto is used to check for any required secondary heavy flavour. 

\begin{description}
\item[Parton-level veto:] If both a $b\bar{b}$ and a $c\bar{c}$ pair is requested (and/or onia containing them), the presence of the lighter of the two flavours is checked for at the end of the parton-level evolution.  
\end{description}
We believe that, by making minor modifications to the \pythia source code it could be possible to allow the event-evolution veto to inspect the event multiple times, at different scales during the event evolution (e.g. at both the $b$ and $c$ quark masses). This would save the time spent evolving the event from the lower mass scale to the hadronisation scale, providing benefits in addition to those demonstrated here. 

%%%%%%%%%%%%%%%%%%%%%%%%%%%%%%%%%%%%%%%%%%%%%%%%%%%%%%%%%%%%%%%%%%%%%%%%%%%%
\subsection{Simulating final states with specific hadrons} 
\label{sec:sim_hadrons}
%%%%%%%%%%%%%%%%%%%%%%%%%%%%%%%%%%%%%%%%%%%%%%%%%%%%%%%%%%%%%%%%%%%%%%%%%%%%
 
\begin{figure}[tbp]
    \centering
    \includegraphics[width=0.6\linewidth]{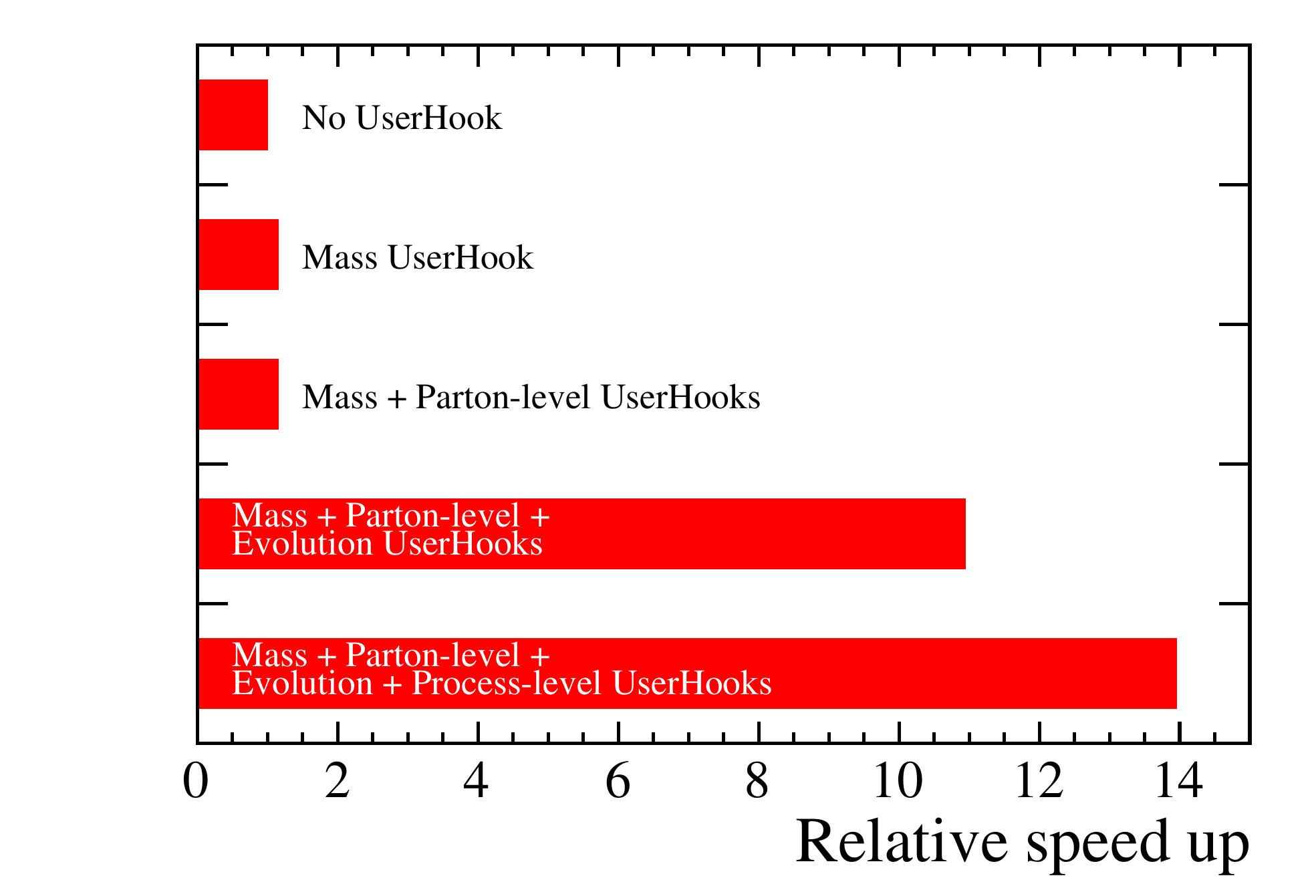}
    \caption{\small Relative speed enhancements of \pythia when generating $B_{c}^{+}$ mesons with the different \texttt{UserHooks} included. Each point includes an additional \texttt{UserHook} as described in the text.}
    \label{fig:Enhancements_Bc}
\end{figure}

Further \texttt{UserHooks} can be placed at the partonic event level to increase the efficiency of generating specific hadrons.  
For example, when generating $B_c^+$ mesons, the invariant mass of all $c$ and $\bar{b}$ pairs (and vice versa) can be calculated to veto events in which there are no combinations that have a small enough invariant mass to form a hadron.  The timing improvements when using the additional levels of \texttt{UserHook} are shown in Fig.~\ref{fig:Enhancements_Bc}, where the first \texttt{UserHook} to be included requires the smallest invariant mass of any $b\bar{c}$ or $\bar{b}c$ pair to be below $10\gevcc$, the second requires the partonic-level event to contain both $b\bar{b}$ and $c\bar{c}$ pairs, and the third and fourth correspond to the hard process and event evolution vetoes already described, here applied just to the $b$ quark. 

Using combinations of the \texttt{UserHooks} described above it is possible to improve the efficiency of generating both singly-heavy and doubly-heavy hadrons with \pythia. Typical times to generate doubly-heavy hadrons are listed in Table~\ref{tab:times_Bc}. 

The \texttt{UserHooks} that inspect the event at the parton-level or after do not result in any further missed hadrons, in contrast to the hard-process level and event-evolution vetoes discussed in Section~\ref{sec:sim_QQ}. Therefore when generating doubly-heavy hadron with these extra requirements, the fraction and kinematic distributions of the missed hadrons will be the same as for singly-heavy hadrons. 

\begin{table}[h]
    \centering
    \begin{tabular}{c c c c c }
         \hline \hline 
         Hadron & \texttt{UserHook} requirement & No \texttt{UH} & With \texttt{UH}  & Speed-up\\
         \hline
         \Bcp & $\bquark\bquarkbar\cquark\cquarkbar $ and $m(\bquark\cquarkbar)<10\gevcc$& 47 & 3.4 & $\sim14$\\ 
         \hline
         \Xiccpp  & $\cquark\cquarkbar\cquark\cquarkbar $ & 84 & 37 &  \\ 
         \Xiccp          & $\cquark\cquarkbar\cquark\cquarkbar $ & 78 & 37& $\sim 2$\\ 
         $\Omega_{cc}^0$  & $\cquark\cquarkbar\cquark\cquarkbar $ & 620 & 260& \\ 
         \hline
         $\Xi_{bc}^{+}$  &$\bquark\bquarkbar\cquark\cquarkbar $ & 830 & 36 & \\
         $\Xi_{bc}^{0}$  &$\bquark\bquarkbar\cquark\cquarkbar $ & 900 & 41 & $\sim 20$\\
         $\Omega_{bc}^{0}$  &$\bquark\bquarkbar\cquark\cquarkbar $ & 3300 &  300 &  \\
         \hline \hline
    \end{tabular}
    \caption{Typical generation times in seconds and the approximate speed-up factors for doubly-heavy hadrons with and without \texttt{UserHooks}. The minimal quark requirements used for each  \texttt{UserHook} are included, along with any additional requirements. Note, the baryonic samples are generated with a different colour reconnection model described later in Section~\ref{sec:Xiccpp_kinematics}. For comparison, generating \Bcp mesons using \bcvegpy takes approximately 4\ms  seconds to generate the hard process and a further 4\ms to perform the full event evolution with \pythia.}
    \label{tab:times_Bc}
\end{table}

%%%%%%%%%%%%%%%%%%%%%%%%%%%%%%%%%%%%%%%%%%%%%%%%%%%%%%%%%%%%%%%%%%%%
\section{Comparison of doubly-heavy hadron kinematics}
\label{sec:single_particle_kinematics}
%%%%%%%%%%%%%%%%%%%%%%%%%%%%%%%%%%%%%%%%%%%%%%%%%%%%%%%%%%%%%%%%%%%%
The efficiency improvements detailed in the previous section enable a range of comparisons that were previously unfeasible. For example, 250\,000 \Bcp mesons have been generated for this study which would take approximately 10 CPU days with the \texttt{UserHooks}, or 140 CPU days without.      
In this section distributions of doubly-heavy hadrons generated with \pythia are compared to other standalone generators.

%%%%%%%%%%%%%%%%%%%%%%%%%%%%%%%%%%
\subsection{\boldmath{$B_{c}^{+}$} mesons}
%%%%%%%%%%%%%%%%%%%%%%%%%%%%%%%%%%

The production of \Bc mesons at hadron colliders is generally calculated in terms of the SPS process $\decay{gg}{\Bc  \bquark \cquarkbar}$ that is assumed to dominate~\cite{Chang:1994aw,Berezhnoy:1994ba,Kolodziej:1995nv}, although there are some suggestions that flavour excitation processes could play an important role~\cite{Chang:2005wd}. Currently, DPS production has not been considered. 
Measurements of the ratio of the \Bc to the 
\Bp cross-section times branching fraction are found to be consistent with predictions of SPS contributions only~\cite{LHCb-PAPER-2014-050}, although the theoretical predictions have large uncertainties.  In contrast, measurements of $\Upsilon D$ cross section are consistent with the presence of DPS~\cite{LHCb:2015wvu}. The $\Upsilon D $ system also requires the creation of a $\bquark\bquarkbar$ and a $\cquark\cquarkbar$ pair, implying that there is the potential for \Bc mesons to be similarly produced.

Samples of \Bc mesons are generated with \pythia and \bcvegpy. The specific generation settings are listed in Table~\ref{tab:generator_settings_Bc}.

\begin{table}[h]
    \centering
    \begin{tabular}[t]{l|c}
    \hline \hline 
    \pythia option     &  Setting \\
    \hline
    Beams:eCM    & 13000 \\
    SoftQCD:nonDiffractive & on \\
    PartonShowers:Model &  1\\   
    PartonLevel:MPI & on\\
    \hline \hline 
    \end{tabular}
    \begin{tabular}[t]{l|c}
    \hline \hline 
    \bcvegpy option     &  Setting \\
    \hline
   \texttt{ENERGYOFLHC}    &13000\\
   naccel       & 2 \\
   i\_mix       & 1\\
   mix\_type    & 1\\
   igenmode     & 1\\
   istate       & 1\\
   numofevents  & 100000\\
   ncall        & 10000\\
   n\_itmx      & 15\\
   $m(\bquark)$  & 5\gev\\
    $m(\cquark)$ & 1.275\gev \\
    \hline \hline 
    \end{tabular}
    \caption{\pythia and \bcvegpy generation settings used to simulate the samples of \Bc mesons. The specific meanings of the parameters can be found in the relevant documentation~\cite{Bierlich:2022pfr,Chang:2015qea}. Additionally, in \pythia all ground-state heavy hadrons are treated as stable. As such, contributions from $X_{c}$ hadrons originating from $X_{b}$ decays are explicitly excluded from the samples.   }
    \label{tab:generator_settings_Bc}
\end{table}

\begin{figure}[h]
    \centering

    \includegraphics[width=\plotwidths\linewidth]{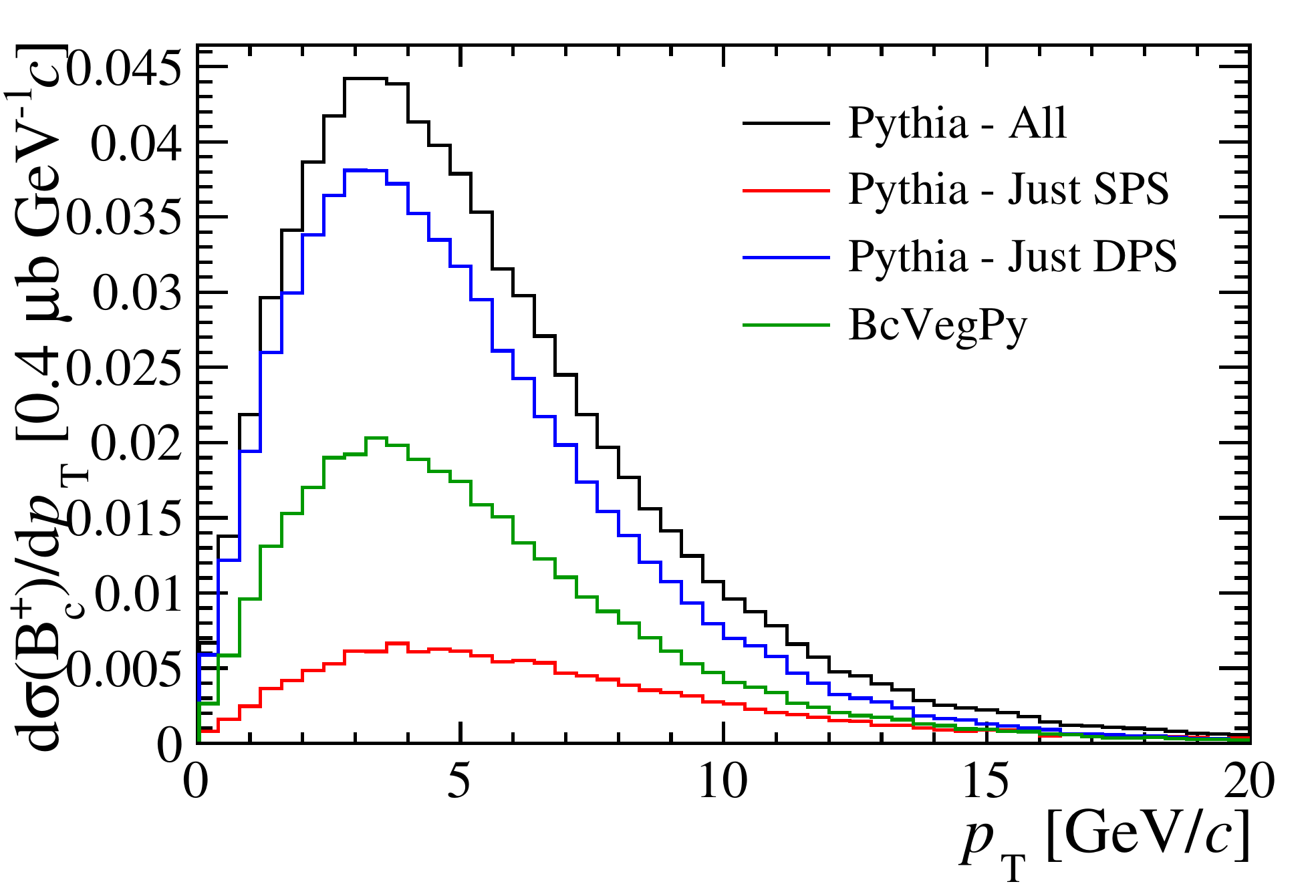}
    \includegraphics[width=\plotwidths\linewidth]{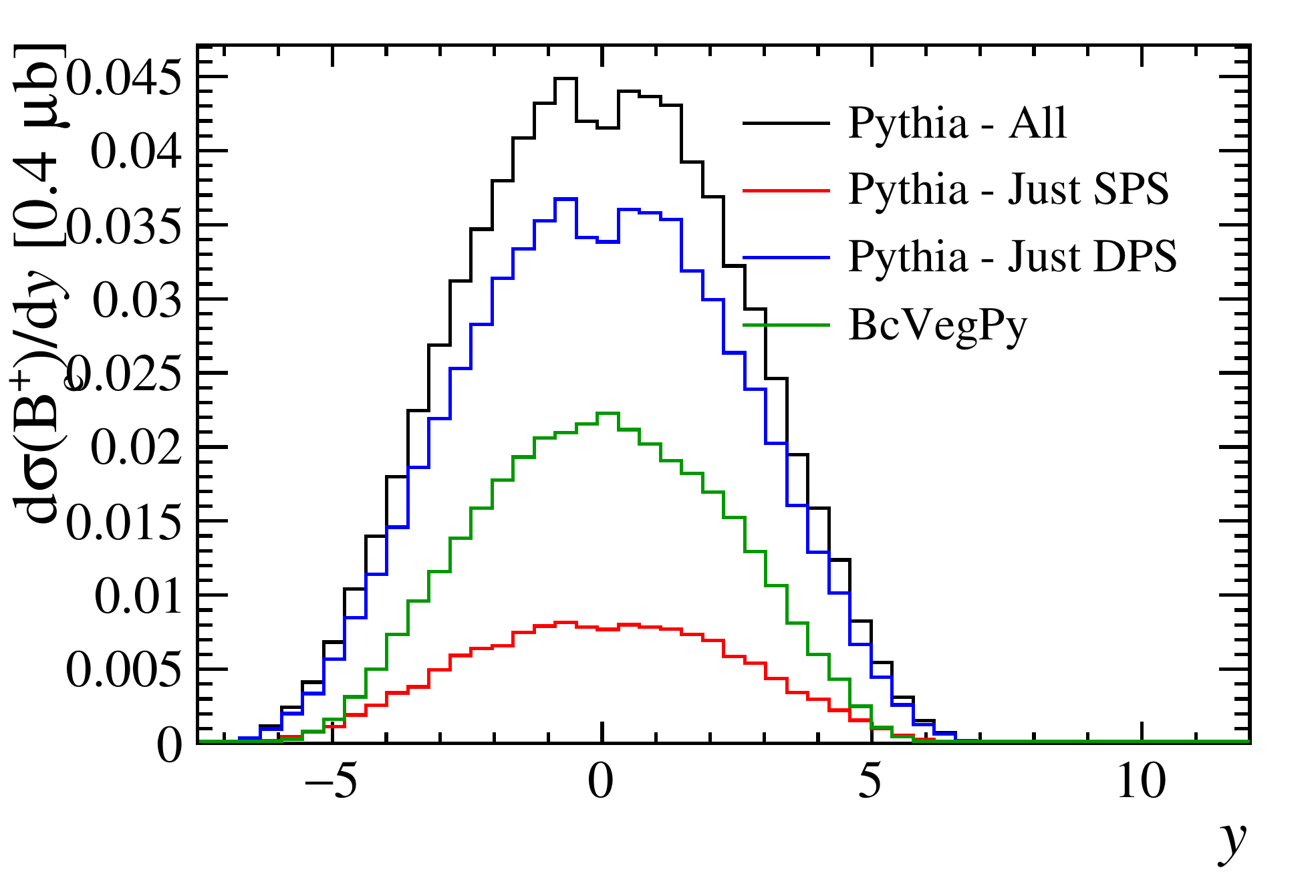}
    \caption{Kinematic distributions of $B_{c}^{+}$ mesons generated with \bcvegpy and \pythia.}
    \label{fig:kinematics_Bc}
\end{figure}

The kinematic distributions and differential cross-sections are compared in Fig.~\ref{fig:kinematics_Bc}. The sample generated with \pythia is split according to the origin of the \cquark-quark and \bquarkbar-quark that formed the \Bcp meson: those that originated in the same parton-parton interaction are categorised as SPS and those from different  parton-parton interactions are categorised as DPS. This is determined by comparing the parent quarks to the record of parton interactions stored in \pythia's \texttt{PartonSystems} class. 
The production cross-section predicted by \pythia receives a significant contribution from DPS and is found to be larger than the production cross-section obtained from \bcvegpy. 

The kinematic distributions of the \Bcp alone do not provide significant discrimination power to the presence of hadron formation in DPS, other than in the absolute normalisation of the cross section. Measurements of the \Bcp cross section  relative to \Bp mesons have been performed~\cite{LHCb-PAPER-2014-050,LHCb:2019tea,CMS:2018ynq}, but extracting the absolute cross section requires theoretical predictions of the branching fractions. 

A better way to investigate the role of SPS and DPS contributions is to measure the production cross section as a function of the number of parton-parton interactions in a collision.  
For hadrons formed in SPS processes, increasing the number of parton interactions would linearly increase the number of opportunities to form the hadron, as each new parton interaction would present one more opportunity for the hadron to form. However, hadrons formed in DPS processes would see the rate of formation increase quadratically with the number of interactions, as each hadron requires two parton interactions to form. 
These different relationships can be exploited to differentiate the components by considering the ratio of doubly-heavy to singly-heavy hadron cross sections, as a function of the number of parton-parton interactions. This ratio would be flat if singly- and doubly-heavy hadrons are produced by the same mechanism --- SPS --- while it would increase linearly if there is a nontrivial DPS component to doubly-heavy hadron production. 
\begin{figure}[tbp]
    \centering

    \includegraphics[width=\plotwidths\linewidth]{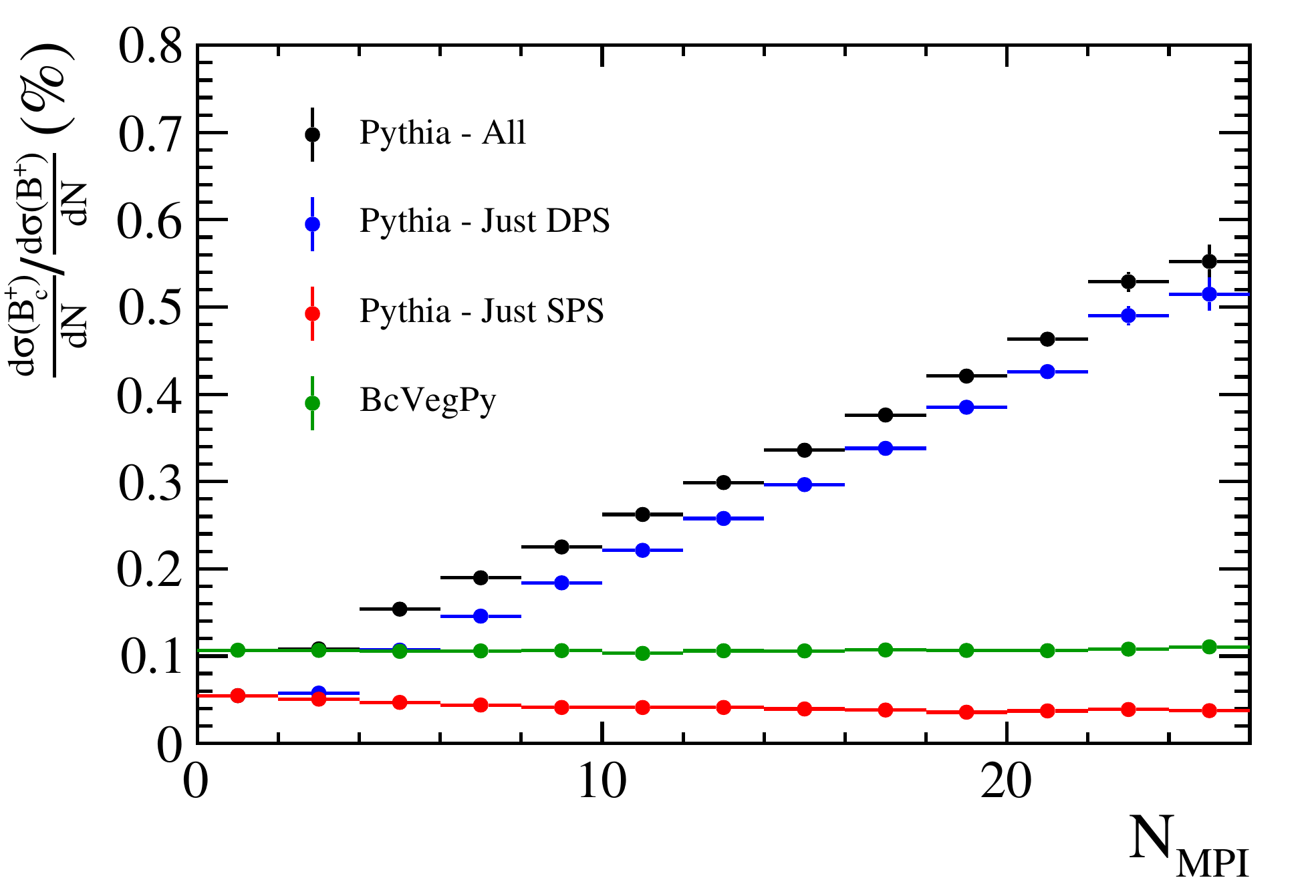}
    \includegraphics[width=\plotwidths\linewidth]{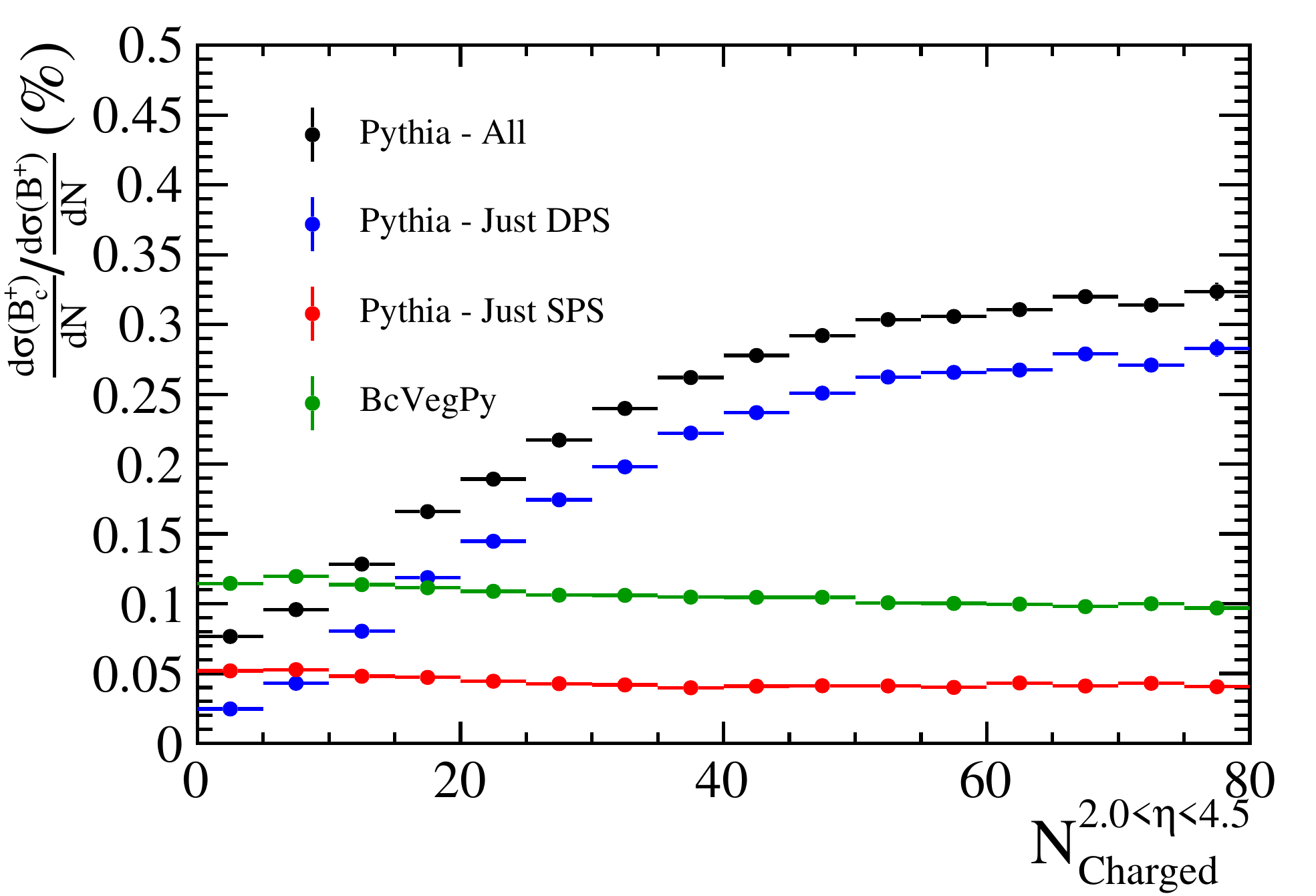}\\
    
    \caption{Ratio of differential cross-sections of $B_{c}^{+}$ and \Bp mesons as a function of (top left) the number of parton-parton interactions in a collision and (top right) the number of charged particles within the pseudo-rapidity region $2.0<\eta<4.5$, as generated with \bcvegpy and \pythia. Uncertainties are from simulation statistics only.}
    \label{fig:multiplicity_Bc}
\end{figure}

In \pythia, both mechanisms are present, while in \bcvegpy, a single $gg\to \Bcp \bquark \cquarkbar$ interaction is produced for each event which is then  passed to \pythia for showering, MPI, and hadronisation. In this case, there is therefore no opportunity for heavy quarks from different parton-parton interactions to form the \Bcp meson and the production is independent of the total number of parton-parton interactions. 

The cross-section ratio of \Bcp to \Bp mesons is compared for \pythia and \bcvegpy in Fig.~\ref{fig:multiplicity_Bc} as a function of the number of parton-parton interactions. In this figure no kinematic requirements have been placed on the rapidity or transverse momentum of the \Bcp meson or final-state particles. As expected, the contribution from DPS varies as a function of the number of parton interactions in the event. A significant enhancement is seen in events with many parton interactions.
This quantity is not directly observable, but it is highly correlated to the number of particles in a collision. A more realistic, experimentally-accessible distribution is therefore the cross-section ratio as a function of the number of charged particles within a given  acceptance. In the right pane of Fig.~\ref{fig:multiplicity_Bc} and throughout we use the acceptance of the LHCb experiment ($2.0<\eta<4.5$), but other ranges would show similar results. 

\begin{figure}[tbp]
    \centering

    \includegraphics[width=\plotwidths\linewidth]{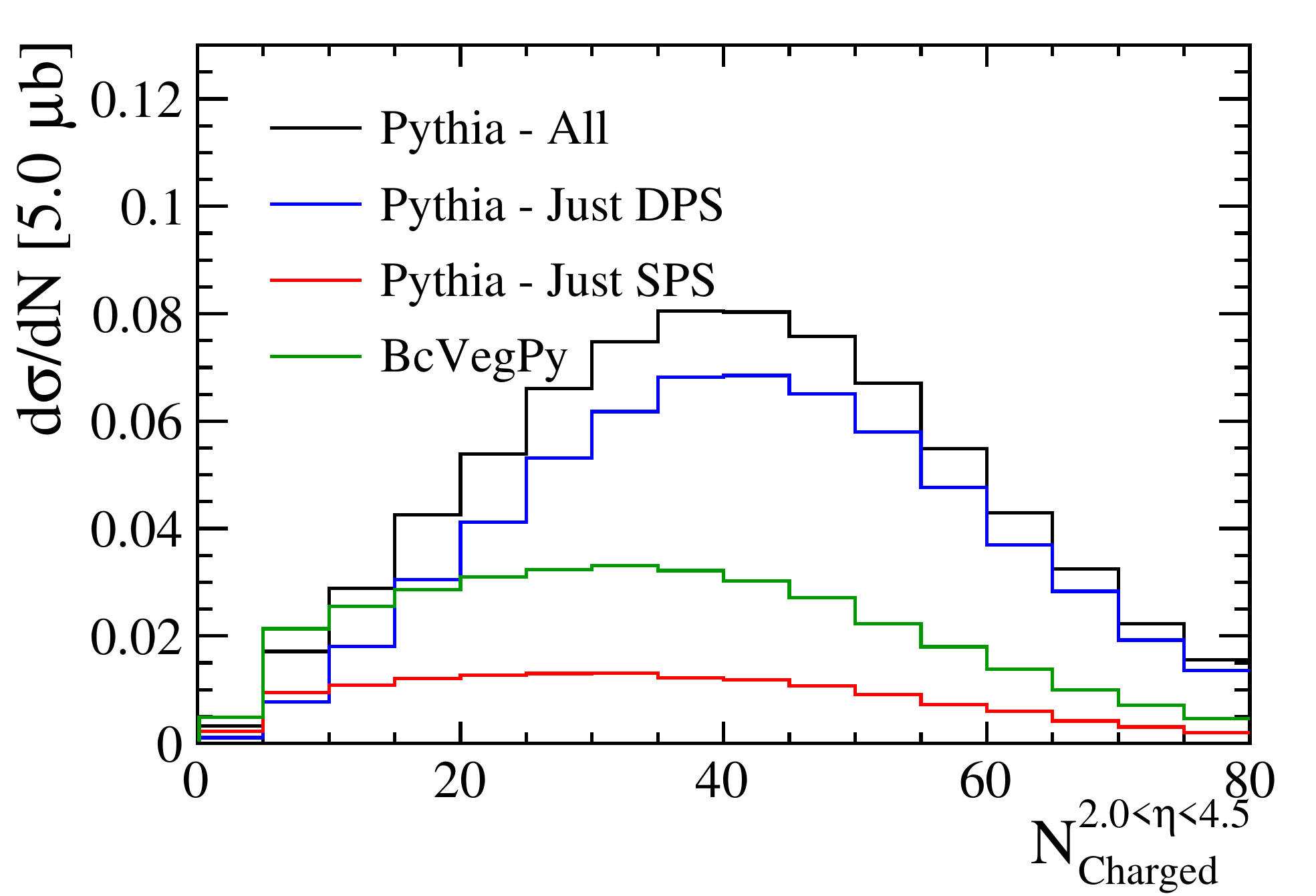}\\
    
    \caption{ Distribution of the number of charged particles within $2.0<\eta<4.5$ in events containing a \Bcp meson as generated with \bcvegpy and \pythia.}
    \label{fig:multiplicity_Bc_nTracks}
\end{figure}
The distribution of the number of charged particles within $2.0<\eta<4.5$ in events containing a \Bcp meson is shown in Fig.~\ref{fig:multiplicity_Bc_nTracks} for the different generation configurations. In the \pythia SPS and \bcvegpy samples there  are means of approximately 34 charged particles per event, whilst the \pythia DPS sample has a higher mean of around 41.   

The significant difference in these distributions would allow measurements of the differential cross section with respect to \Bp mesons to differentiate between \Bcp mesons produced in SPS and DPS.

%%%%%%%%%%%%%%%%%%%%%%%%%%%%%%%%%%
\subsection{\boldmath{$\Xi_{cc}^{++}$} baryons}
\label{sec:Xiccpp_kinematics}
\begin{figure}[tpb]
    \centering
    \includegraphics[width=\plotwidths\linewidth]{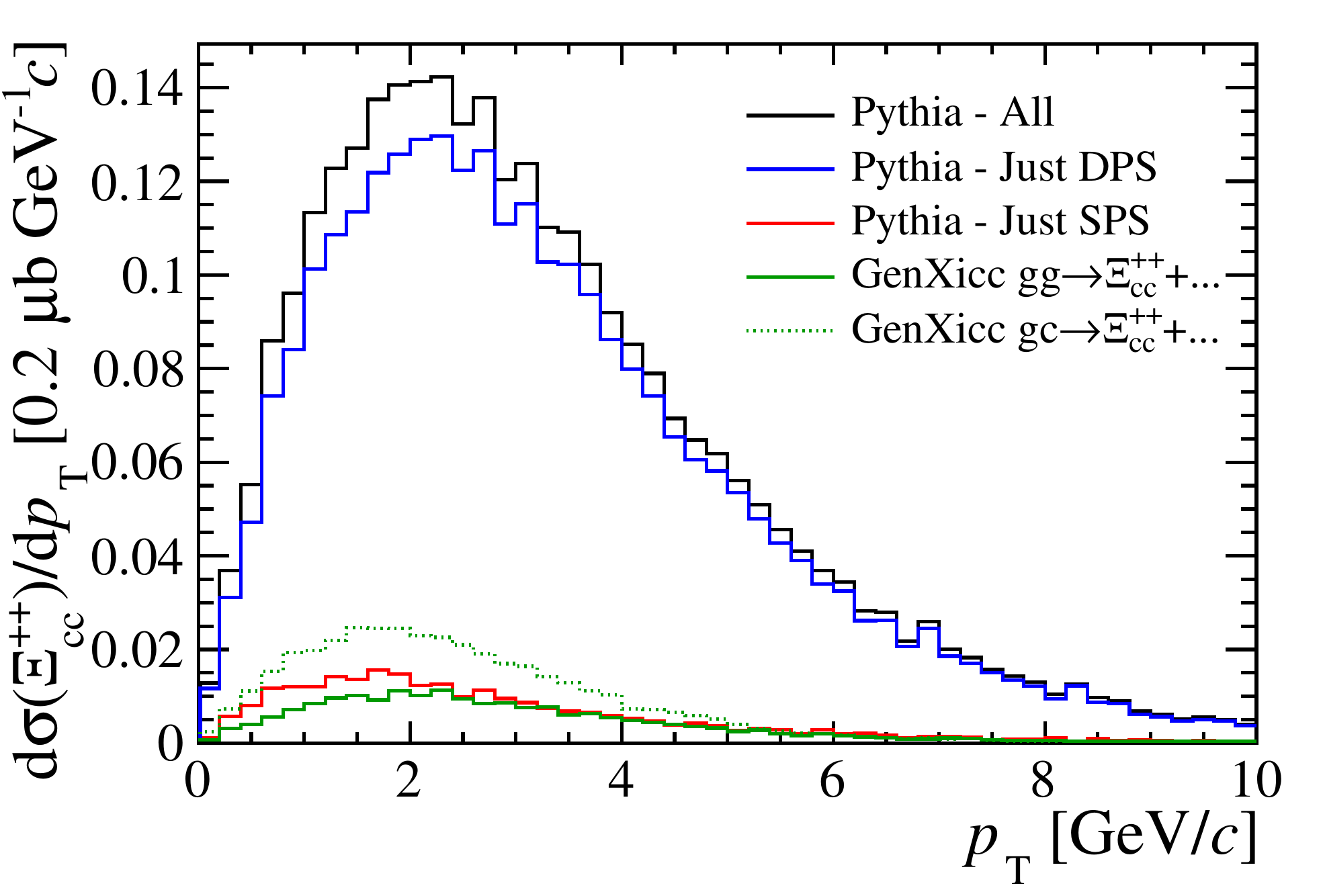}
    \includegraphics[width=\plotwidths\linewidth]{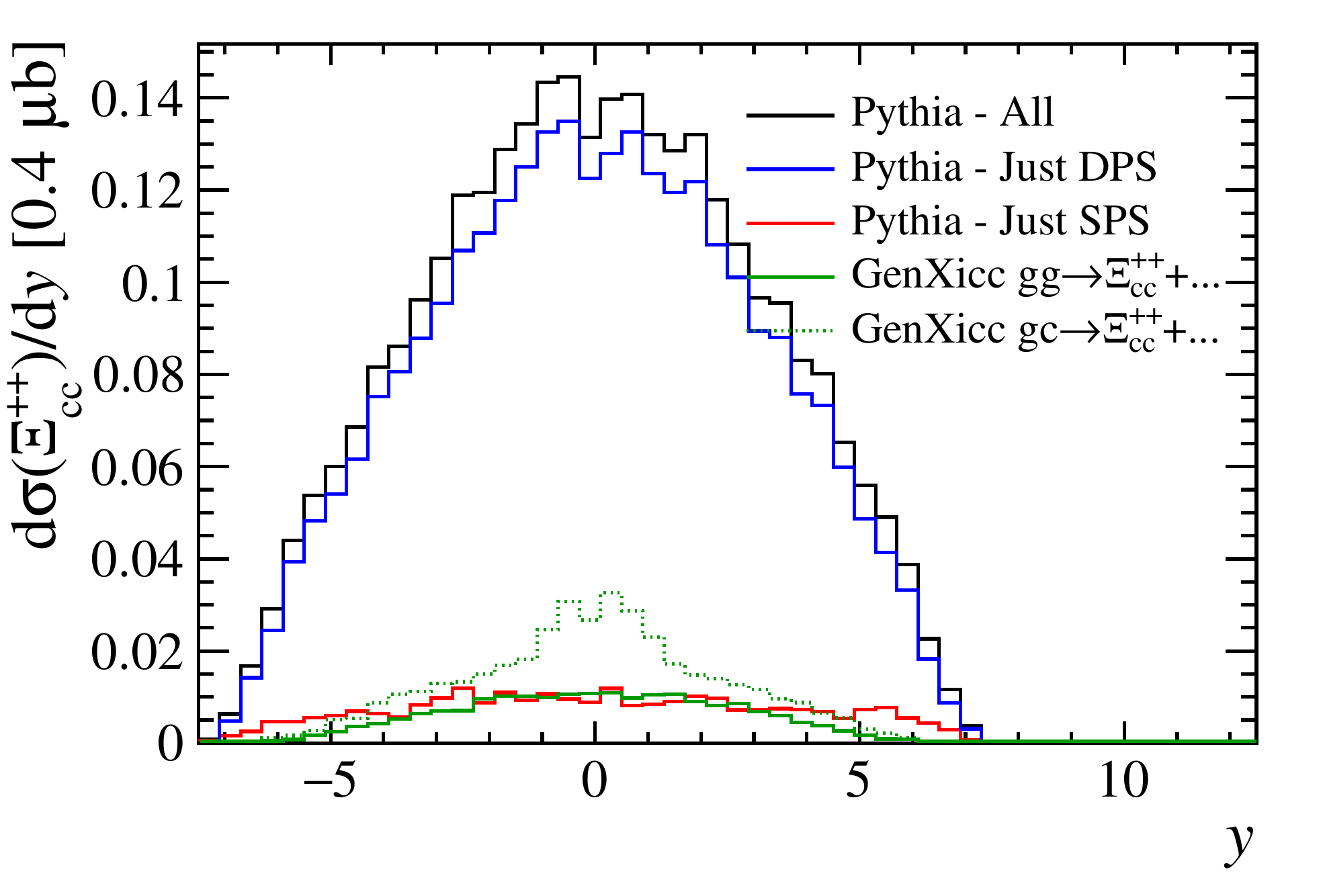}
    \caption{Kinematic distributions of $\Xi_{cc}^{++}$ baryons generated with \pythia and \genxicc.}
    \label{fig:kinematics_Xicc}
\end{figure}

%%%%%%%%%%%%%%%%%%%%%%%%%%%%%%%%%%
The doubly-charmed baryon \Xiccpp has been observed~\cite{SELEX:2002wqn,LHCb-PAPER-2017-018}, and could similarly receive contributions from DPS production mechanisms.
Samples of $\Xi_{cc}^{++}$ baryons are simulated with both \pythia and \genxicc. 
In \pythia, the formation of baryons with multiple heavy quarks is only possible with the so-called QCD colour-reconnection option~\cite{Christiansen:2015yqa} which allows for the formation of doubly-heavy diquarks via string junctions~\cite{Sjostrand:2002ip}.  The specific \pythia settings are listed in Table~\ref{tab:pythia_settings_Xicc}. 
 \begin{table}[h]
    \centering
    \begin{tabular}{l|c}
    \hline \hline 
    \pythia option     &  Setting \\
    \hline
    Beams:eCM    & 13000 \\
    SoftQCD:nonDiffractive & on \\
    PartonShowers:Model &  1\\   
    PartonLevel:MPI & on\\
    \hline 
StringPT:sigma                        &  0.335                       \\
StringZ:aLund                         &  0.36                        \\
StringZ:bLund                         &  0.56                        \\
StringFlav:probQQtoQ                  &  0.078                       \\
StringFlav:ProbStoUD                  &  0.2                         \\
StringFlav:probQQ1toQQ0join           &  0.0275, 0.0275, 0.0275, 0.0275 \\
MultiPartonInteractions:pT0Ref        &  2.15                        \\
BeamRemnants:remnantMode              &  1                           \\
BeamRemnants:saturation               &  5                           \\
ColourReconnection:mode               &  1                           \\
ColourReconnection:allowDoubleJunRem  &  off                         \\
ColourReconnection:m0                 &  0.3                         \\
ColourReconnection:allowJunctions     &  on                          \\
ColourReconnection:junctionCorrection &  1.20                        \\
ColourReconnection:timeDilationMode   &  2                           \\
ColourReconnection:timeDilationPar    &  0.18                        \\
    \hline \hline 
    \end{tabular}
    \caption{\pythia generation settings used to simulate the samples of \Xiccpp mesons.}
    \label{tab:pythia_settings_Xicc}
\end{table}
The \genxicc generator can produce different configurations of the initial charm hadrons. The processes $\decay{gg}{\Xiccpp \cquarkbar  \cquarkbar }$ and  $\decay{gc}{\Xiccpp \cquarkbar }$ are simulated. 
The kinematic distributions of the \Xiccpp samples are shown in Fig~\ref{fig:kinematics_Xicc}, where again the \pythia samples have been split according to the whether the two heavy quarks originated from the same or different parton-parton interactions.  Analogously to the case for \Bcp, a significantly larger cross section is predicted by \pythia, due to the presence of the DPS component.  

\begin{figure}[tbp]
    \centering
    \includegraphics[width=\plotwidths\linewidth]{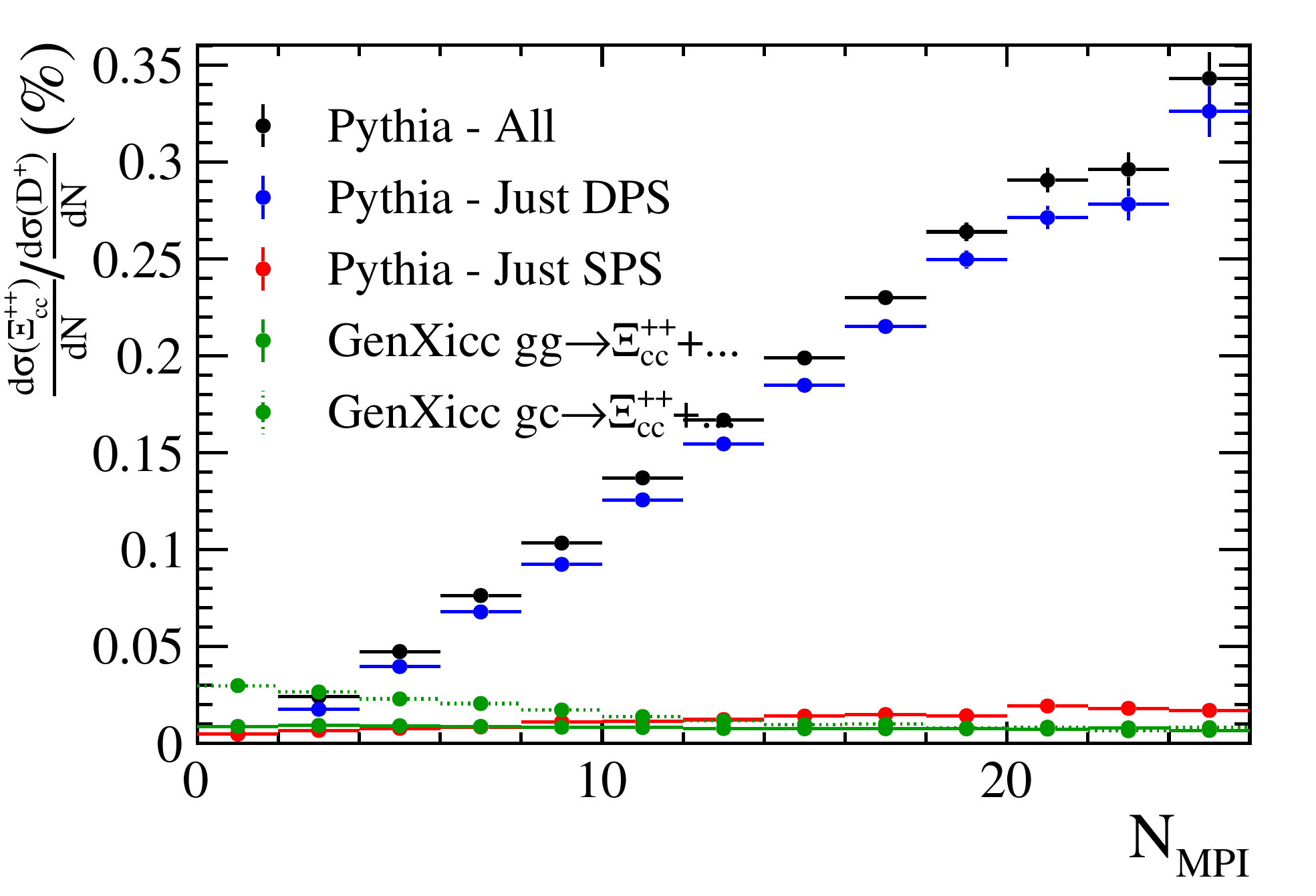}
    \includegraphics[width=\plotwidths\linewidth]{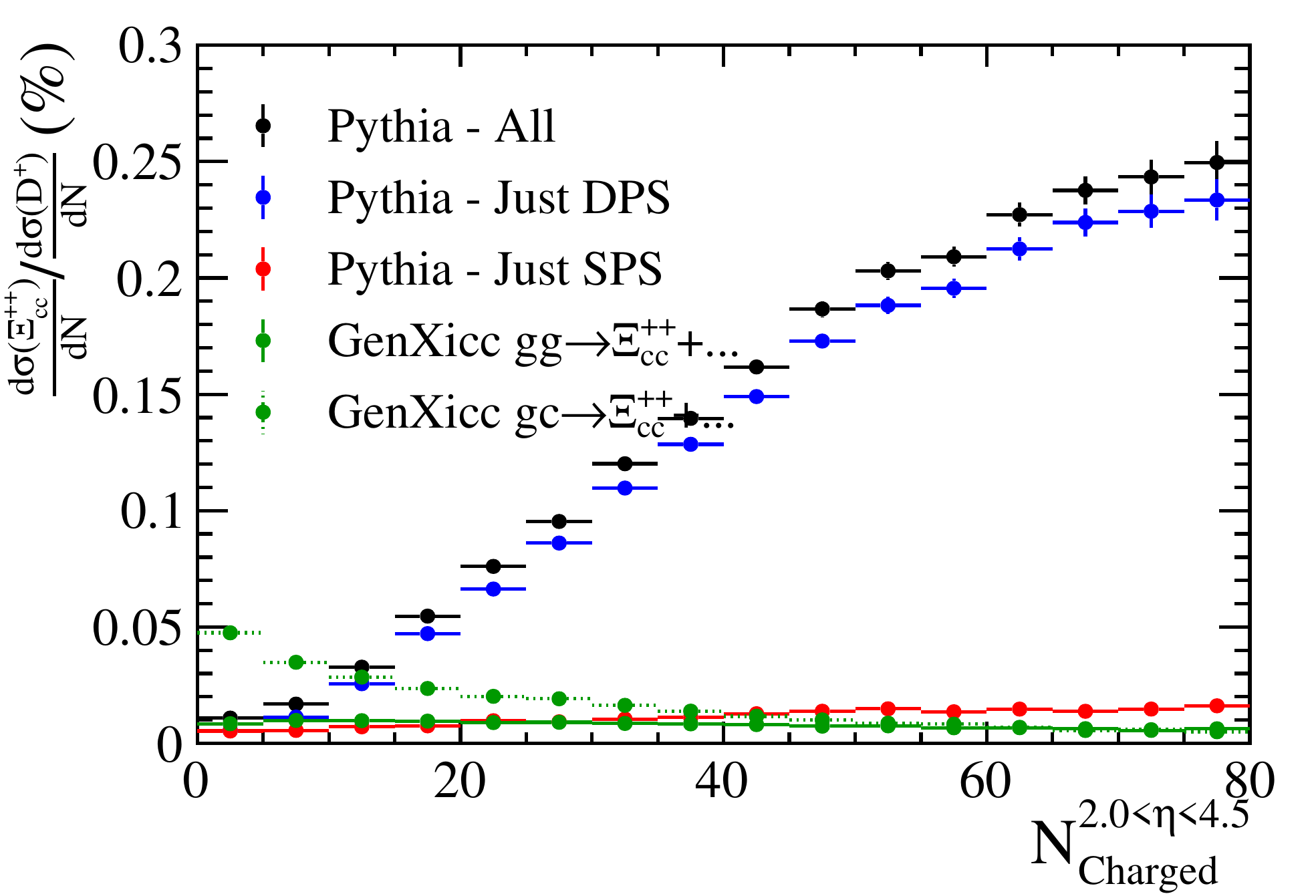}
    \caption{Ratio of differential cross-section of \Xiccpp and \Dp hadrons as a function of (left) the number of parton interactions in a collision and (right) the number of charged particles within the pseudo-rapidity region $2.0<\eta<4.5$, as generated with \genxicc and \pythia.}
    \label{fig:multiplicity_Xicc}
\end{figure}

The ratio of the differential cross-section between the \Xiccpp and \Dp hadrons are shown as a function of the number of parton interactions and number of charged particles within the pseudo-rapidity region $2.0<\eta<4.5$ in Fig.~\ref{fig:multiplicity_Xicc}. The cross-section ratio is clearly able to differentiate between DPS and SPS production mechanisms. The difference in the cross-section ratios for the $\decay{gg}{\Xiccpp \cquarkbar  \cquarkbar }$ and  $\decay{gc}{\Xiccpp \cquarkbar }$ processes simulated by \genxicc may be caused by the different typical momentum transfers in the two processes. When simulating the rest of the underlying event with \pythia,  this quantity determines the maximum scale of the event evolution, therefore the typically lower momentum transfers in $\decay{gc}{\Xiccpp \cquarkbar }$ processes may cause fewer parton-parton interactions to be generated in the subsequent event evolution.

%%%%%%%%%%%%%%%%%%%%%%%%%%%%%%%%%%
\subsection{\boldmath{$\Xi_{bc}^{+}$} baryons}
\begin{figure}[tpb]
    \centering
    \includegraphics[width=\plotwidths\linewidth]{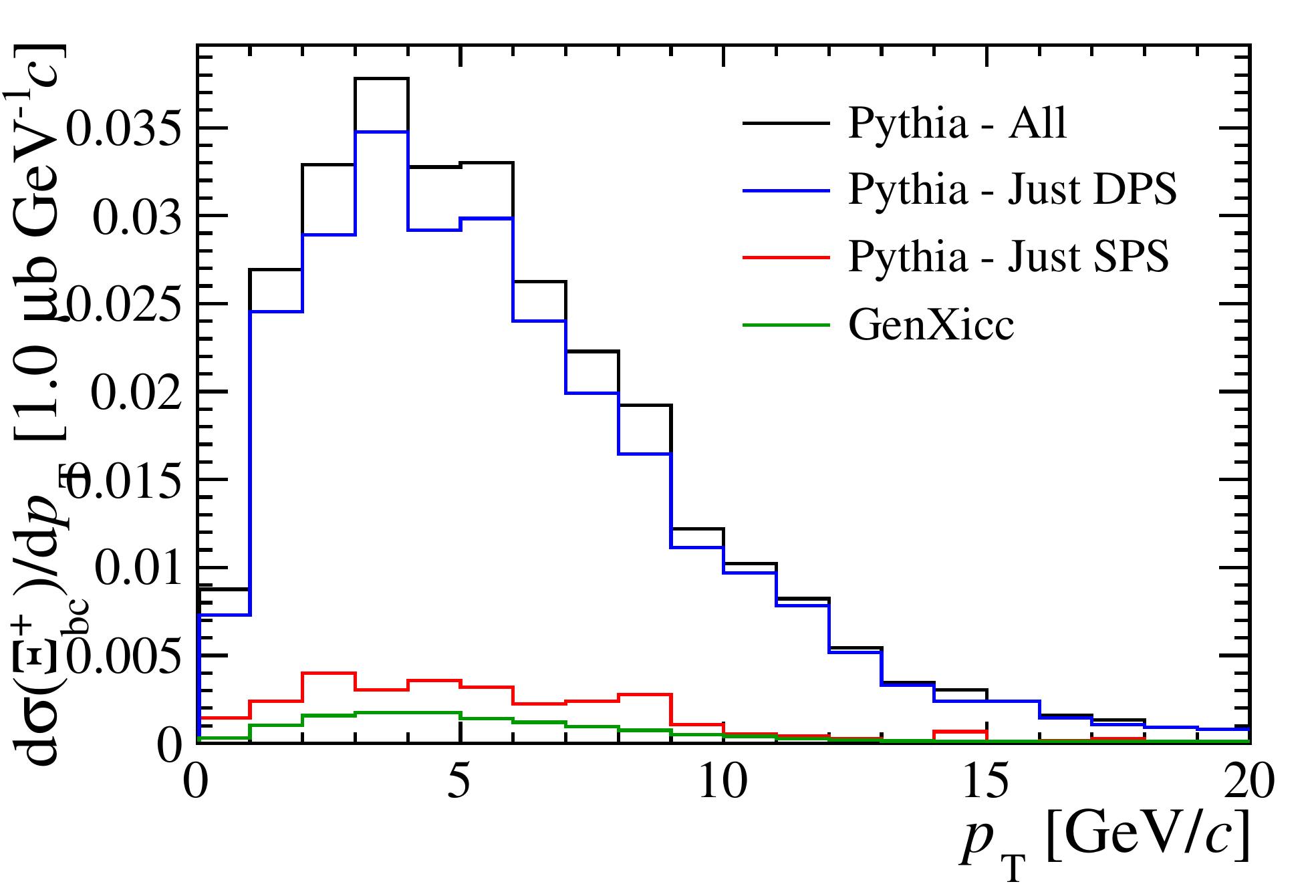}
    \includegraphics[width=\plotwidths\linewidth]{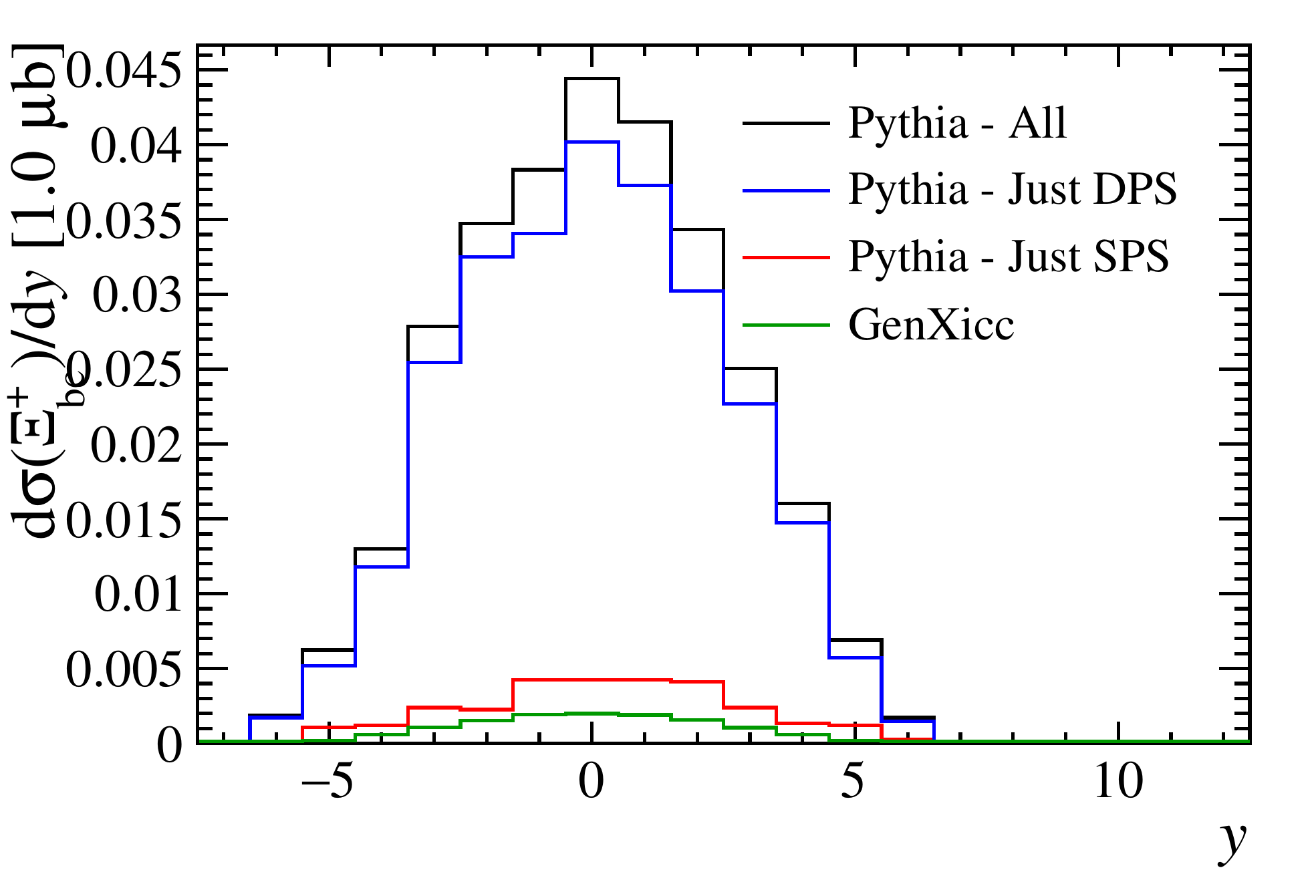}
    \caption{Kinematic distributions of $\Xi_{bc}^{+}$ baryons generated with \pythia and \genxicc.}
    \label{fig:kinematics_Xibc}
\end{figure}

\begin{figure}[tbp]
    \centering
    \includegraphics[width=\plotwidths\linewidth]{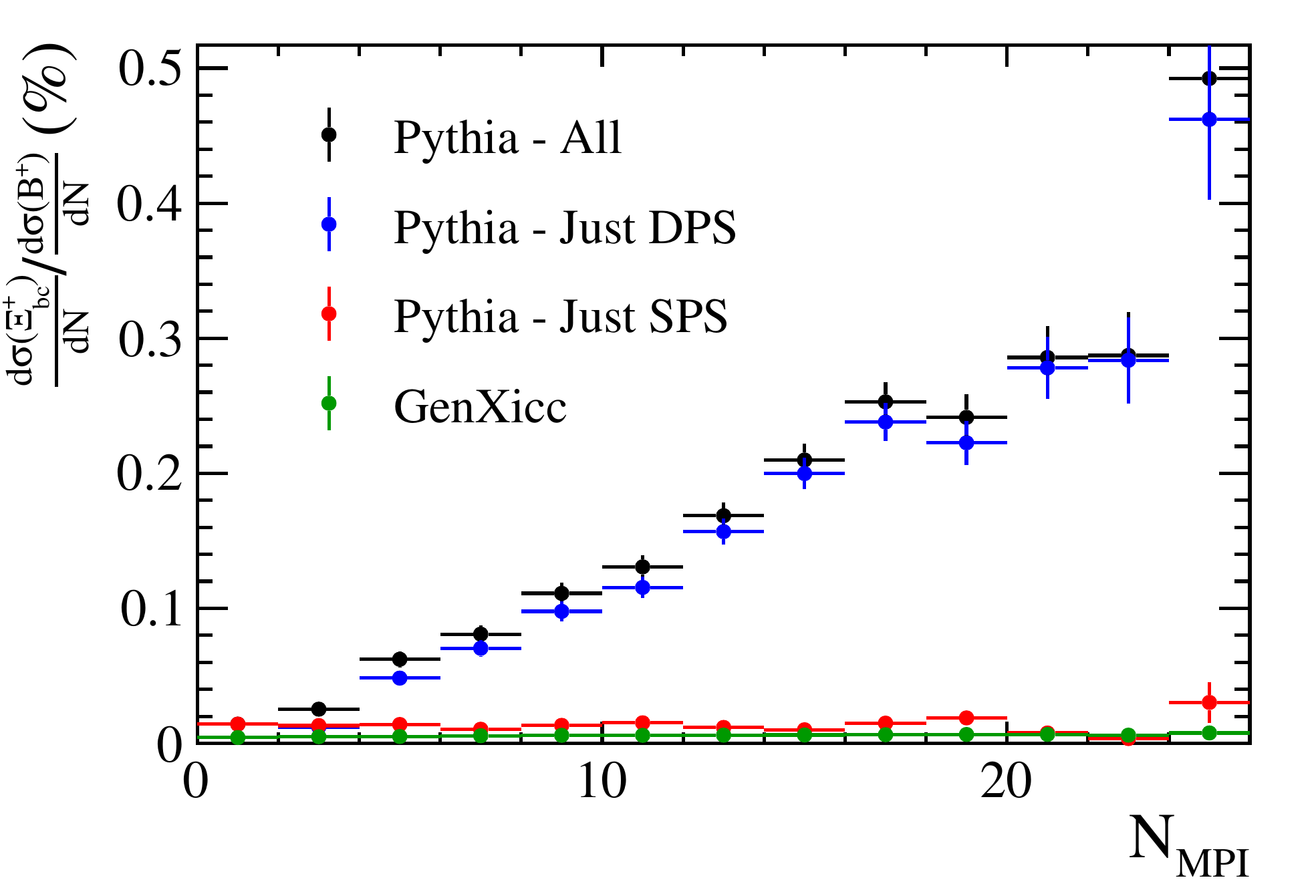}
    \includegraphics[width=\plotwidths\linewidth]{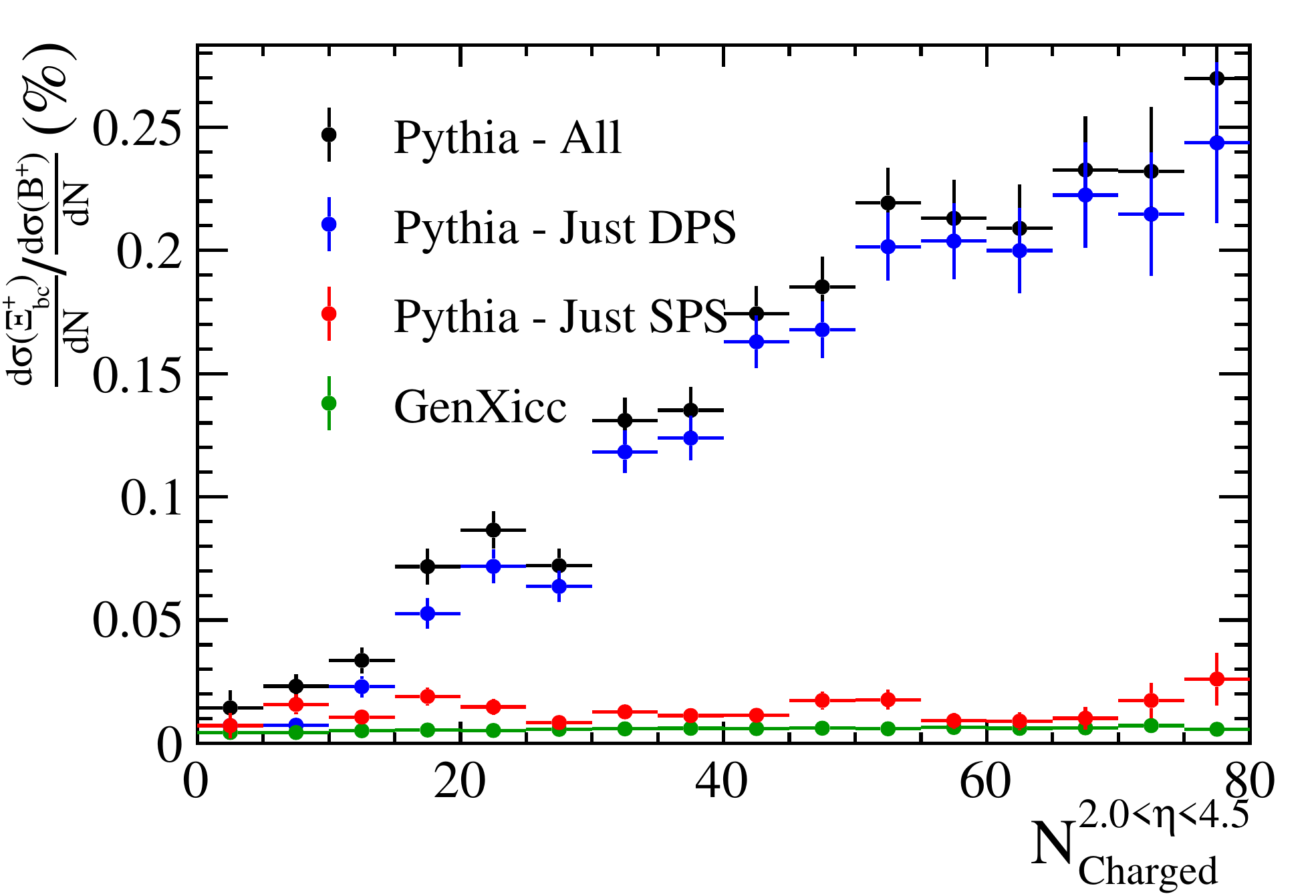}
    \caption{Differential cross-section of $\Xi_{bc}^{+}$ baryons as a function of (left) the number of parton interactions in a collision and (right) the number of charged tracks within the rapidity region $2.0<y<4.5$, as generated with \genxicc and \pythia.}
    \label{fig:multiplicity_Xibc}
\end{figure}

%%%%%%%%%%%%%%%%%%%%%%%%%%%%%%%%%%
Similarly, samples of $\Xi_{bc}^{+}$ can be produced with \pythia and \genxicc. These baryons have not yet been observed. The kinematic distributions of the $\Xi_{bc}^{+}$ baryons in samples generated with \pythia and \genxicc are compared in Fig~\ref{fig:kinematics_Xibc}, and the ratios of the differential cross section with respect to the \Bp meson are shown in Fig.~\ref{fig:multiplicity_Xibc} as a function of both the number of parton interactions and number of charged tracks within $2.0<\eta<4.5$. The production in \pythia is again dominated by a significant DPS contribution. 

%%%%%%%%%%%%%%%%%%%%%%%%%%%%%%%%%%
\subsection{Quarkonia}
\label{sec:just_quarkonia}
%%%%%%%%%%%%%%%%%%%%%%%%%%%%%%%%%%

\begin{figure}[tbp]
    \centering
    \includegraphics[width=\plotwidths\linewidth]{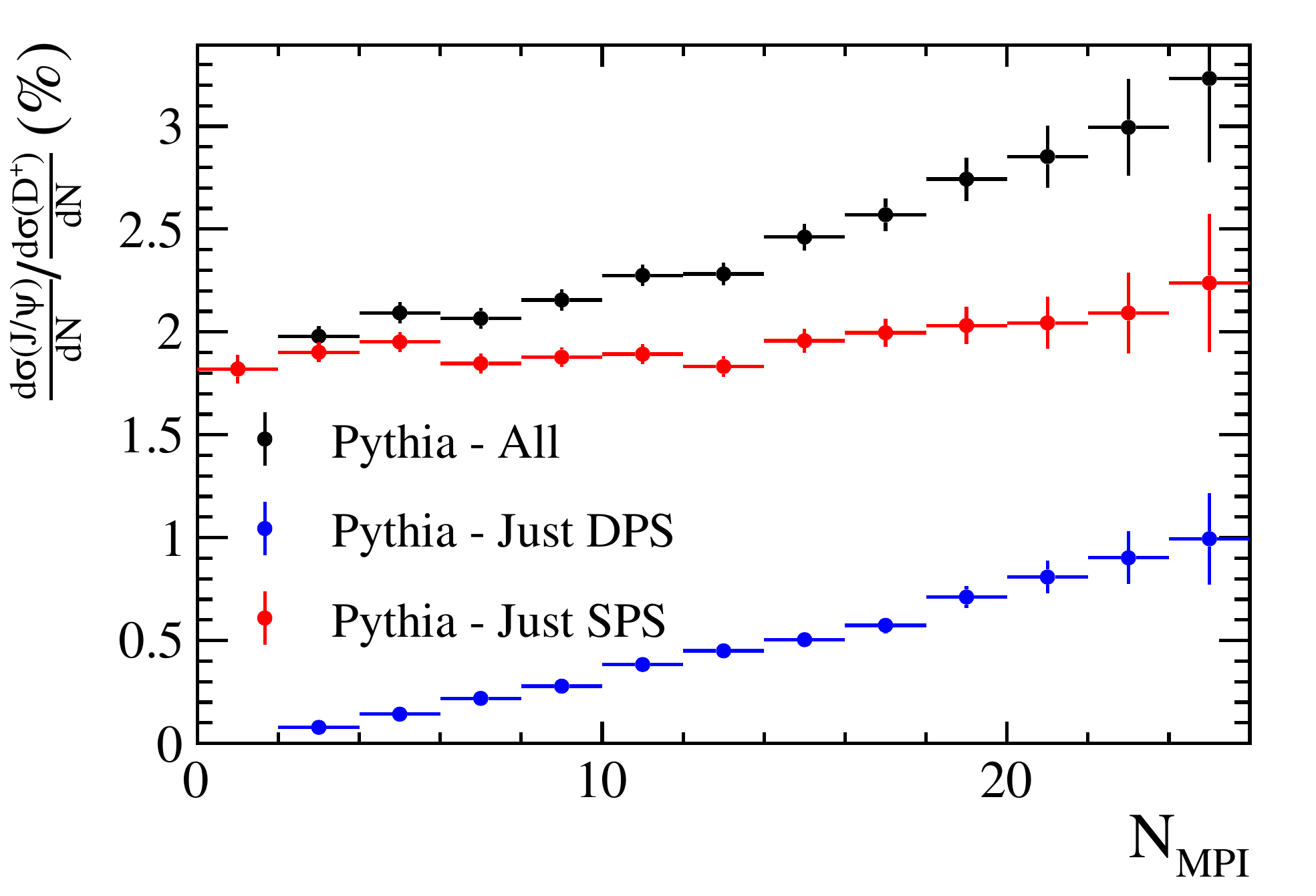}
    \includegraphics[width=\plotwidths\linewidth]{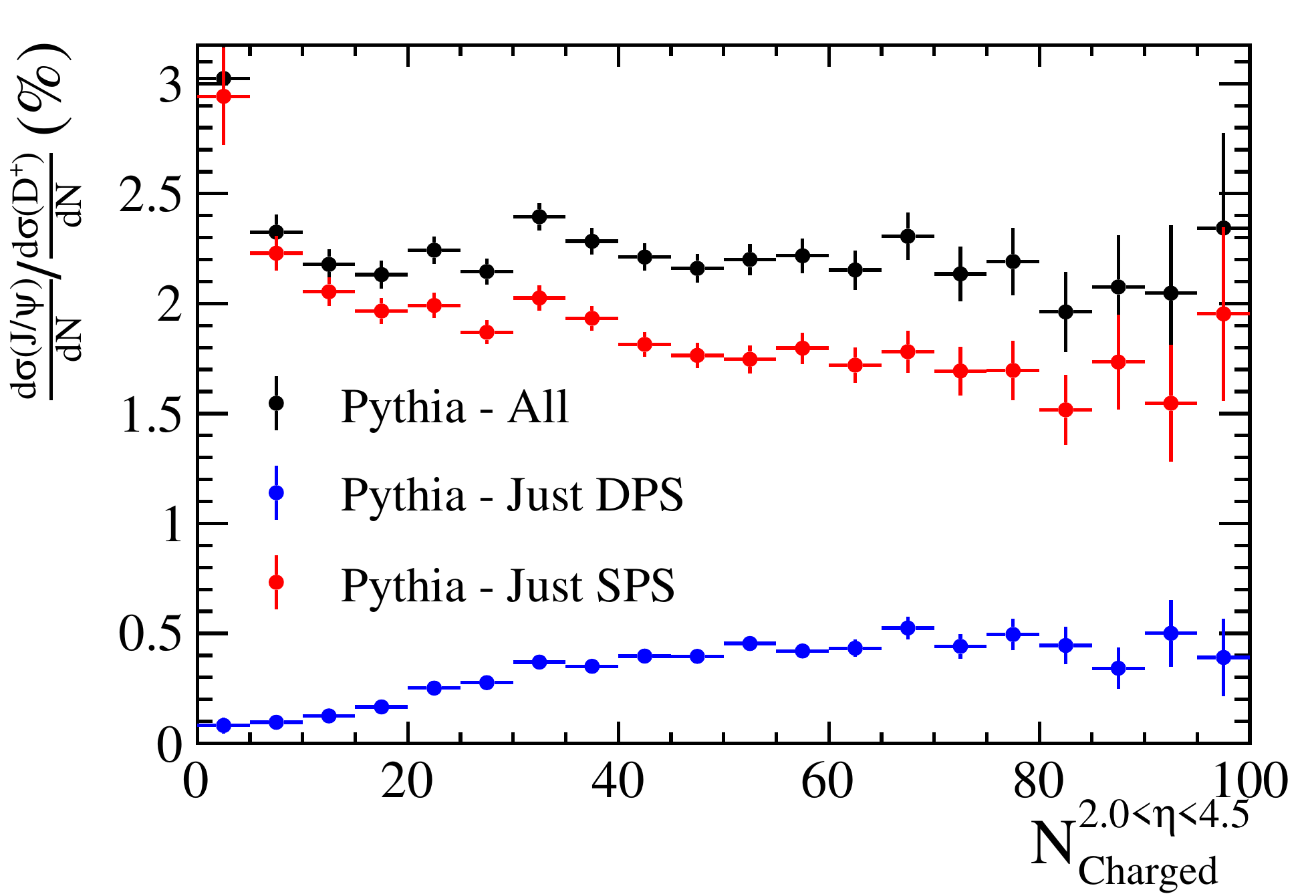}
    \caption{Ratio of differential cross-section of $\jpsi$ and \Dp hadrons as a function of (left) the number of parton interactions in a collision and (right) the number of charged particles within the pseudo-rapidity region $2.0<\eta<4.5$, as generated with \pythia.}
    \label{fig:multiplicity_Jpsi}
\end{figure}

For the purposes of this study, quarkonia are considered doubly-heavy hadrons as they contain two heavy quarks. However, unlike the previously discussed doubly-heavy hadrons, quarkonia can be formed from a single heavy-quark pair. As a result, contributions to quarkonium production from heavy quarks in different parton interactions will only give a  subleading contribution to the total production rate. This can be observed in Fig.~\ref{fig:multiplicity_Jpsi} where the ratio of \jpsi to \Dp production cross sections are shown as a function of the number of parton interactions and the number of charged tracks within $2.0<\eta<4.5$. There is a contribution from \jpsi mesons formed from \cquark and \cquarkbar quarks from different parton interactions, however this is much smaller than the total rate of \jpsi production.
To increase the fraction of events with DPS contributions, events with both a \jpsi meson and two additional charm hadrons can be reconstructed. This removes events in which there is only a single quark pair and leads to a different set of measurements that can be made as discussed in the next Section. 

%%%%%%%%%%%%%%%%%%%%%%%%%%%%%%%%%%%%%%%%%%%%%%%%%%%%%%%%%%%%%%%%%%%%
\section{Associated production of singly- and doubly-heavy hadrons in events with multiple \boldmath{$Q\bar{Q}$} pairs}
\label{sec:associated_production}
%%%%%%%%%%%%%%%%%%%%%%%%%%%%%%%%%%%%%%%%%%%%%%%%%%%%%%%%%%%%%%%%%%%%

In proton-proton collisions that produce two pairs of heavy quarks, \textit{i.e.} $c\bar{c}c\bar{c}$,  $c\bar{c}b\bar{b}$ or $b\bar{b}b\bar{b}$,  information about the production mechanisms can be inferred from the relative properties of a doubly-heavy hadron and two singly-heavy hadrons that can be formed from the additional heavy quarks.
Examples of the different combinations of doubly- and singly-heavy associated productions are listed in Table~\ref{tab:doubly_singly_associated_combs}, where singly-heavy hadrons containing a heavy quark are represented by $X_{Q}$. Only the \Bcp meson, \bquark- and \cquark-flavoured quarkonia and \Xiccpp baryon have currently been observed~\cite{E598:1974sol,PhysRevLett.33.1406,PhysRevLett.39.252,CDF:1998ihx,SELEX:2002wqn,LHCb-PAPER-2017-018}.
\begin{table}[h]
    \centering
    \begin{tabular}{l c c c }
    \hline \hline
           & $c\bar{c}c\bar{c}$ & $c\bar{c} b\bar{b}$ &  $b\bar{b} b\bar{b}$ \\
           \hline
            Doubly-heavy meson &  &$B_{c}^+ X_b X_{\bar{c}}$ & \\
            (excluding quarkonia) \\
            \\
            Doubly-heavy baryons & $\Xi_{cc} X_{\bar{c}} X_{\bar{c}}$ &$\Xi_{bc} X_{\bar{b}} X_{\bar{c}}$  & $\Xi_{bb} X_{\bar{b}} X_{\bar{b}}$\\
              & $\Omega_{cc} X_{\bar{c}} X_{\bar{c}}$ &$\Omega_{bc} X_{\bar{b}} X_{\bar{c}}$  & $\Omega_{bb} X_{\bar{b}} X_{\bar{b}}$\\
              \\
            Quarkonia & $\psi(nS) X_{c} X_{\bar{c}}$ & $\psi(nS) X_b X_{\bar{b}}$ & $\Upsilon(nS) X_{b} X_{\bar{b}}$ \\
            & &$\Upsilon(nS) X_{c} X_{\bar{c}}$ & \\
            %   &  & \\
    \hline \hline
    \end{tabular}
    \caption{\small Examples of combinations of doubly- and singly-heavy hadrons in processes with two pairs of heavy quarks. }
    \label{tab:doubly_singly_associated_combs}
\end{table}

\begin{figure}[h]
    \centering
    \begin{subfigure}{.32\textwidth}
        \includegraphics[width=\linewidth]{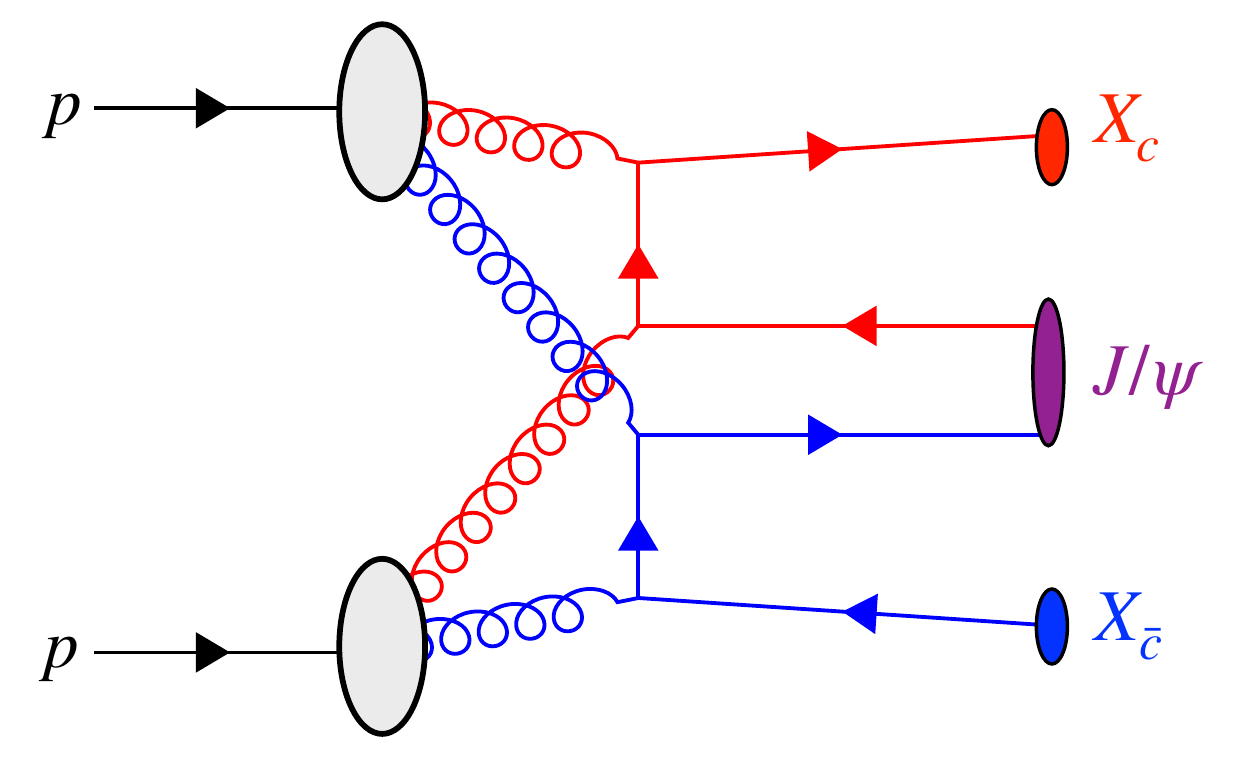}
        \caption{\small \textit{Mixed}}
        \label{fig:mixed_feynman_mixed}
    \end{subfigure}%
    \begin{subfigure}{.32\textwidth}
        \includegraphics[width=\linewidth]{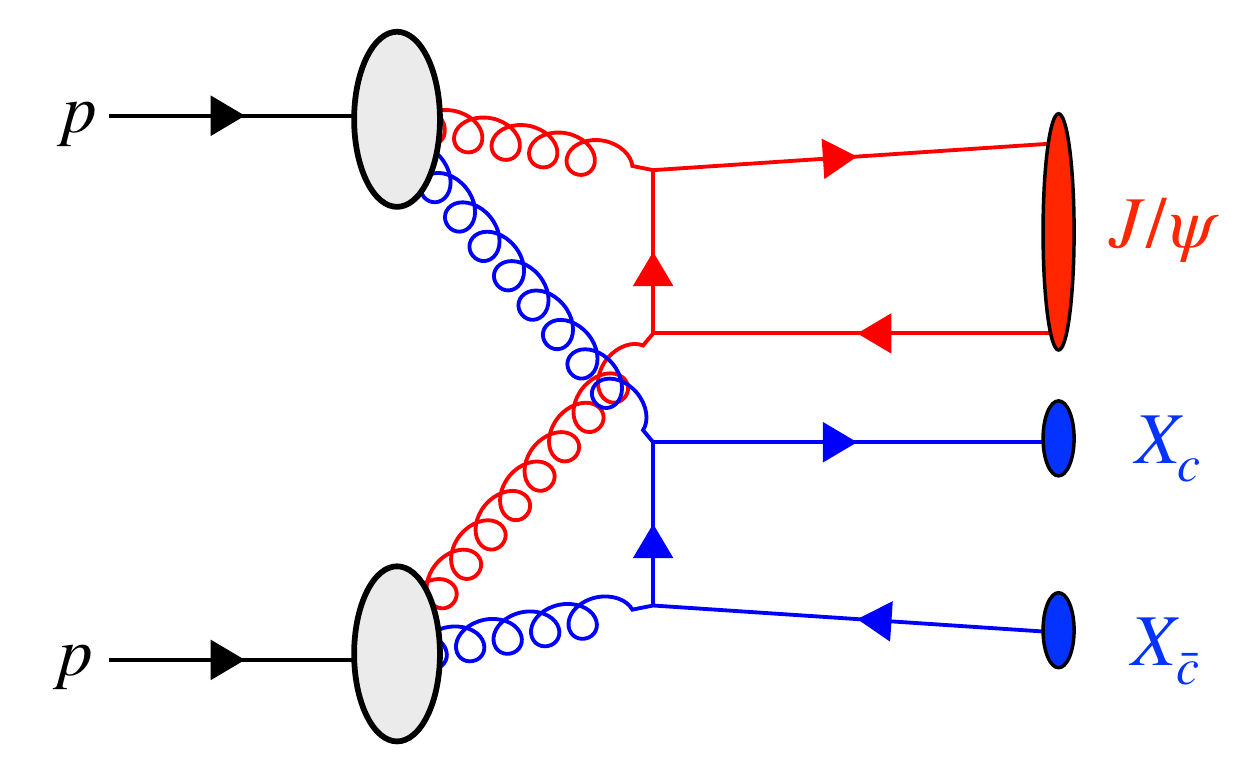}
        \caption{\small \textit{Unmixed}}
        \label{fig:mixed_feynman_unmixed}
    \end{subfigure}%
    \caption{Possible configurations of quark lines in DPS events with two pairs of charm quarks.}
    \label{fig:mixed_feynman_definition}
\end{figure}

Of the processes listed in Table~\ref{tab:doubly_singly_associated_combs}, those categorised as doubly-heavy mesons and doubly-heavy baryons require quarks from different $Q\bar{Q}$ pairs to combine in order to form. This configuration is referred to as \textit{mixed}. In contrast, the quarkonium in $\psi(nS) X_b X_{\bar{b}}$ or $\Upsilon(nS) X_{c} X_{\bar{c}}$ events contain differently flavoured quarks compared to the associated singly-heavy hadrons, and therefore must be formed from an individual heavy-quark pair. This configuration is referred to as \textit{unmixed}. In events containing a quarkonium, $X_Q$ and $ X_{\bar{Q}}$ hadrons where the flavours of the singly-heavy hadrons match the doubly-heavy flavour,
 the quarkonium could have been formed from a single $Q\bar{Q}$ pair, \ie \textit{unmixed}, or from the combination of a $Q$ and $\bar{Q}$ from two different $Q\bar{Q}$ pairs, \ie \textit{mixed}.  This is demonstrated for the charm quark in Fig.~\ref{fig:mixed_feynman_definition}.   

Therefore, for the \Bcp, $\Xi_{cc}$ and $\Xi_{bc}$ hadrons any indication that DPS processes contribute to the production would indicate that heavy quarks from different parton interactions can form hadrons. In contrast, the indications that DPS processes contribution to $\psi(nS) X_b X_{\bar{b}}$ or $\Upsilon(nS) X_{c} X_{\bar{c}}$ events implies that there are events with multiple heavy-quark pairs from different parton interactions, but doesn't give any information about the hadronisation of those different pairs. 
The quarkonia final states $\psi(nS) X_{c} X_{\bar{c}}$ and $\Upsilon(nS) X_{b} X_{\bar{b}}$ can give information about both the presense of multiple heavy quark pairs and hadronisation in DPS, but care must be taken to separate the \textit{mixed} and \textit{unmixed} contributions.

The kinematic distributions of the singly- and doubly-heavy hadrons depend on the specific production mechanism and differ for those possible within SPS or DPS.
Collisions in which both heavy hadron pairs originate from a SPS, for example the process shown in Fig.~\ref{fig:feyn_MPI_shower}, have correlations between the kinematics of at least one pair of heavy quarks. Processes that involve DPS introduce the possibility of double flavour excitation processes, such as Fig.~\ref{fig:feyn_MPI_bx_cx}, in which the kinematics of the heavy quarks that form the hadron can be relatively uncorrelated to the remaining two companion heavy quarks.\footnote{The small fraction of hadrons that are missed by the \texttt{UserHooks} (Sec.~\ref{sec:sim_QQ}) are not found to significantly alter the relative kinematic correlations.}

The production of quarkonia is complicated further by contributions from colour-octet mechanisms~\cite{Cho:1995vh,Cho:1995ce,Yuan:1998gr}, which are included by default in the \pythia simulation samples.

%%%%%%%%%%%%%%%%%%%%%%%%%%%%%%%%%%%%%%%%
\subsection{Studies with \boldmath{$B_c^{+}X_b X_{\bar{c}}$} events}
%%%%%%%%%%%%%%%%%%%%%%%%%%%%%%%%%%%%%%%%
\label{sec:Bc_Xb_Xc_studies}

\begin{figure}[h]
    \centering
    \begin{subfigure}{.32\textwidth}
        \includegraphics[width=\linewidth]{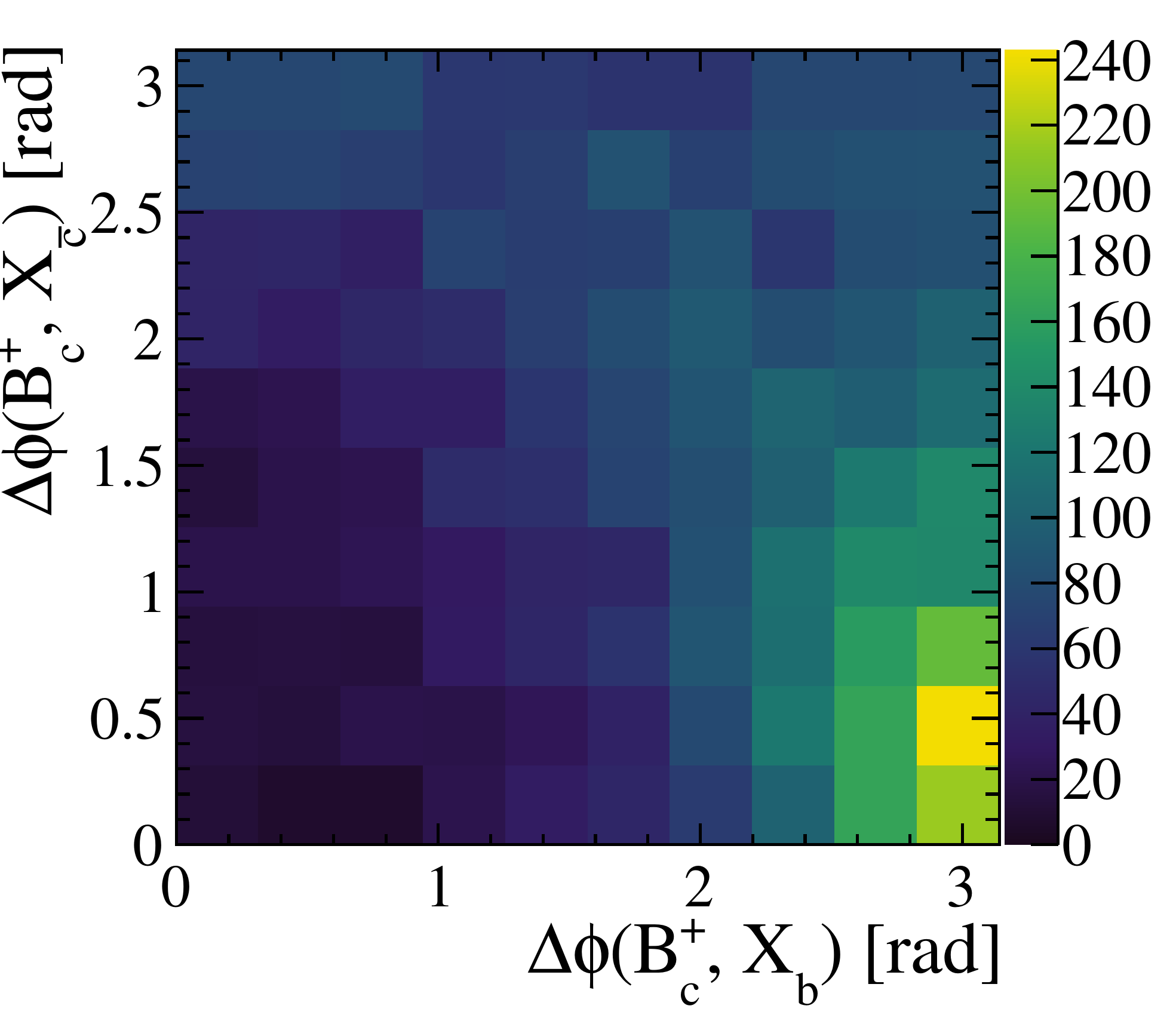}
        \caption{\bcvegpy}
        \label{fig:deltaphi_bb}
    \end{subfigure}%
    \begin{subfigure}{.32\textwidth}
        \includegraphics[width=\linewidth]{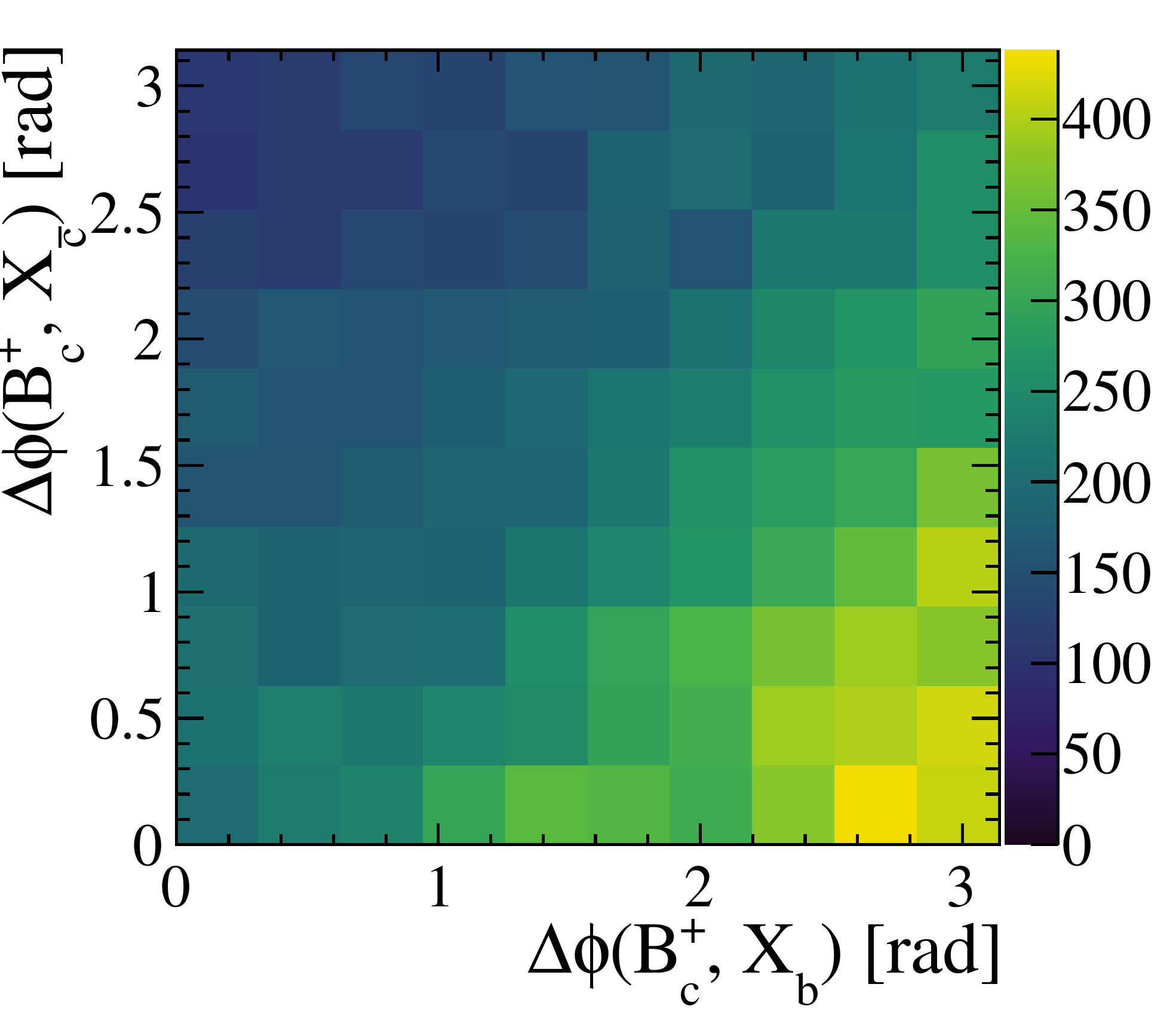}
        \caption{ \pythia - Just DPS}
        \label{fig:deltaphi_pythia}
    \end{subfigure}%
    \begin{subfigure}{.32\textwidth}
        \includegraphics[width=\linewidth]{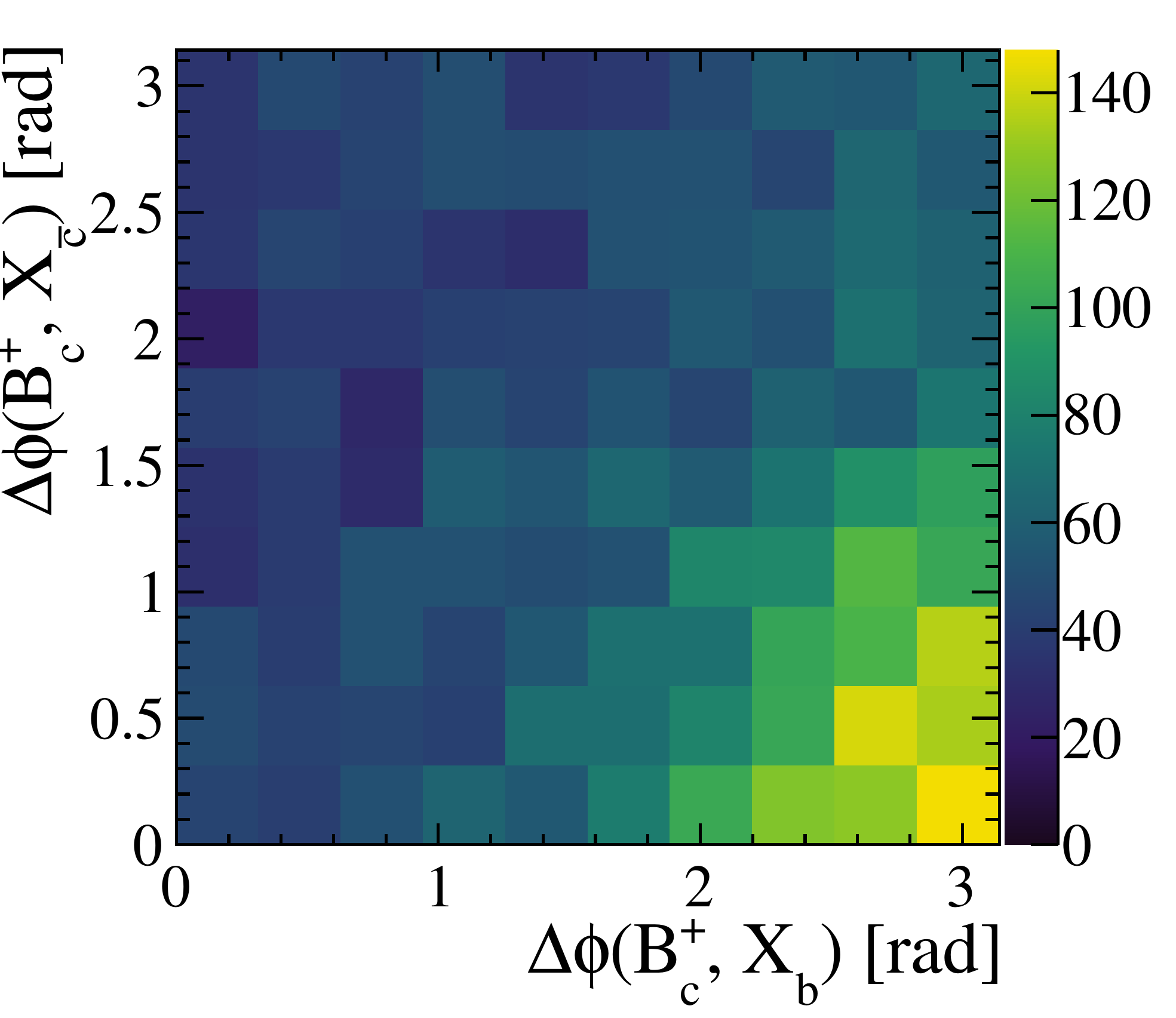}
        \caption{\pythia - Just SPS}
        \label{fig:deltaphi_pythia_nompi}
    \end{subfigure}%
    \caption{\small Angular separation in the transverse view between $B_{c}^{+}$ mesons and the associated $X_{b}$ or $X_{\bar{c}}$ hadron in events generated with the \bcvegpy and \pythia generators.}
    \label{fig:delta_phi_correlations}
\end{figure}

The relative kinematic distribution in $B_c^{+}X_b X_{\bar{c}}$ events are studied to determine if there are differences between the DPS and SPS sub-samples for the \textit{mixed} configuration, and to compare with the standalone generator \bcvegpy. 
The angular separation in the transverse view between the \Bcp and the associated $X_{b}$ hadron $\Delta\phi(\Bcp,X_{b})$ is plotted against the same quantity between the \Bcp and associated $X_{\cquarkbar}$ hadron $\Delta\phi(\Bcp,X_{\bar{c}})$ in Fig.~\ref{fig:delta_phi_correlations} for the different generators and configurations. To ensure an unambiguous association between the heavy quarks, only events with a total of one \Bc meson, one $X_{\bar{c}}$ hadron and one $X_b$ hadron are used. 

In the sample generated with \bcvegpy the generated events are found predominately with large angles between the $B_{c}^{+}$ meson and $X_{b}$ hadron, and small angles between the $B_{c}^{+}$ and $X_{\bar{c}}$ hadrons. This is consistent with the topology shown in Fig.~\ref{fig:feyn_MPI_shower} in which the $b\bar{b}$ pair are produced in the hardest interaction, and therefore back-to-back in the transverse plane. The $c$-quark resulting from a gluon splitting would then be produced in a cone around the $B_{c}^{+}$ direction. 
In contrast, the DPS sample produced by \pythia introduces the possibility of additional production mechanisms including those shown in Fig.~\ref{fig:feyn_MPI_bb_cc} and Fig.~\ref{fig:feyn_MPI_bx_cx}. As such the distribution of events in the 2D plane is less localised as a result of the contributions from many different associated production correlations.

%%%%%%%%%%%%%%%%%%%%%%%%%%%%%%%%%%%%%%%%
\subsection{Studies with \boldmath{$\OneS X_c X_{\bar{c}}$} events}
%%%%%%%%%%%%%%%%%%%%%%%%%%%%%%%%%%%%%%%%

Events containing both an $\OneS$ meson and $X_c X_{\bar{c}}$ pair can only receive contributions from the configuration referred to as \textit{unmixed}. It therefore provides a suitable system to test whether MPIs contribute significantly to events with multiple pairs of heavy quarks, but cannot provide insight into the hadronisation of heavy quarks from different parton interactions.   
\begin{figure}[h]
    \centering%
    \begin{subfigure}{.32\textwidth}
        \includegraphics[width=\linewidth]{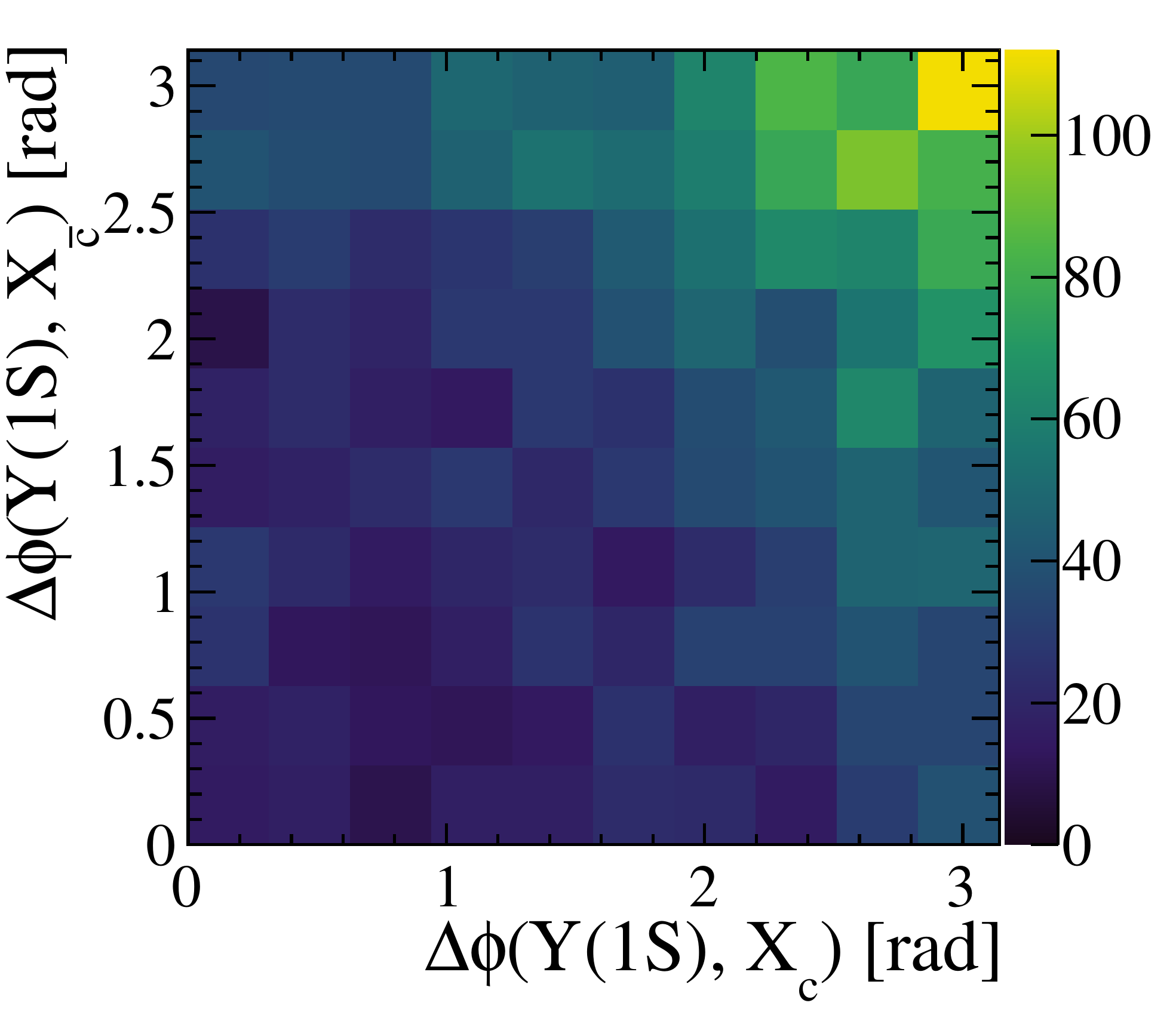}
        \caption{\pythia - Just SPS}
        \label{fig:UpsilonXcXc_SPS}
    \end{subfigure}
    \begin{subfigure}{.32\textwidth}
        \includegraphics[width=\linewidth]{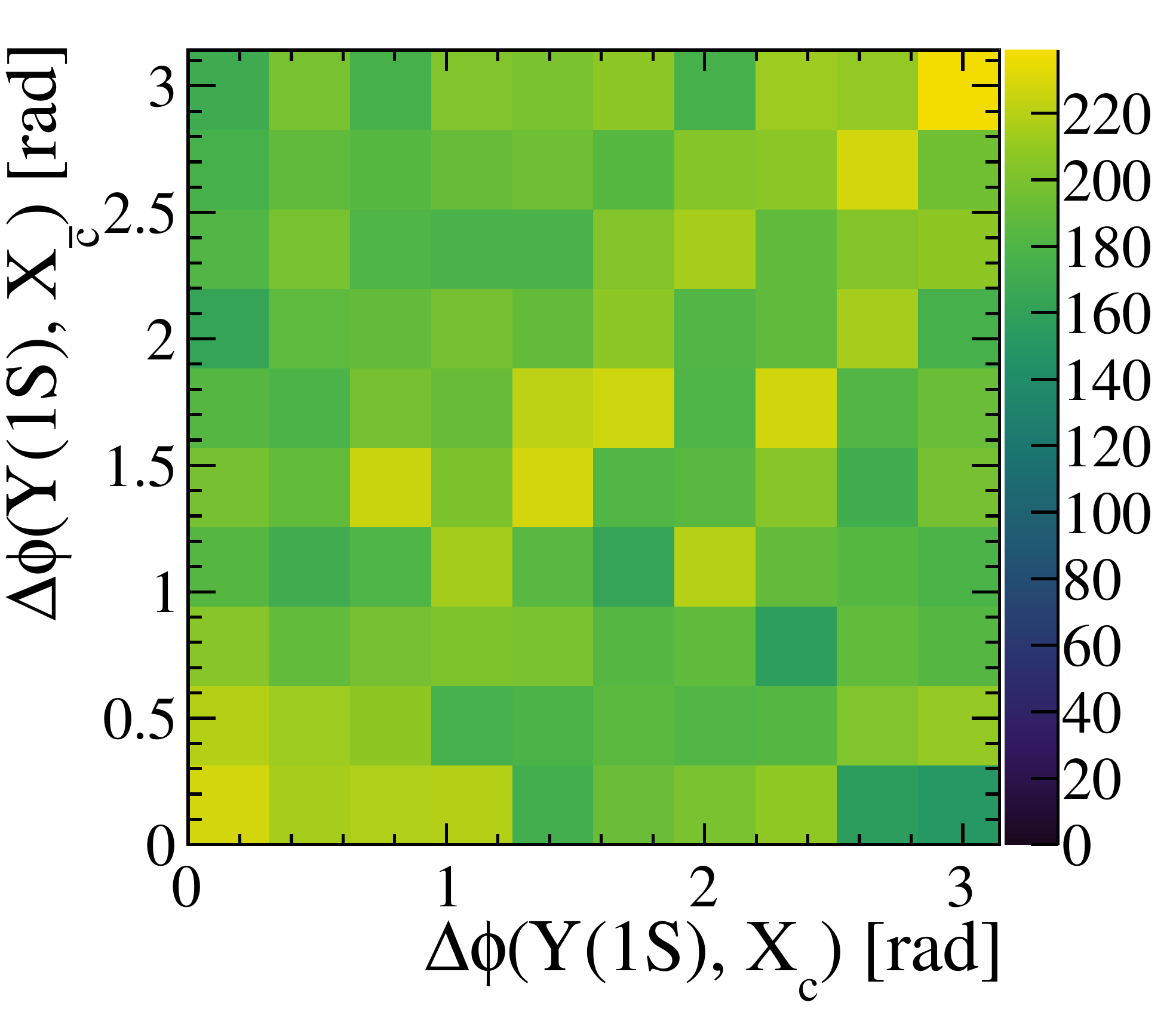}
        \caption{\pythia - Just DPS}
        \label{fig:UpsilonXcXc_DPS}
    \end{subfigure}\\
    \begin{subfigure}{.40\textwidth}
        \includegraphics[width=\linewidth]{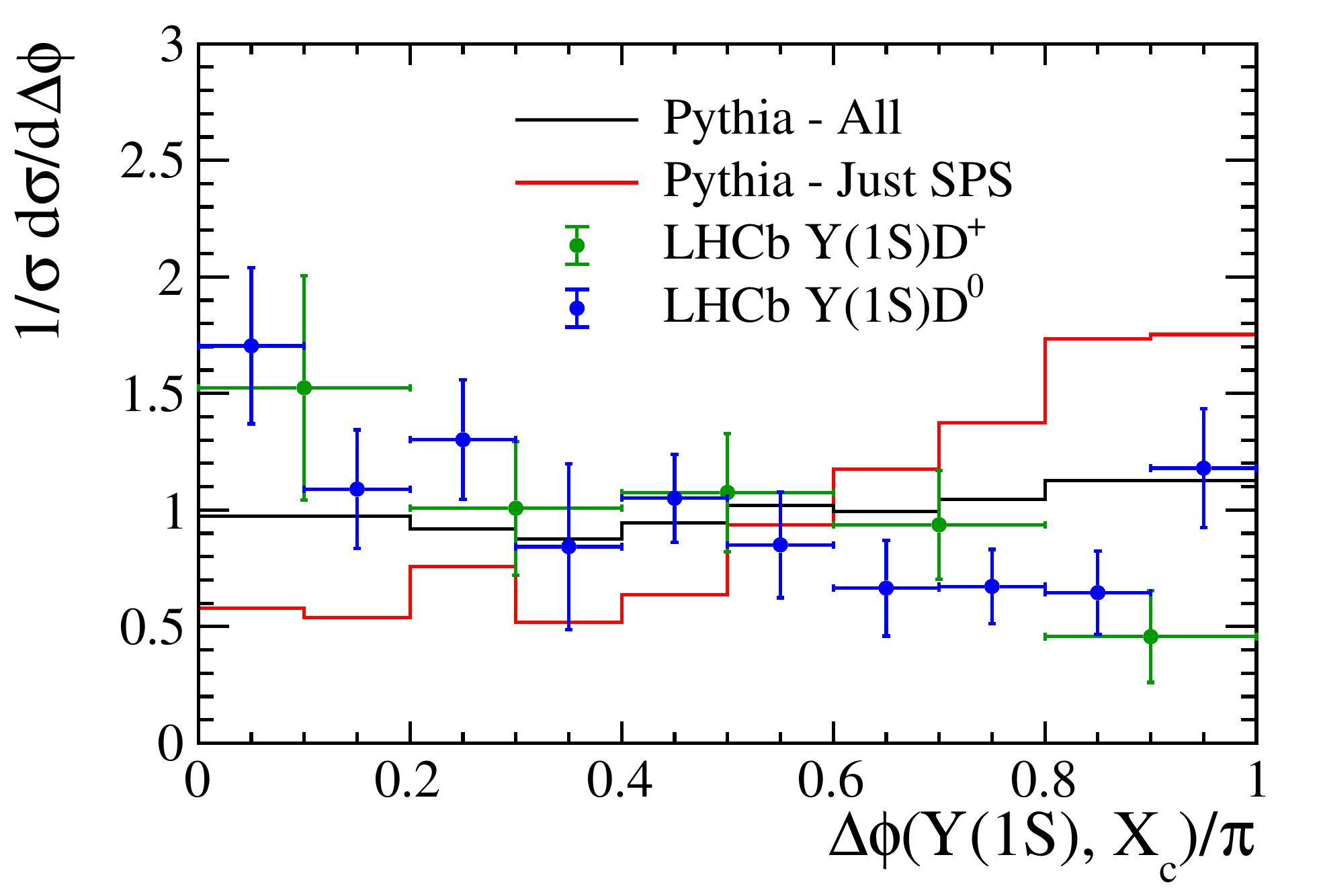}
        \caption{}
        \label{fig:UpsilonXcXc_lhcb}
    \end{subfigure}%
    \begin{subfigure}{.40\textwidth}
        \includegraphics[width=\linewidth]{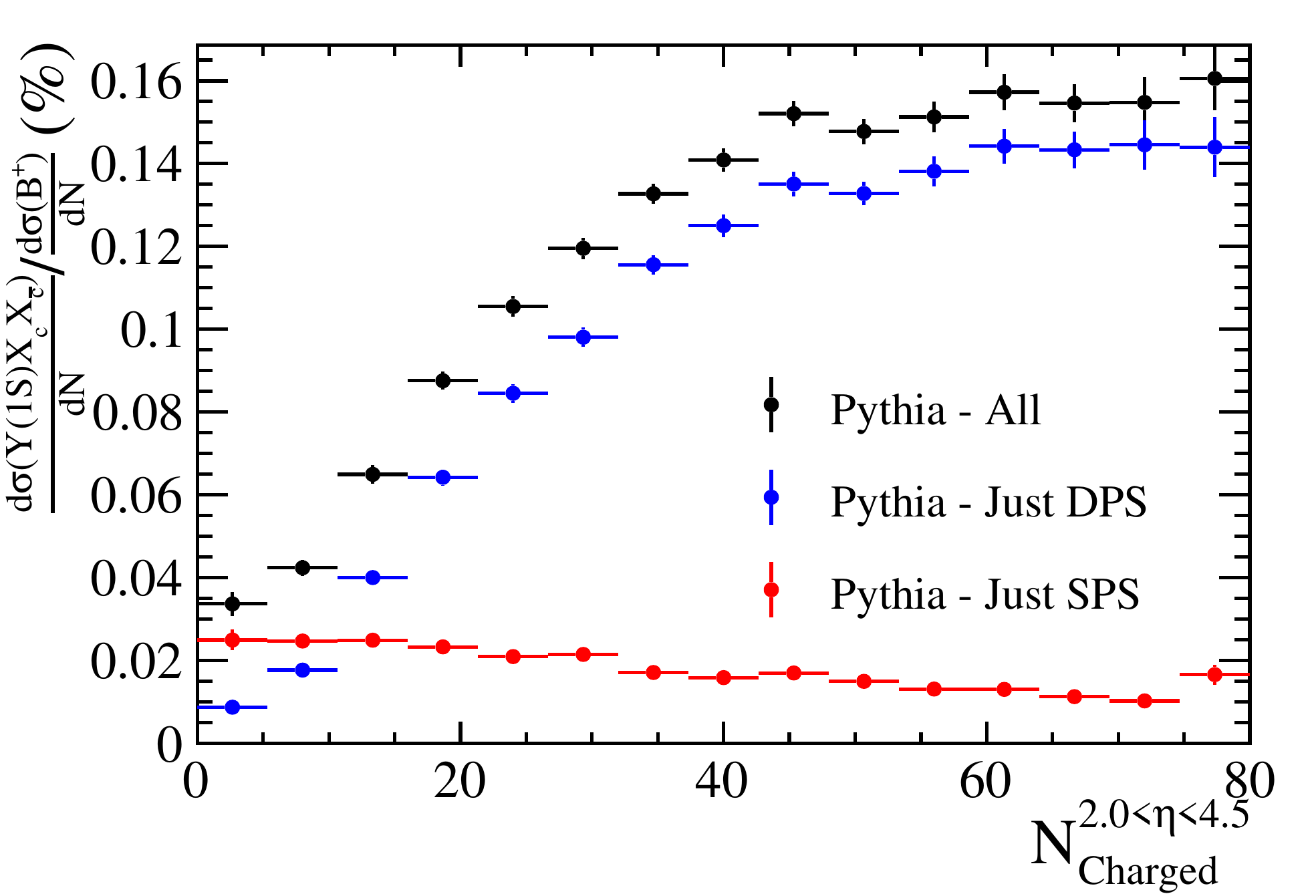}
        \caption{}
        \label{fig:UpsilonXcXc_multi}
    \end{subfigure}%
    \caption{(Top) Relative transverse distributions in $\OneS X_c X_{\bar{c}}$ events for SPS and DPS processes. (Bottom left) One-dimensional projections of the relative angular distributions, compared to measurements from Ref.~\cite{LHCb:2015wvu}. (Bottom right) Relative differential cross-section of $\OneS X_c X_{\bar{c}}$ events with respect to \Bp as a function of the number of charged particles within the pseudo-rapidity region $2.0<\eta<4.5$.}
    \label{fig:Upslion_Xc_Xc}
\end{figure}
The relative transverse distributions between the $\OneS$ meson and the $X_c$ and $X_{\bar{c}}$ hadrons are shown in Fig.~\ref{fig:UpsilonXcXc_SPS} and~\ref{fig:UpsilonXcXc_DPS} . The two-dimensional distributions show a clear difference in the relative distributions of the two types of process. In the SPS process the $\OneS$ meson has a strong tendency to be produced back-to-back to both charm hadrons in the transverse plane. This could result from $gg\to\OneS g$ parton interactions in which the outgoing gluon subsequently splits $g\to\cquark\cquarkbar$. In contrast, in DPS processes there is little correlation between the transverse directions. The predictions can be compared to measurements by LHCb~\cite{LHCb:2015wvu} by making a one-dimensional projection, as shown in Fig.~\ref{fig:UpsilonXcXc_lhcb}. Only events in which both the $\OneS$ and one of the corresponding $X_c$ hadron are within the LHCb acceptance are compared in this figure.  
The data are consistent with the predictions including DPS, but the significant difference between the SPS and DPS samples is diluted when projected onto one dimension. This strongly motivates performing measurements in which both associated hadrons are reconstructed, such that the two dimensional distributions can be determined. 
Finally, the ratio of the differential cross section relative to the \Bp meson is shown in Fig.~\ref{fig:UpsilonXcXc_multi} as a function of the number of charged particles within $2.0<\eta<4.5$. It similarly shows a strong separation power between the two process types.

%%%%%%%%%%%%%%%%%%%%%%%%%%%%%%%%%%%%%%%%
\subsection{Studies with \boldmath{$J/\psi X_c X_{\bar{c}}$} events}
%%%%%%%%%%%%%%%%%%%%%%%%%%%%%%%%%%%%%%%%
\label{sec:Jpsi_Xc_Xc}

Samples containing a \jpsi meson and two charm hadrons are generated with \pythia. The samples are split according to whether the \jpsi meson was created in a single interaction, or from quarks in different parton interactions, as before. Additionally, the ancestors of the charm quarks are studied to determine if the two quark lines are \textit{mixed} or \textit{unmixed}, as defined in Fig.~\ref{fig:mixed_feynman_definition}. \begin{figure}[h]
    \centering
    \includegraphics[width=\plotwidths\linewidth]{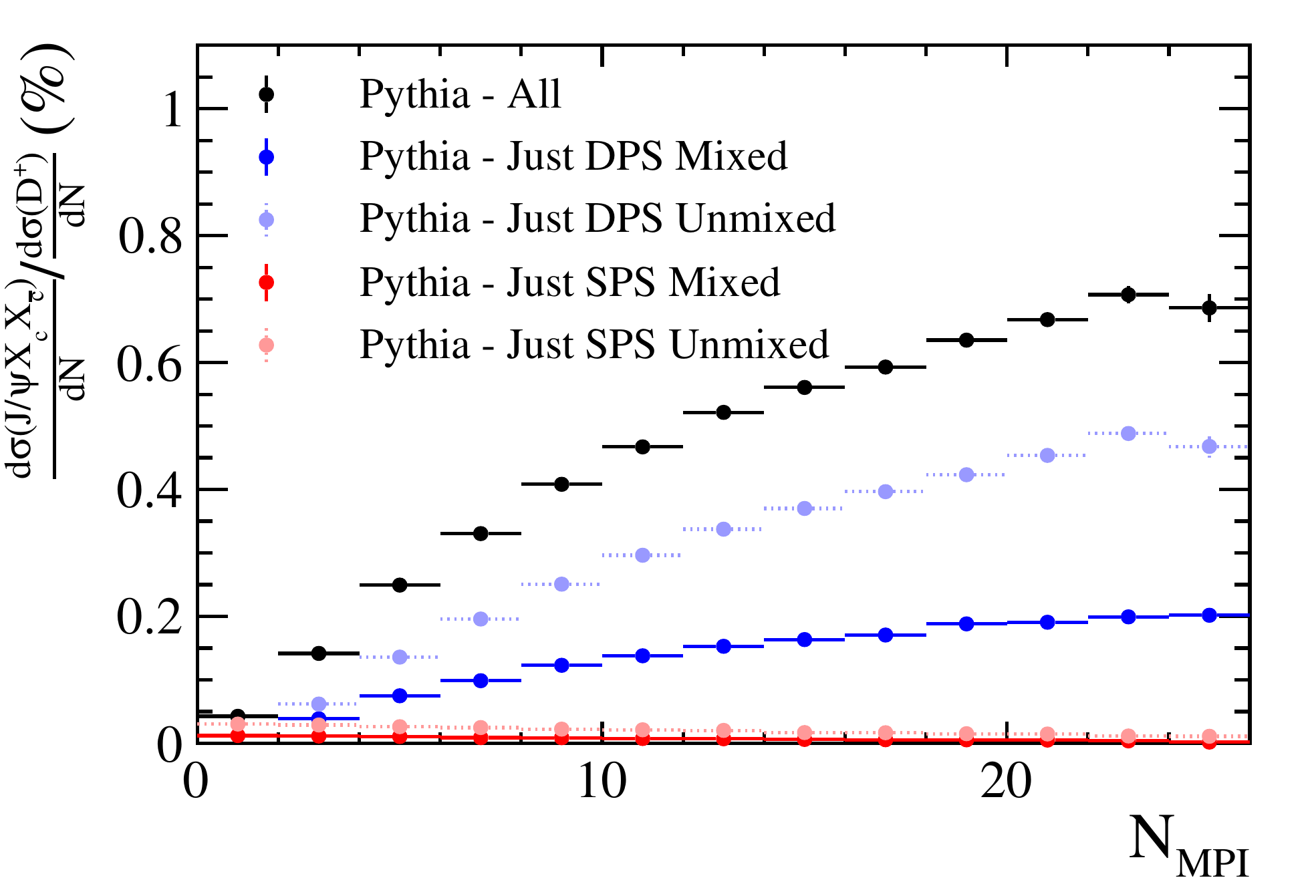}
    \includegraphics[width=\plotwidths\linewidth]{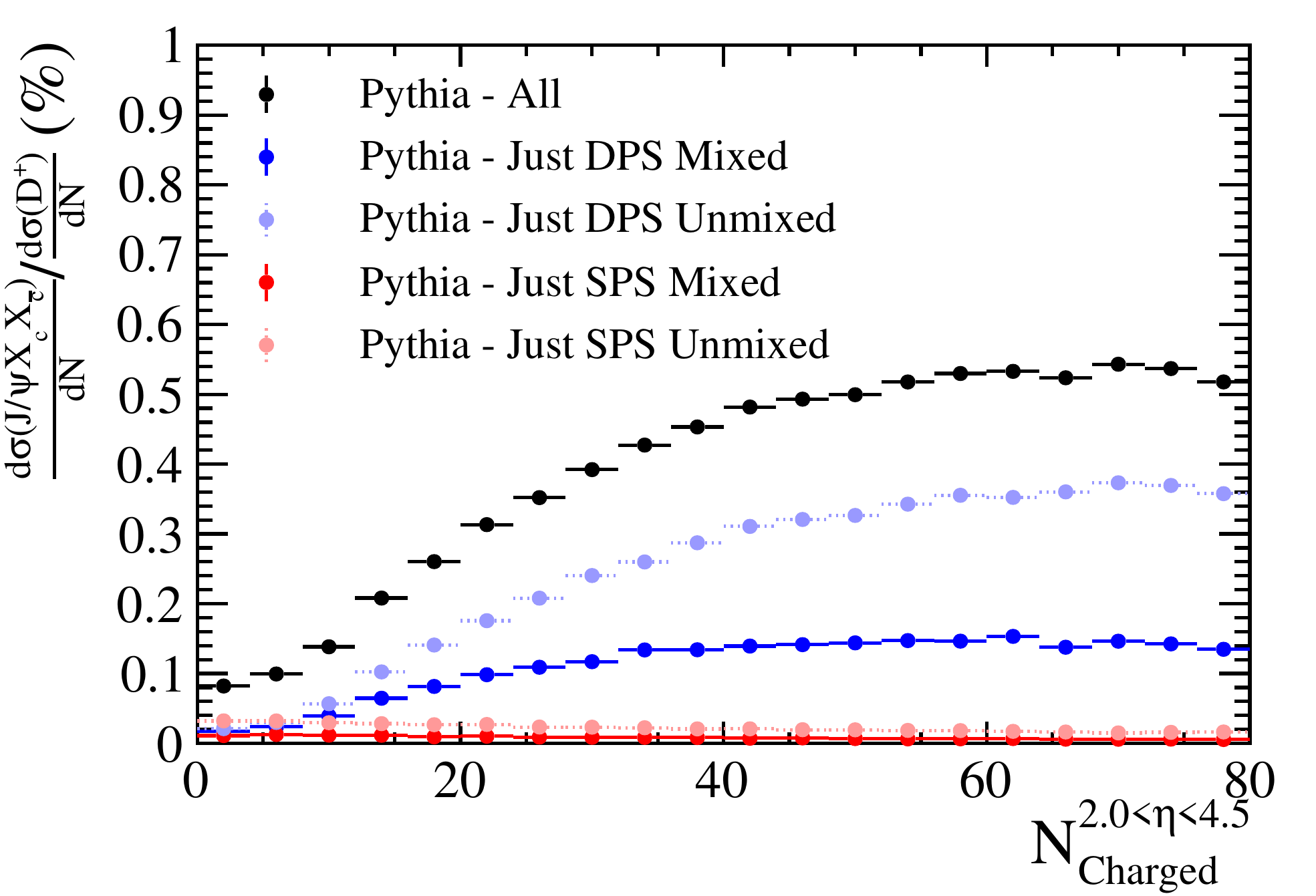}
    \caption{Ratio of differential cross-section of $\jpsi X_{\cquark} X_{\cquarkbar}$ and \Dp hadrons as a function of (left) the number of parton interactions in a collision and (right) the number of charged particles within the pseudo-rapidity region $2.0<\eta<4.5$, as generated with \pythia.}
    \label{fig:JpsiXcXc_multi_noColRec}
\end{figure}
The ratio of the differential cross sections with respect to the \Dp meson is compared for the four categories, as shown in Fig.~\ref{fig:JpsiXcXc_multi_noColRec}. The two DPS categories show an increasing trend with both the number of parton interactions and charged particles within $2.0<\eta<4.5$.  The fraction of events produced in DPS processes is significantly increased with respect to the samples in which just the \jpsi was reconstructed in Section~\ref{sec:just_quarkonia}.

\begin{figure}[h]
    \centering
    \begin{subfigure}{.32\textwidth}
        \includegraphics[width=\linewidth]{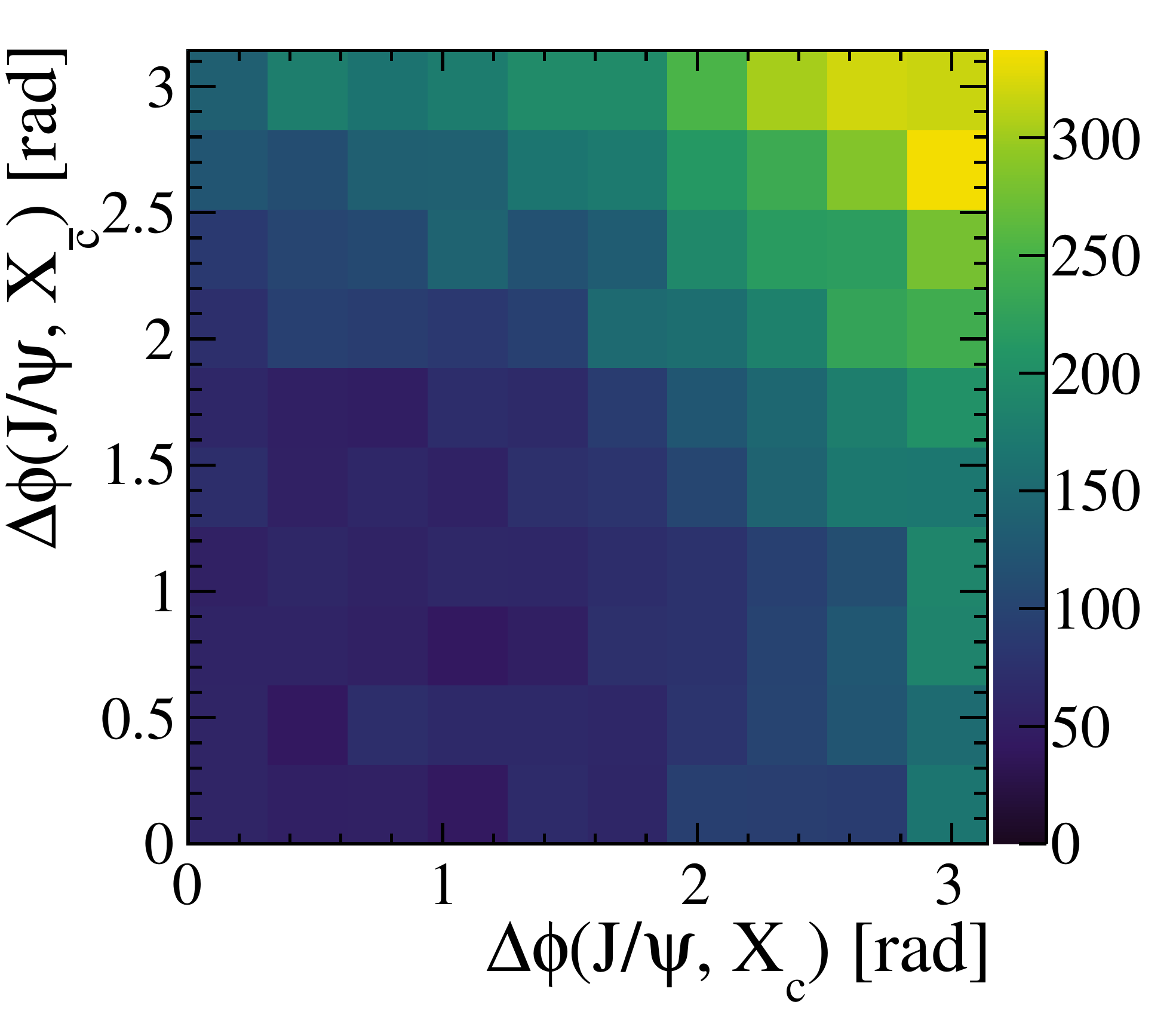}
        \caption{\pythia - SPS \textit{unmixed}}
        \label{fig:JpsiXcXc_SPS_Unmix}
    \end{subfigure}%
    \begin{subfigure}{.32\textwidth}
        \includegraphics[width=\linewidth]{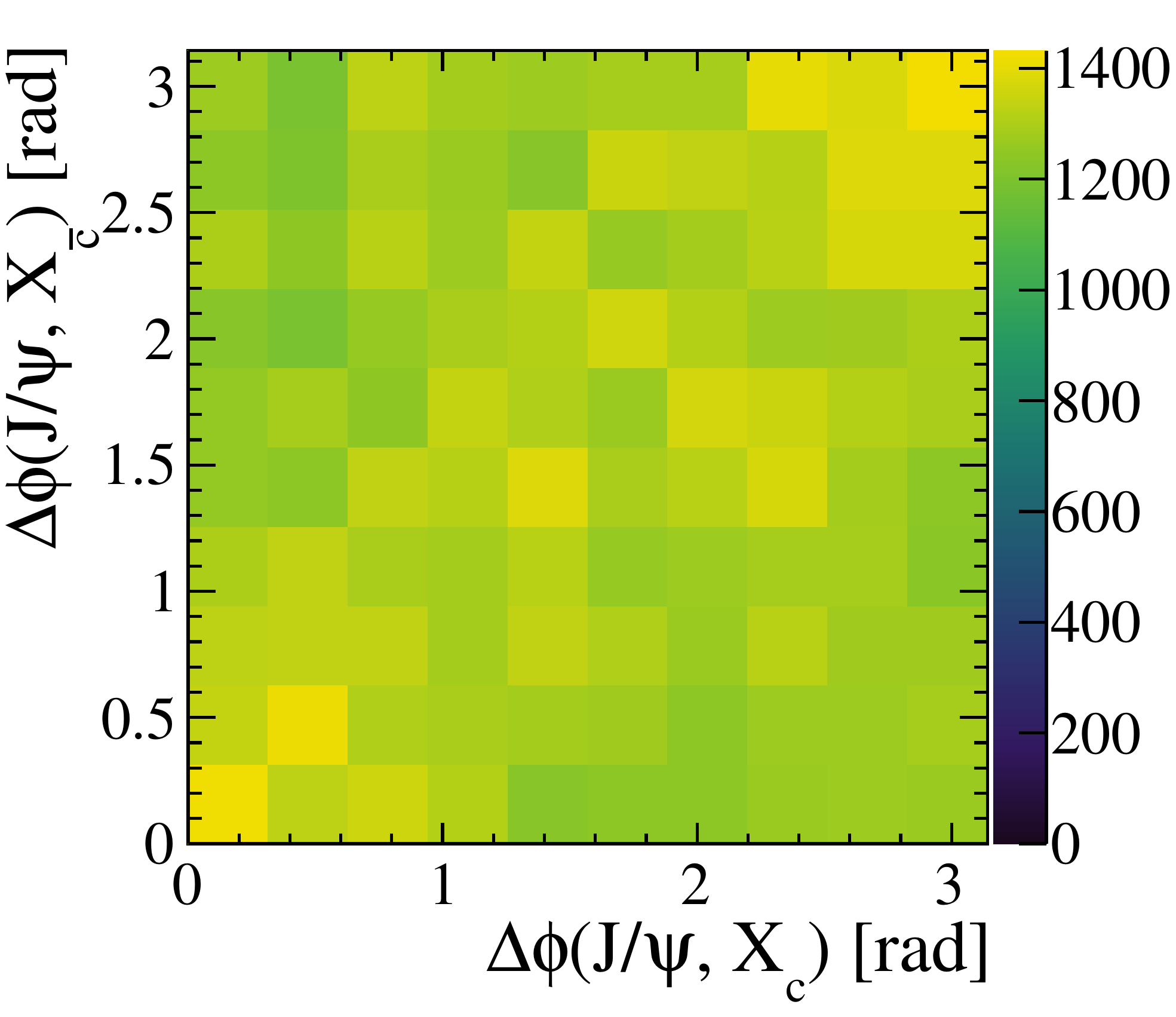}
        \caption{\pythia - DPS \textit{unmixed}}
        \label{fig:JpsiXcXc_DPS_Unmix}
    \end{subfigure}\\
    \begin{subfigure}{.32\textwidth}
        \includegraphics[width=\linewidth]{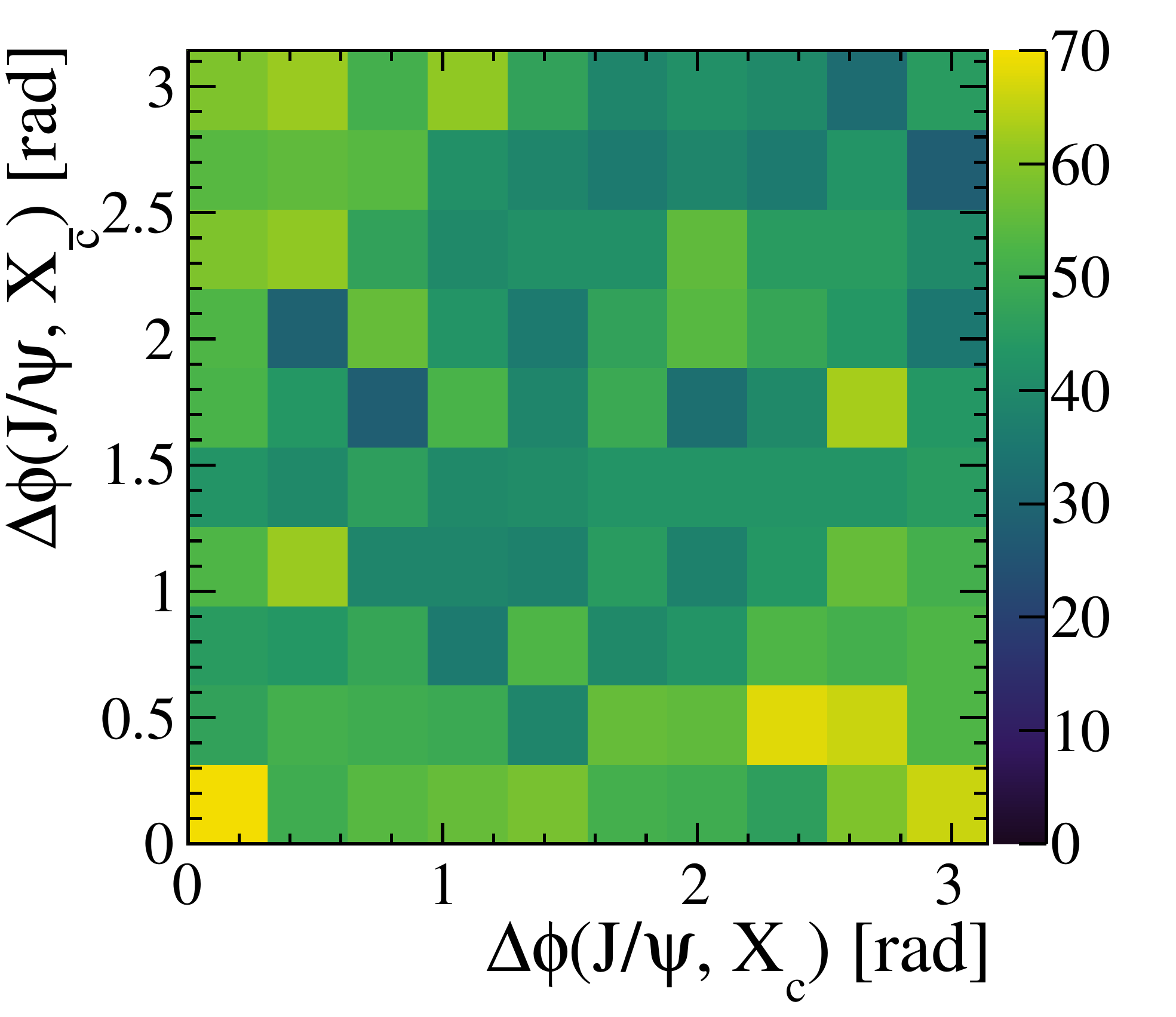}
        \caption{\pythia - SPS \textit{mixed}}
        \label{fig:JpsiXcXc_SPS_Mix}
    \end{subfigure}%
    \begin{subfigure}{.32\textwidth}
        \includegraphics[width=\linewidth]{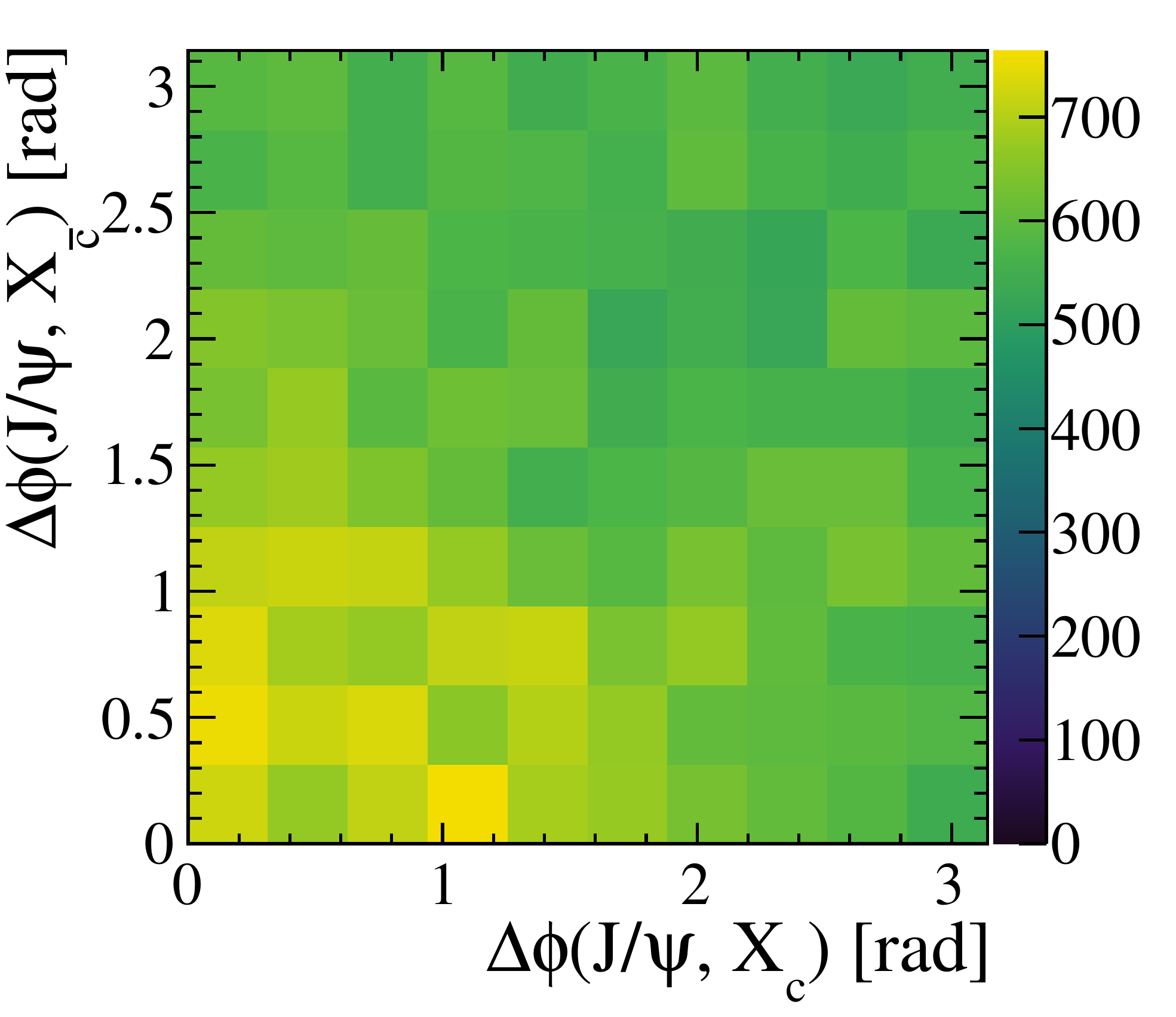}
        \caption{\pythia - DPS \textit{mixed}}
        \label{fig:JpsiXcXc_DPS_Mix}
    \end{subfigure}%
    \caption{Relative transverse distributions in $\jpsi X_c X_{\bar{c}}$ events for SPS and DPS processes in both the \textit{mixed} and \textit{unmixed} configurations.}
    \label{fig:JpsiXcXc_2D_noColRec}
\end{figure}

The relative transverse directions of the three particles is shown for the four categories in Fig.~\ref{fig:JpsiXcXc_2D_noColRec}. The two \textit{unmixed} categories show similar features to those already seen in $\OneS X_{\cquark}X_{\cquarkbar}$ events.  In contrast the two \textit{mixed} categories have distinct distributions, with the DPS \textit{mixed} category having a slight tendency to have the two charm quarks produced in the same direction as the \jpsi, a topology that may arise due to events with two flavour excitation interactions.

The presence of these events categorised as DPS \textit{mixed} would indicate that heavy quarks from different parton interactions can hadronise to form a $J/\psi$ meson. Using a combination information from the relative cross section ratio and relative transverse directions, the presence of this contribution could be determined.

%%%%%%%%%%%%%%%%%%%%%%%%%%%%%%%%%%%%%%%%
\subsection{Studies with \boldmath{$\OneS X_b X_{\bar{b}}$} events}
%%%%%%%%%%%%%%%%%%%%%%%%%%%%%%%%%%%%%%%%

\begin{figure}[h]
    \centering
    \includegraphics[width=\plotwidths\linewidth]{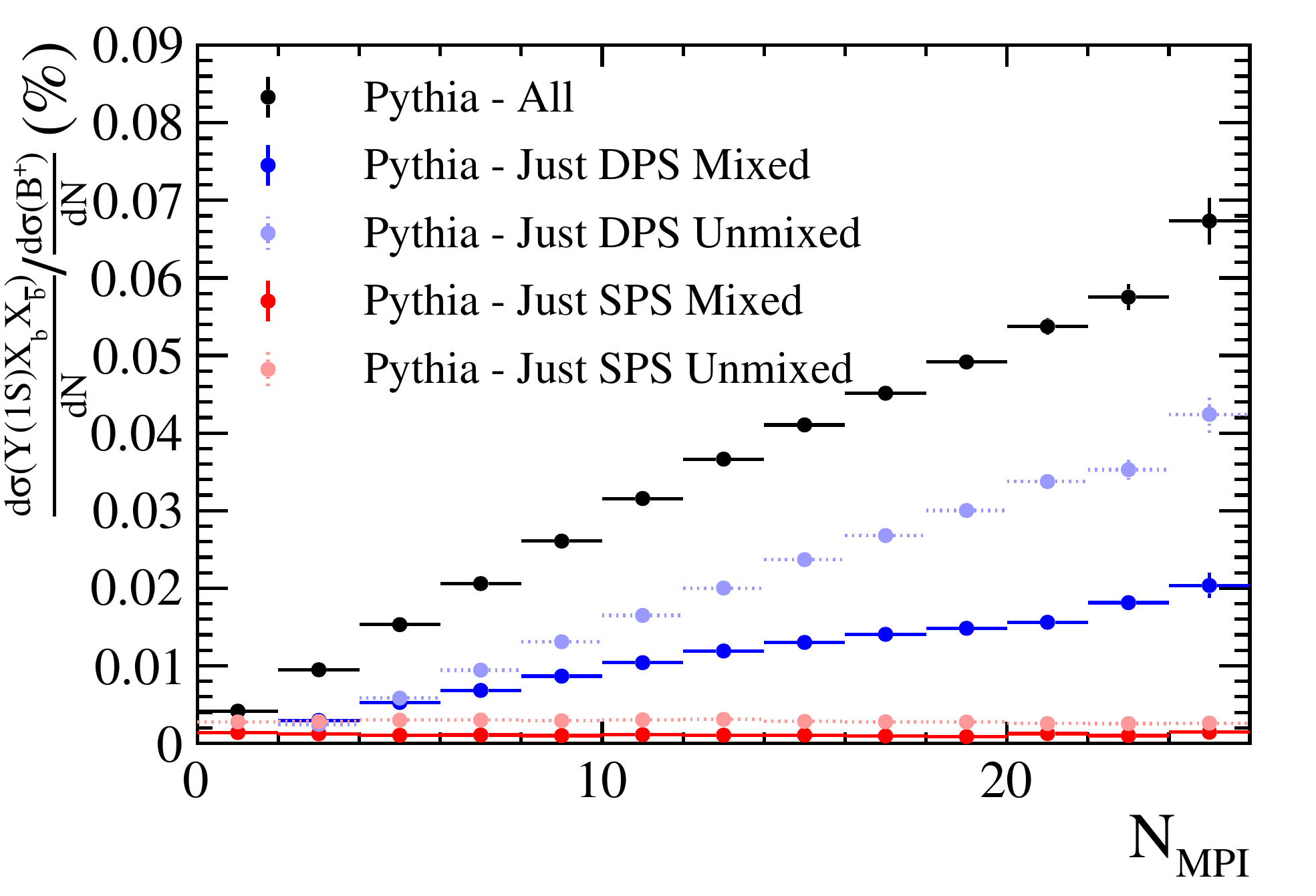}
    \includegraphics[width=\plotwidths\linewidth]{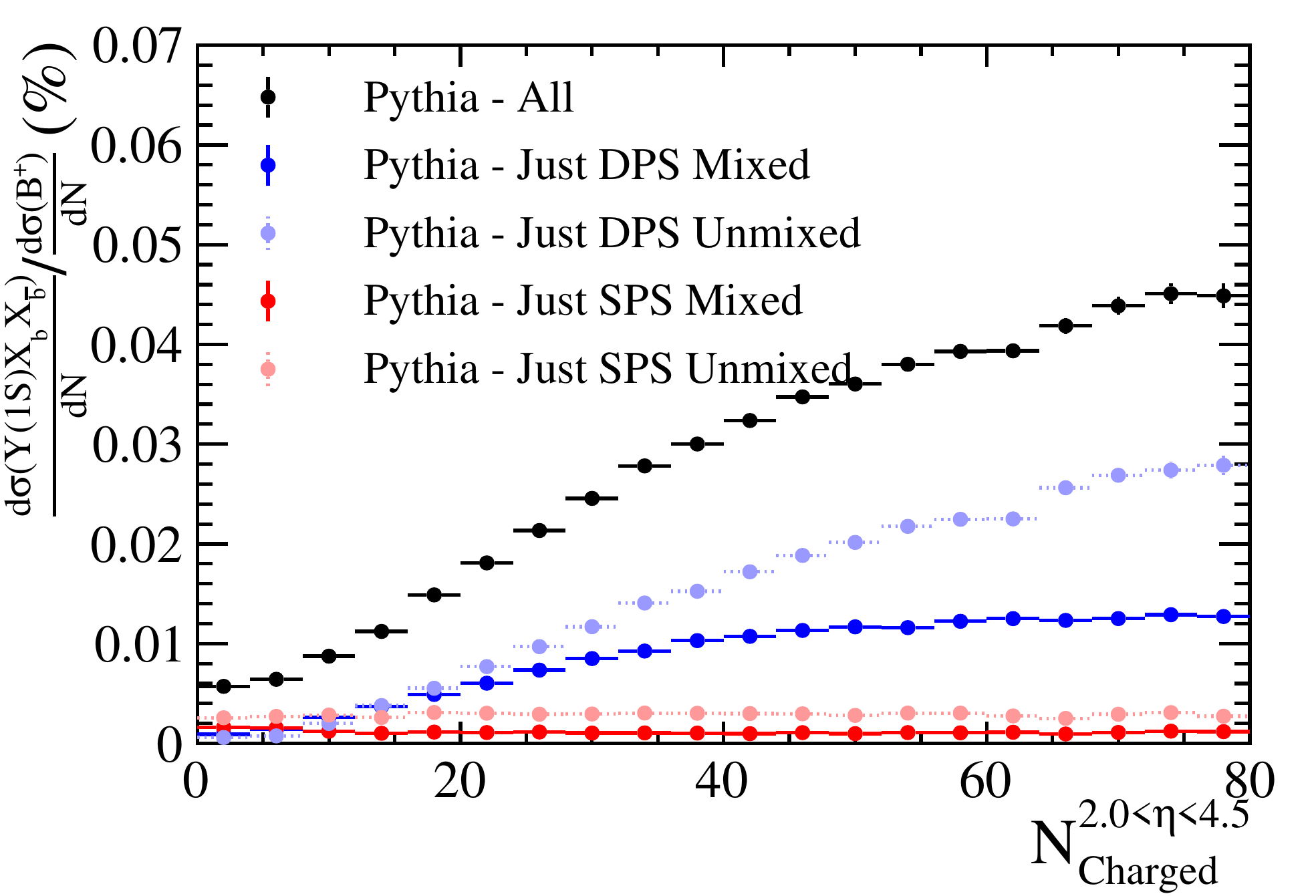}
    \caption{Ratio of differential cross-section of $\OneS X_{\bquark} X_{\bquarkbar}$ and \Bp hadrons as a function of (left) the number of parton interactions in a collision and (right) the number of charged particles within the pseudo-rapidity region $2.0<\eta<4.5$, as generated with \pythia.}
    \label{fig:UpsilonXbXb_multi_noColRec}
\end{figure}

Similar studies can be performed on the $\OneS X_{\bquark} X_{\bquarkbar}$ system. Large samples are generated with \pythia requiring a minimal event content of $\bquark\bquarkbar\bquark\bquarkbar$. Events with one $\OneS$ meson and a $X_\bquark X_\bquarkbar$ pair are selected, and the relative distributions studied. The events are split according to whether the \bquark and \bquarkbar that formed the $\OneS$ were from the same or different parton systems. As before, the ancestry of the \bquark and \bquarkbar quarks is studied to determine if the configuration is \textit{mixed} or \textit{unmixed}. The relative differential cross-section relative to the \Bp meson is shown in Fig.~\ref{fig:UpsilonXbXb_multi_noColRec} as a function of the number of parton interactions and the number of charged particles within $2.0<\eta<4.5$. A similar trend is observed in this system. 

\begin{figure}[h]
    \centering
    \begin{subfigure}{.32\textwidth}
        \includegraphics[width=\linewidth]{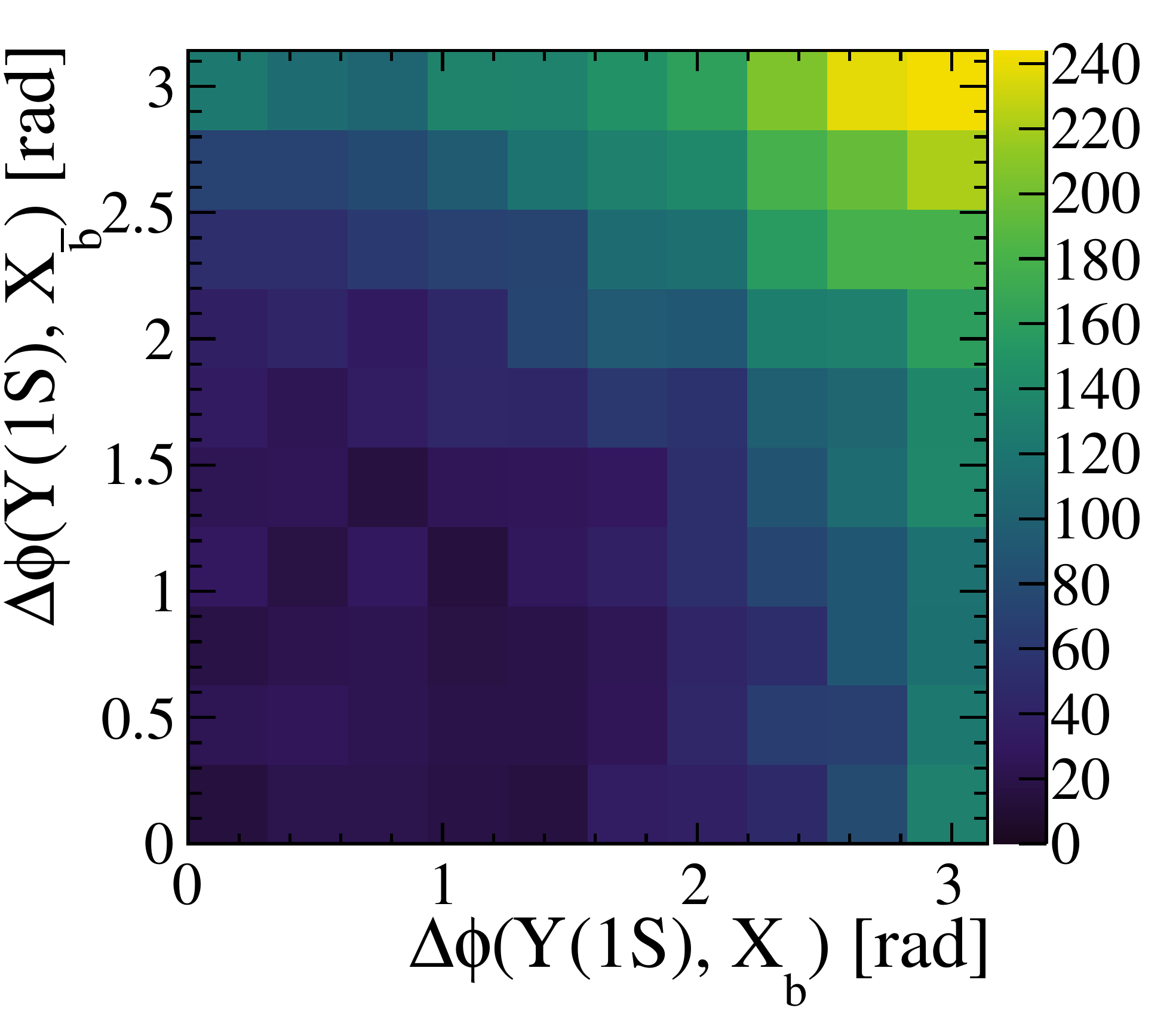}
        \caption{\pythia - SPS \textit{unmixed}}
        \label{fig:UpsilonXbXb_SPS_Unmix}
    \end{subfigure}%
    \begin{subfigure}{.32\textwidth}
        \includegraphics[width=\linewidth]{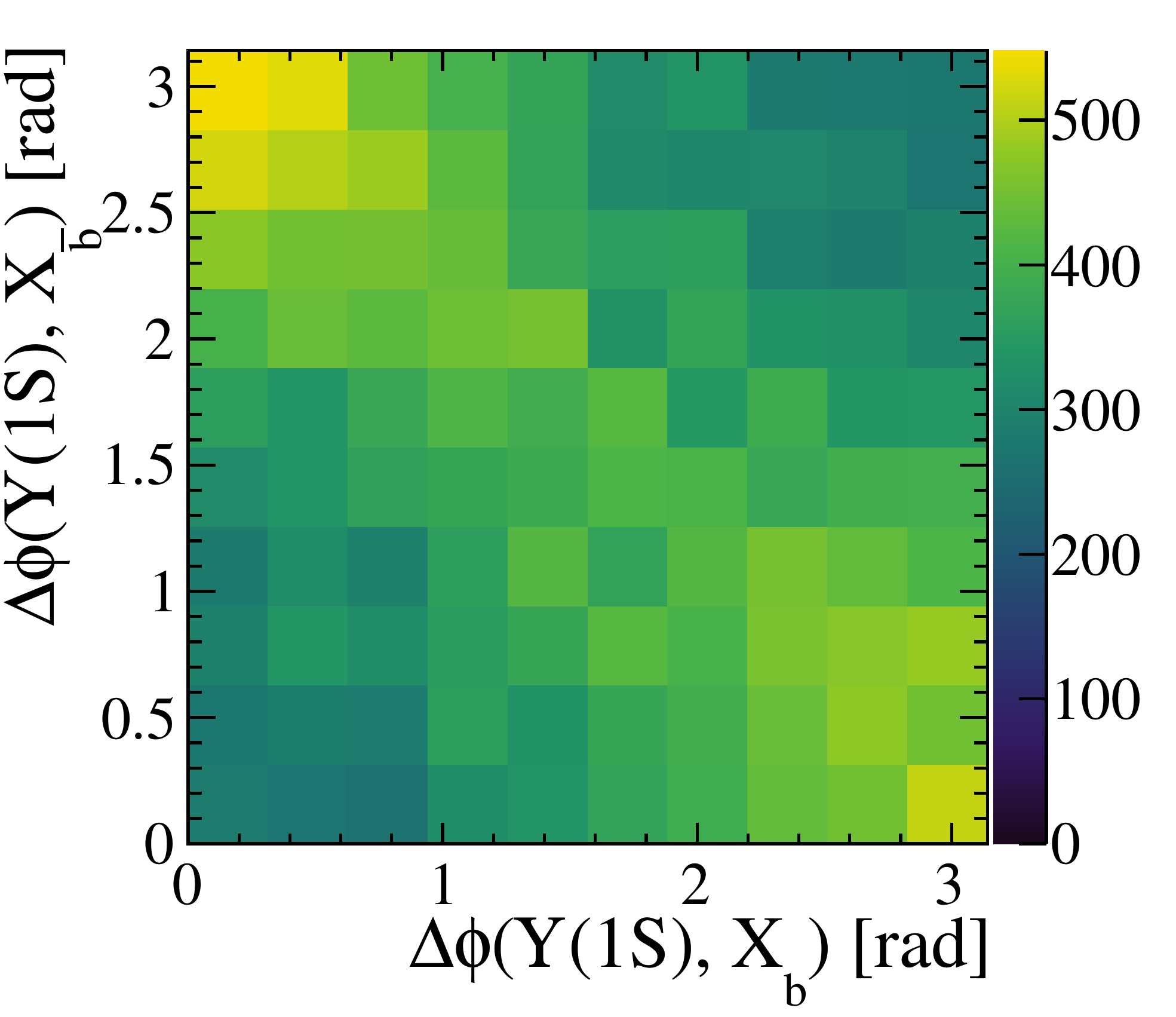}
        \caption{\pythia - DPS \textit{unmixed}}
        \label{fig:UpsilonXbXb_DPS_Unmix}
    \end{subfigure}\\
    \begin{subfigure}{.32\textwidth}
        \includegraphics[width=\linewidth]{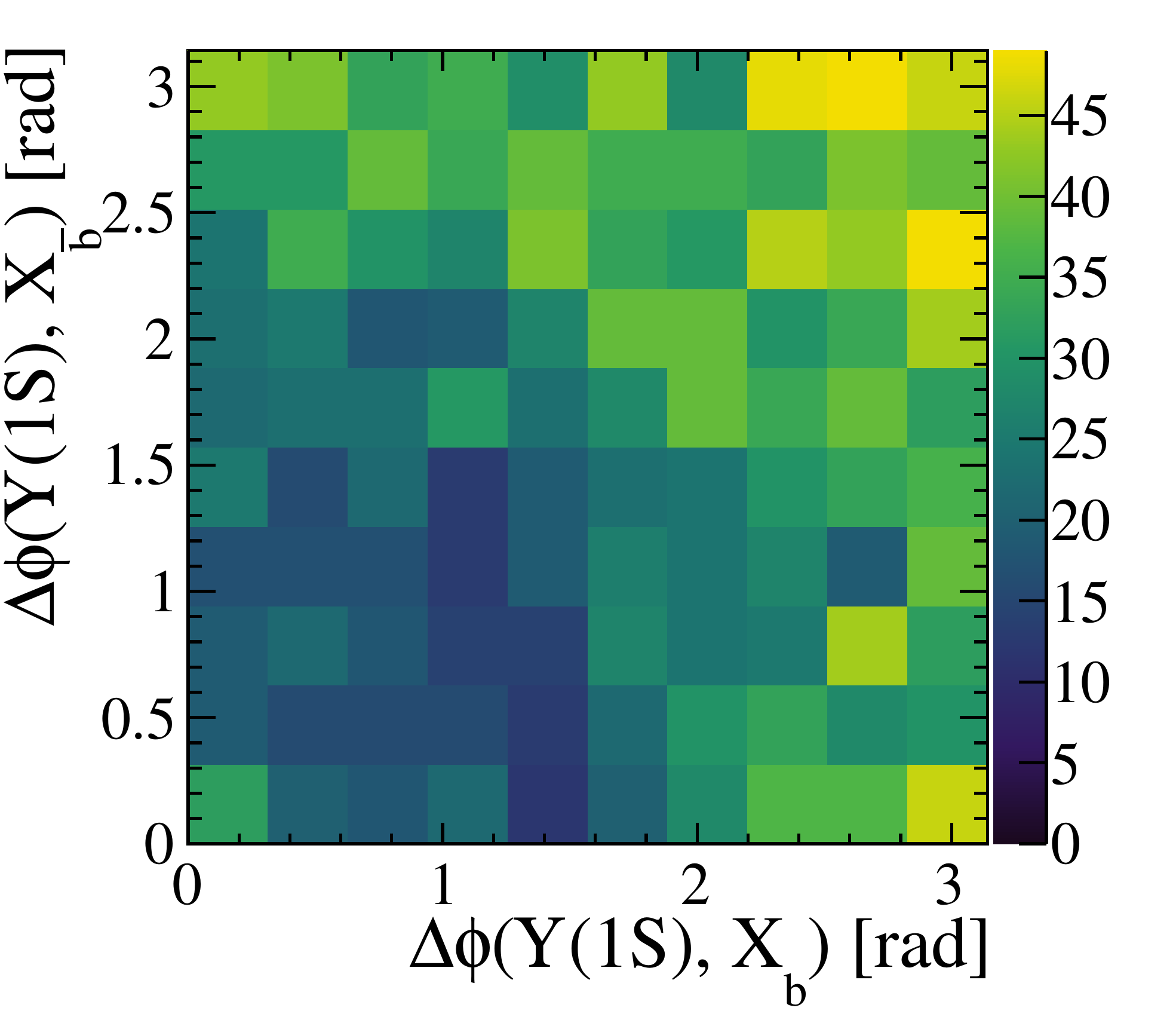}
        \caption{\pythia - SPS \textit{mixed}}
        \label{fig:UpsilonXbXb_SPS_Mix}
    \end{subfigure}%
    \begin{subfigure}{.32\textwidth}
        \includegraphics[width=\linewidth]{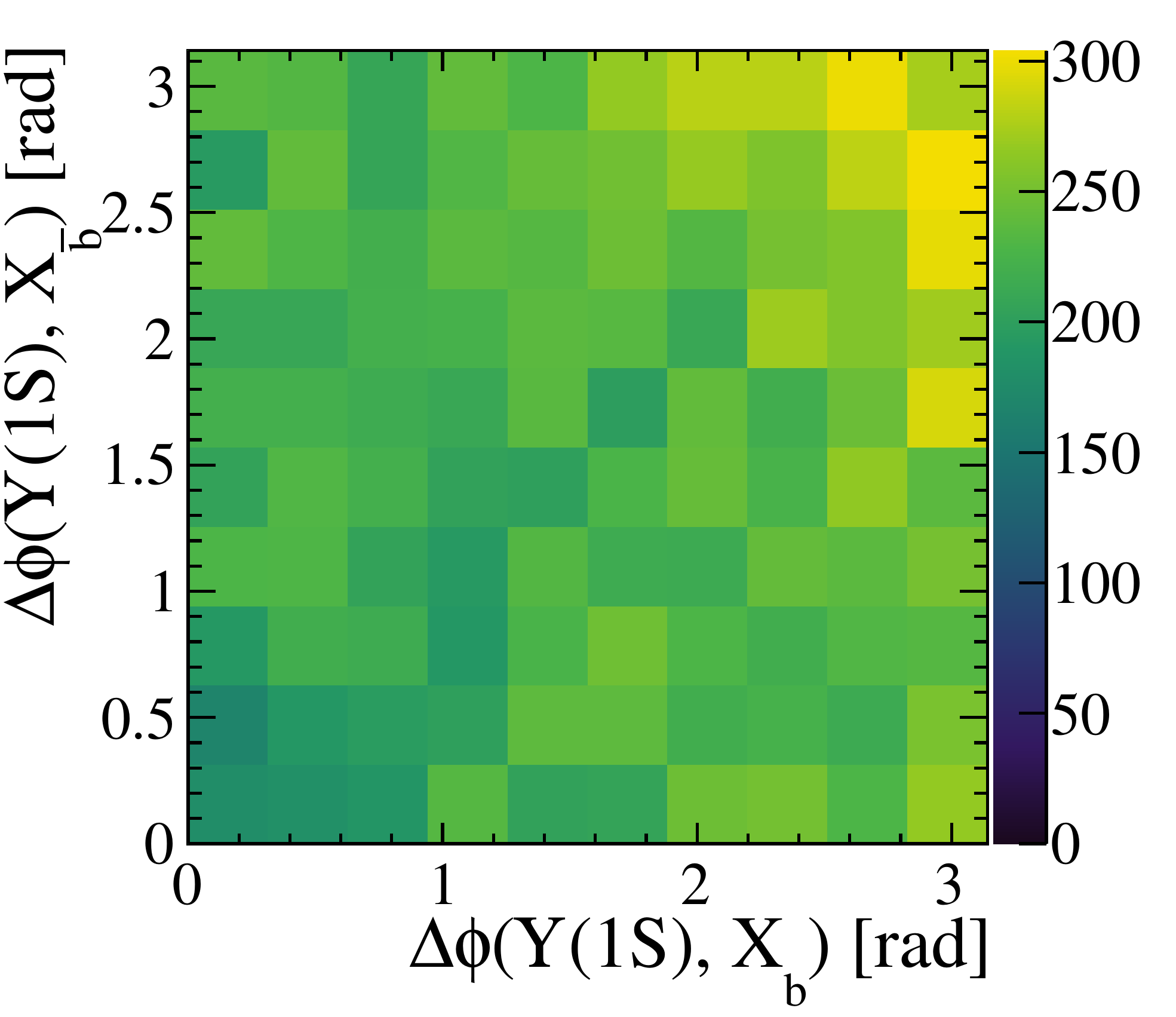}
        \caption{\pythia - DPS \textit{mixed}}
        \label{fig:UpsilonXbXb_DPS_Mix}
    \end{subfigure}%
    \caption{Relative transverse distributions in $\OneS X_b X_{\bar{b}}$ events for SPS and DPS processes in both the \textit{mixed} and \textit{unmixed} configurations.}
    \label{fig:UpsilonXbXb_2D_noColRec}
\end{figure}

Additionally, the relative transverse directions of the $\OneS$, $X_{\bquark}$ and $X_{\bquarkbar}$ hadrons is studied for the four production categories. The distributions, shown in Fig.~\ref{fig:UpsilonXbXb_2D_noColRec} show a similar distribution for the SPS \textit{unmixed} as before. In this system, events categorised as DPS \textit{unmixed} have a slight tendency to have an anti-correlation between $\Delta\phi(\OneS,X_{b})$ and $\Delta\phi(\OneS,X_{\bar{b}})$. This could correspond to events in which one parton interaction produces the $\OneS$ meson and another produces a \bquark\bquarkbar pair from pair production that are back-to-back in the transverse plane. Therefore as $\Delta\phi(\OneS,X_{b})$ increases, the corresponding value of $\Delta\phi(\OneS,X_{\bar{b}})$ decreases. The DPS \textit{unmixed} distribution has a distinct distribution, potentially allowing discrimination.   

\clearpage
%%%%%%%%%%%%%%%%%%%%%%%%%%%%%%%%%%%%%%%%
\subsection{Impact of the Colour Reconnection model}
%%%%%%%%%%%%%%%%%%%%%%%%%%%%%%%%%%%%%%%%

As mentioned briefly in Sec.~\ref{sec:Xiccpp_kinematics} and discussed in more detail in \cite{Christiansen:2015yqa,Bierlich:2015rha,Bierlich:2022pfr}, the QCD CR model of \cite{Christiansen:2015yqa} allows for coherent combinations of three SU(3) triplets (or antitriplets) into colour-singlet states, connected via Y-shaped ``string junctions''~\cite{Sjostrand:2002ip}. In the limit of a small invariant mass between two heavy quarks, the junction topology (when allowed by the CR selection rules~\cite{Christiansen:2015yqa}) reduces to a doubly-heavy diquark. In \pythia, this mechanism is essentially the sole means by which doubly-heavy baryons can be produced at all, making the rates and spectra of such baryons --- and their dependence on event characteristics such as $N_\mathrm{Charged}$ --- particularly sensitive probes of this type of colour-space ambiguities. (Accordingly, we note that all of our \pythia results for doubly-heavy baryons in this work were produced with the QCD CR option; otherwise the rates would be essentially zero.)

\begin{figure}[h]
    \centering
    \includegraphics[width=\plotwidths\linewidth]{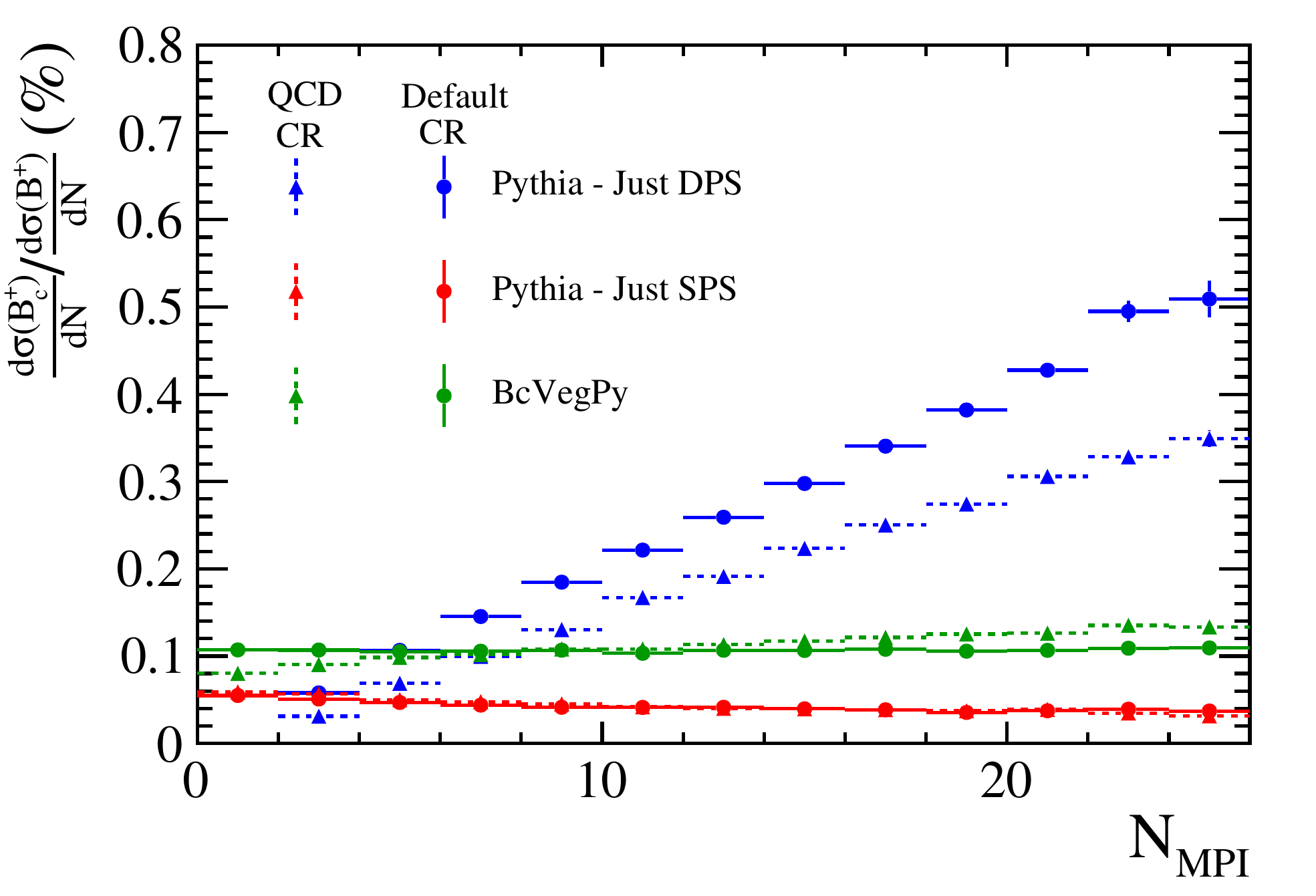}
    \includegraphics[width=\plotwidths\linewidth]{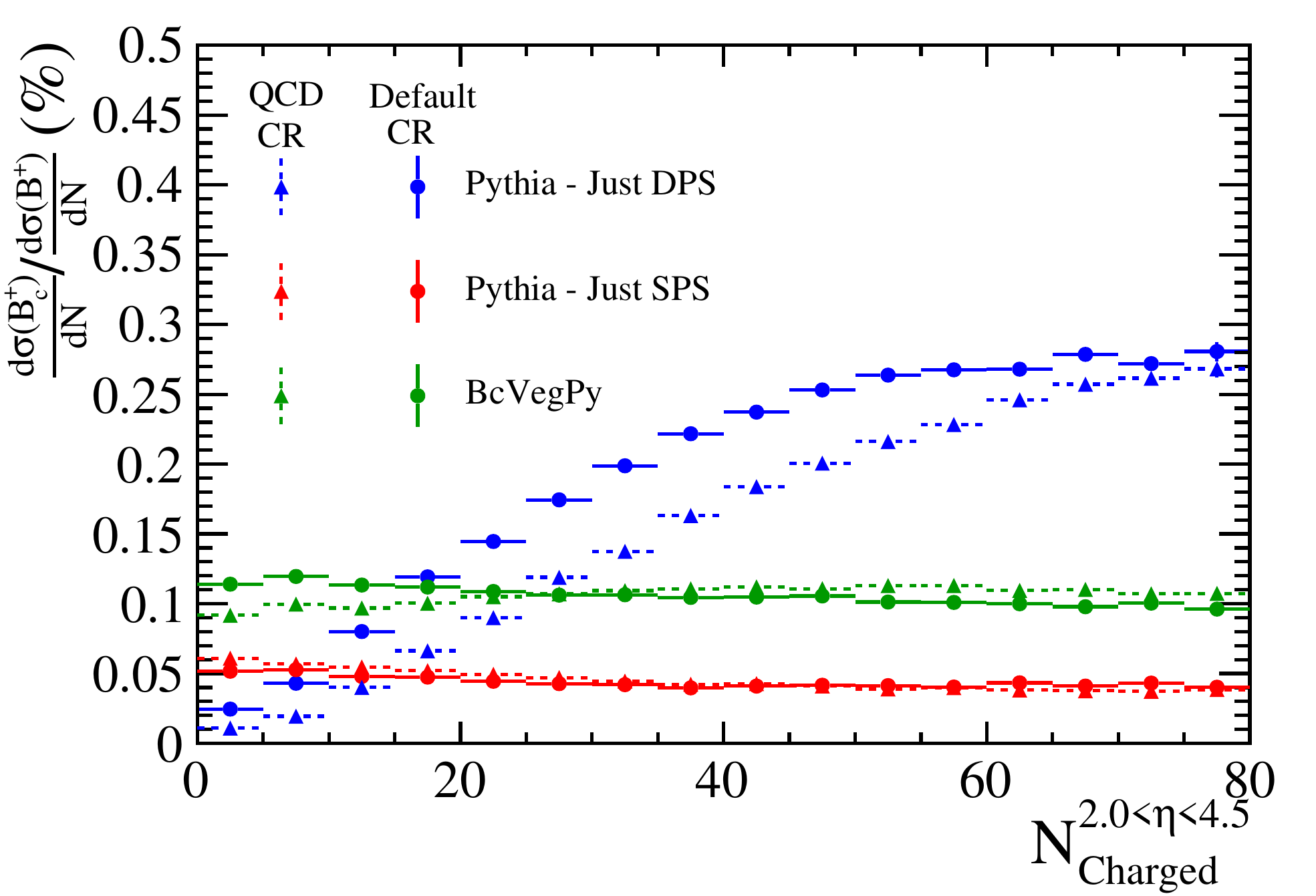}
    \caption{Ratio of differential cross-section of \Bcp and \Bp hadrons as a function of (left) the number of parton interactions in a collision and (right) the number of charged particles within the pseudo-rapidity region $2.0<\eta<4.5$, as generated with \pythia both with the alternative CR options specified in Table~\ref{tab:pythia_settings_Xicc} (triangles) and with the default CR options (circles). }
    \label{fig:Bc_CR_compare}
\end{figure}

The choice of CR model also has an impact on the
relative size of the DPS contribution to doubly-heavy meson production. 
The effect of using the QCD CR model described in Table~\ref{tab:pythia_settings_Xicc} on the slope of the ratio of \Bcp to \Bp differential cross sections is shown in Fig~\ref{fig:Bc_CR_compare}. With the QCD CR model, the DPS contribution gets smaller, altering the slope of the total distribution. As such, measurements of this distribution may help differentiate between the different models. 

\begin{figure}[h]
    \centering
    \includegraphics[width=\plotwidths\linewidth]{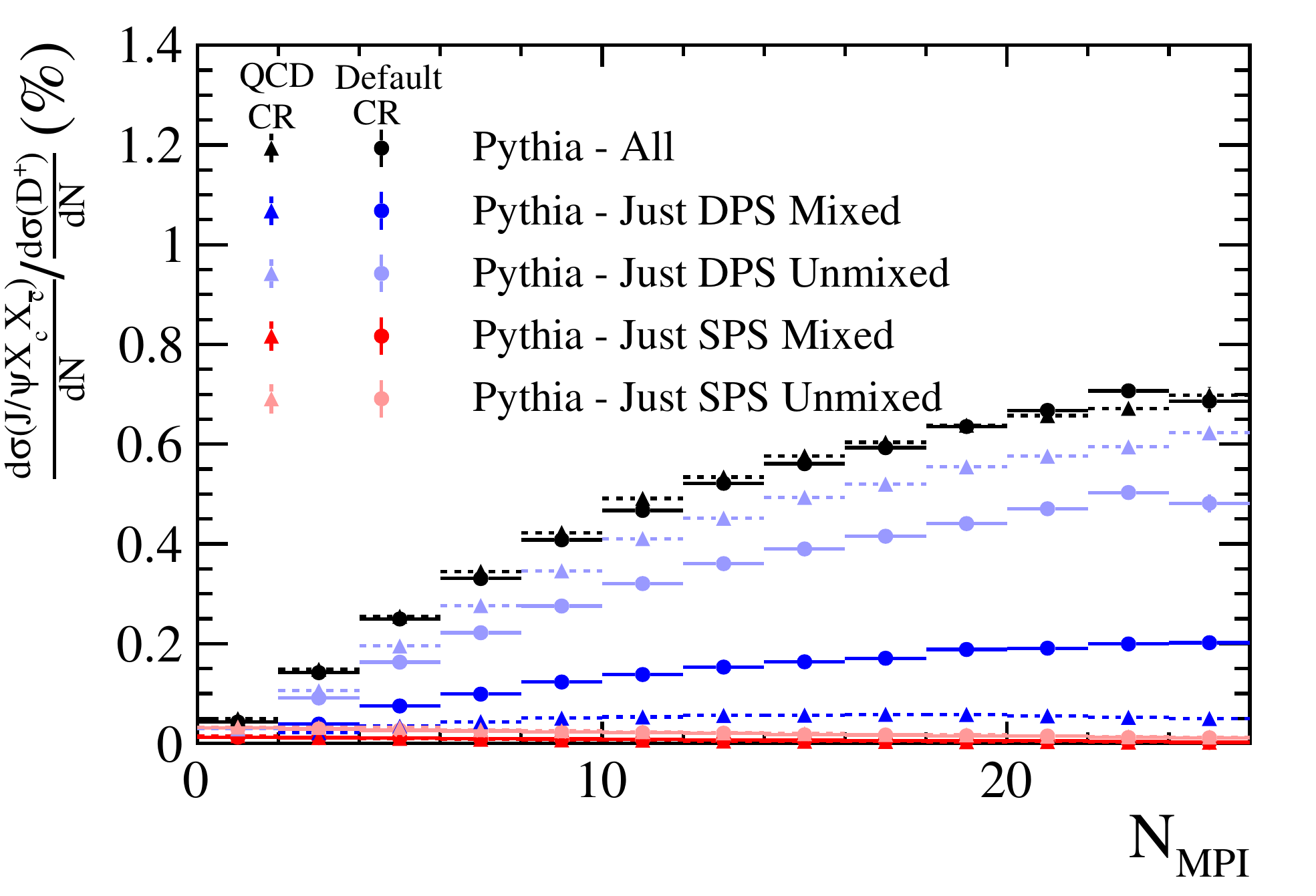}
    \includegraphics[width=\plotwidths\linewidth]{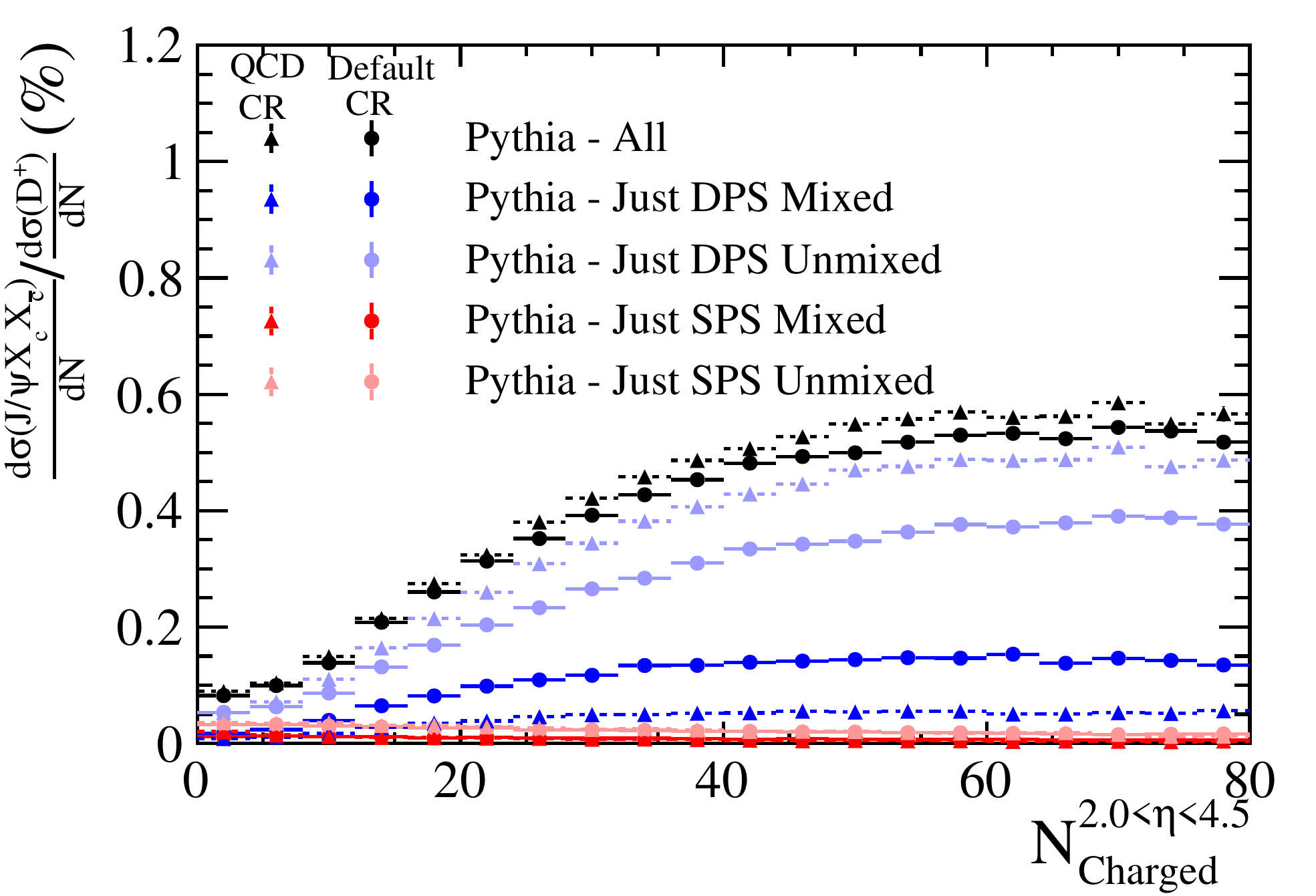}
    \caption{Ratio of differential cross-section of $\jpsi X_{\cquark} X_{\cquarkbar}$ and \Dp hadrons as a function of (left) the number of parton interactions in a collision and (right) the number of charged particles within the pseudo-rapidity region $2.0<\eta<4.5$, as generated with \pythia both with the alternative CR options specified in Table~\ref{tab:pythia_settings_Xicc} (triangles) and with the default CR options (circles).}
    \label{fig:JpsiXcXc_multi_withColRec}
\end{figure}

\begin{figure}[h]
    \centering
    \begin{subfigure}{.32\textwidth}
        \includegraphics[width=\linewidth]{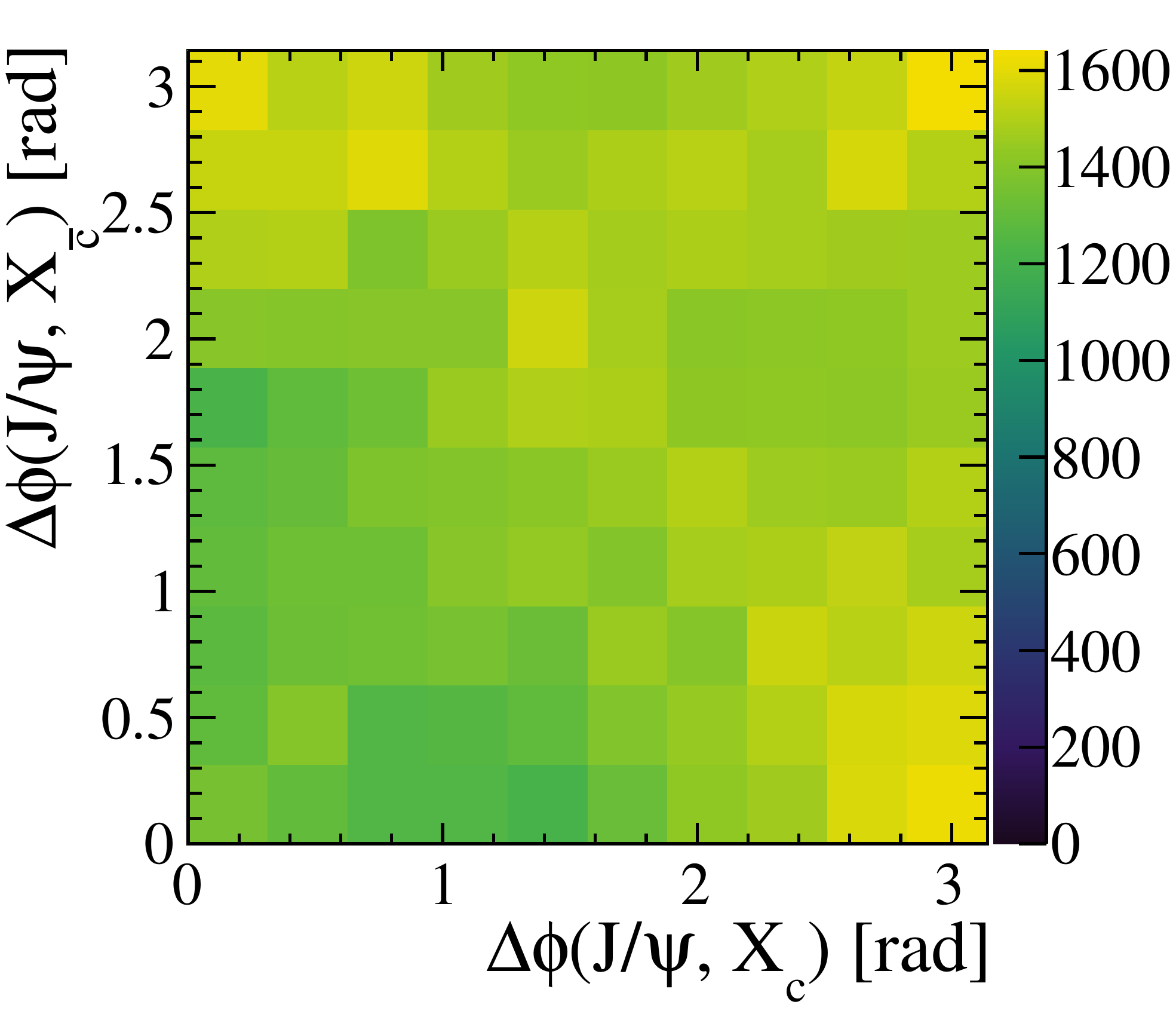}
        \caption{\pythia - DPS \textit{unmixed}}
        \label{fig:JpsiXcXc_DPS_Unmix_withColRec}
    \end{subfigure}
    \begin{subfigure}{.32\textwidth}
        \includegraphics[width=\linewidth]{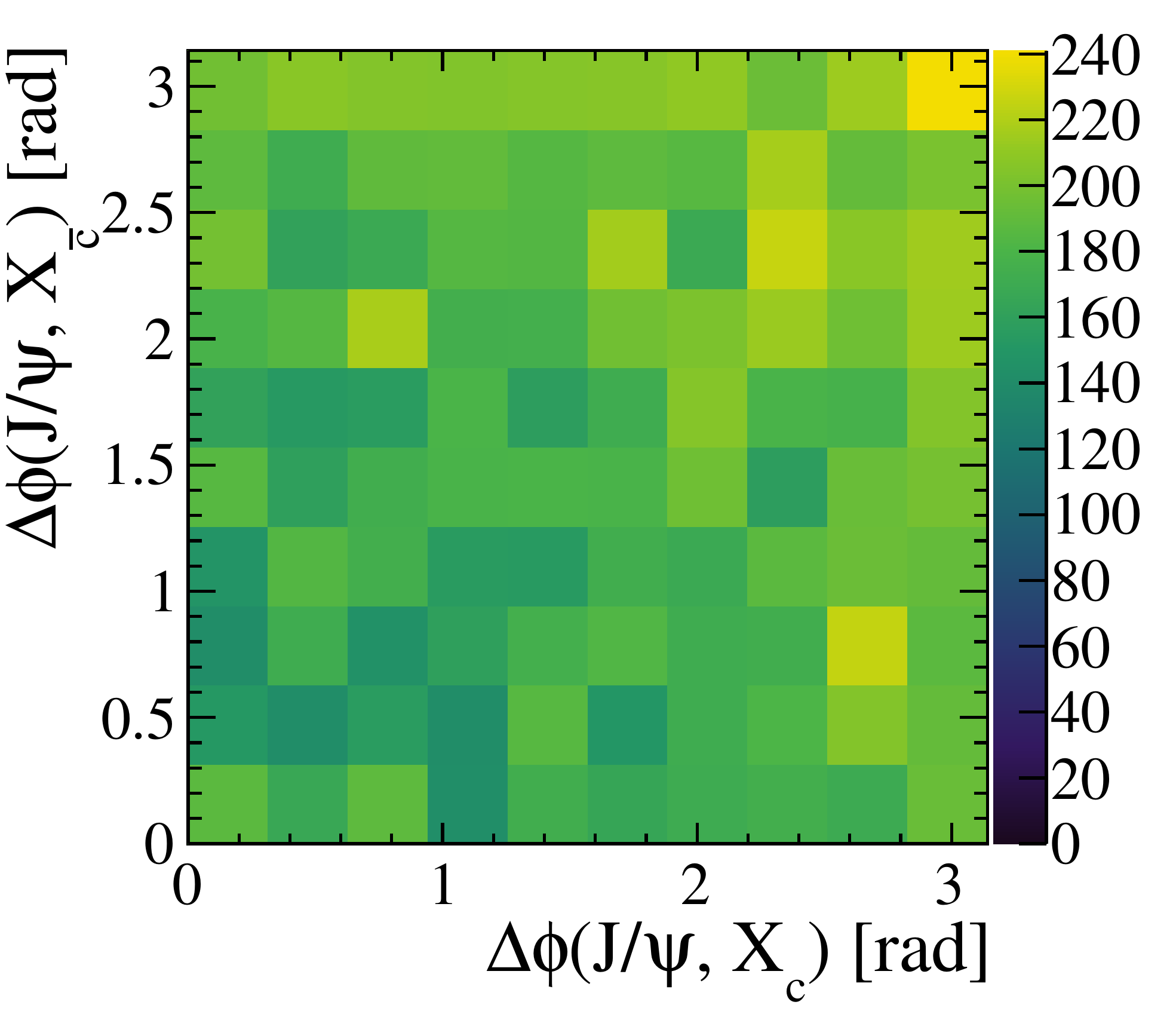}
        \caption{\pythia - DPS \textit{mixed}}
        \label{fig:JpsiXcXc_DPS_Mix_withColRec}
    \end{subfigure}%
    \caption{Relative transverse distributions in $\jpsi X_c X_{\bar{c}}$ events for DPS processes in both the \textit{mixed} and \textit{unmixed} configurations, as generated with \pythia with the colour reconnection options specified in Table~\ref{tab:pythia_settings_Xicc}.}
    \label{fig:JpsiXcXc_2D_withColRec}
\end{figure}

Similarly, the relative cross section distributions are found to vary in $\jpsi X_c X_{\bar{c}}$ samples generated with the QCD CR options enabled, as shown in Fig.~\ref{fig:JpsiXcXc_multi_withColRec}. The contribution from the DPS \textit{mixed} configuration decreases whilst DPS \textit{unmixed} increases, implying that the likelihood for the heavy quarks from different parton-parton interactions to be combined into a single hadron is sensitive to the choice of CR scheme. 
Additionally, the relative transverse direction distributions are found to differ for the DPS configurations. The corresponding distributions are shown in Fig.~\ref{fig:JpsiXcXc_2D_withColRec}. Measurements of $J\/\psi X_{c} X_{\bar{c}}$ events may additionally help to differentiate between the different models of colour reconnection. 
 
\clearpage 

%%%%%%%%%%%%%%%%%%%%%%%%%%%%%%%%%%%%%%%%%%%%%%%%%%%%%%%%%%%%%%%%%%%%
\section{Experimental measurements and feasibility}
\label{sec:lhcb}
%%%%%%%%%%%%%%%%%%%%%%%%%%%%%%%%%%%%%%%%%%%%%%%%%%%%%%%%%%%%%%%%%%%%

In the \pythia simulation studies performed for this paper, the production of doubly-heavy hadrons is predicted to have a significant contribution from DPS production processes. 
New measurements of the relative cross section for the doubly-heavy hadrons with respect to singly-heavy hadron as a function of the collision multiplicity would help identify if such contributions are present in nature, as proposed in Section~\ref{sec:single_particle_kinematics}. Unlike recent observations of strangeness enhancements in the ratio of \Bs to \Bz cross sections~\cite{LHCb:2022syj}, the enhancements from DPS are not expected to be localised.   
The most suitable doubly-heavy hadron for this would be the \Bcp meson. The significant yields reported in a selection of different papers are listed in Table~\ref{tab:bc_yields}. Studies may also be feasible for \Xiccpp baryons. 

\begin{table}[h]
    \centering
    \begin{tabular}{l l r c c }
    \hline \hline
        Experiment &  Mode &   Yield  & Dataset & Ref.\\
                  \hline
         LHCb & $B_{c}^{+} \to J/\psi \mu^+\nu$           & 19\,000 & Run1 & \cite{LHCb:2017vlu}\\
         LHCb & $B_{c}^{+} \to J/\psi \pi^+$           & 25\,181 & Run1+Run2 & \cite{LHCb:2020ayi}\\ 
         LHCb & $B_{c}^{+} \to J/\psi \pi^+\pi^-\pi^+$ & 9\,497  & Run1+Run2 & \cite{LHCb:2020ayi}\\
         LHCb & $B_{c}^{+} \to J/\psi D_{s}^{+}$       & 1\,135  & Run1+Run2 & \cite{LHCb:2020ayi}\\
         LHCb & $B_{c}^{+} \to B_{s}^{0}\pi^+$         &    316  & Run1+Run2 & \cite{LHCb:2020ayi}\\ 
         CMS  & $B_{c}^{+} \to J/\psi \pi^+$ &7629&Run2 &\cite{CMS:2020rcj}\\
         \hline 
         LHCb & $\Xiccpp \to \Lc \Km \pip \pip$        &1598      & Run2 & \cite{LHCb:2019epo} \\
         LHCb & $\Xiccpp \to \Xicp \pip$         &  616& Run2 & \cite{LHCb:2019epo} \\ 
         
    \hline \hline
    \end{tabular}
    \caption{Yields of doubly-heavy hadrons reconstructed in recent measurements at the LHC. Run1 corresponds to $\sqs = 7$~TeV and/or 8~TeV, while Run2 corresponds to $\sqs=13$~TeV.}
    \label{tab:bc_yields}
\end{table}

\begin{figure}[h]
    \centering
    \includegraphics[width=0.4\linewidth]{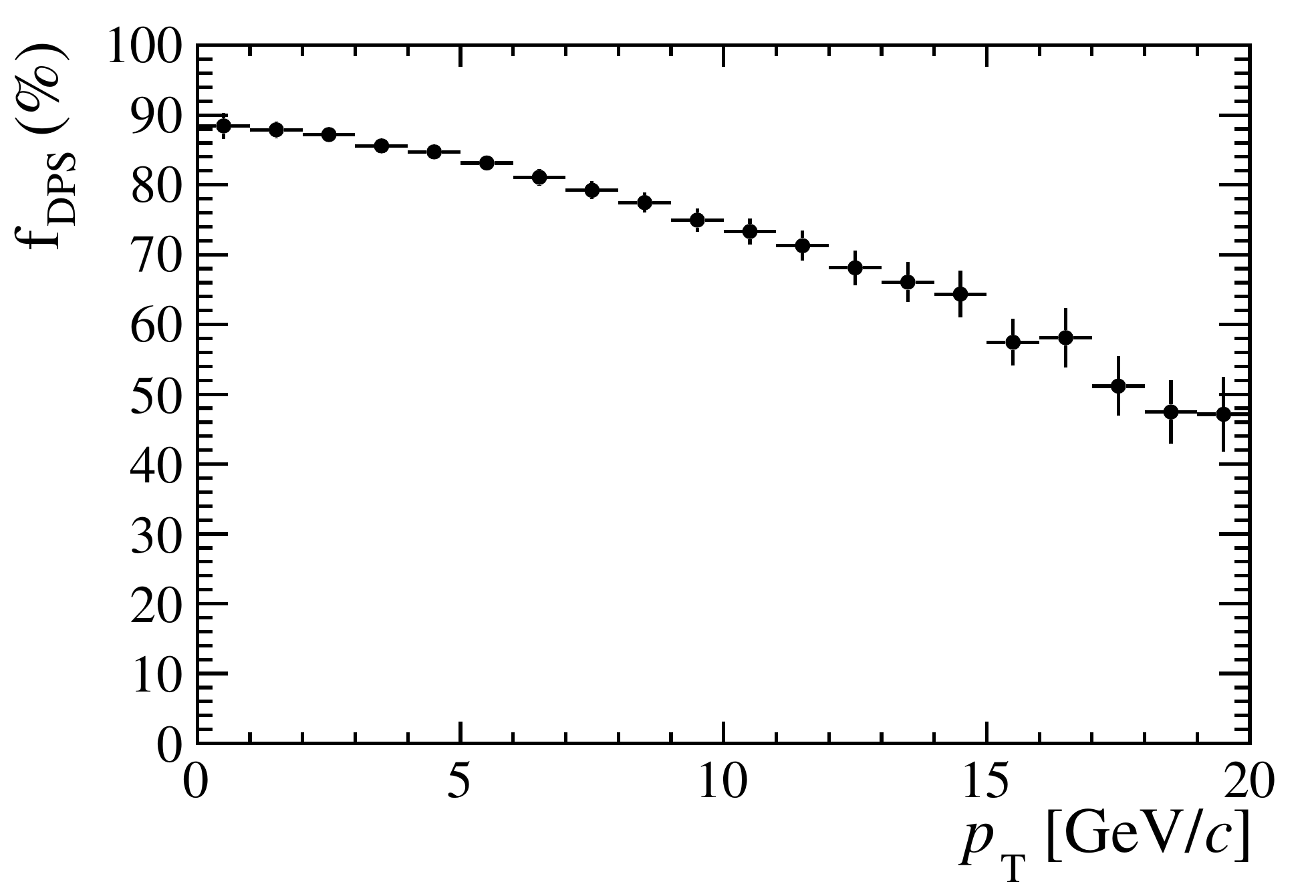}
    \includegraphics[width=0.4\linewidth]{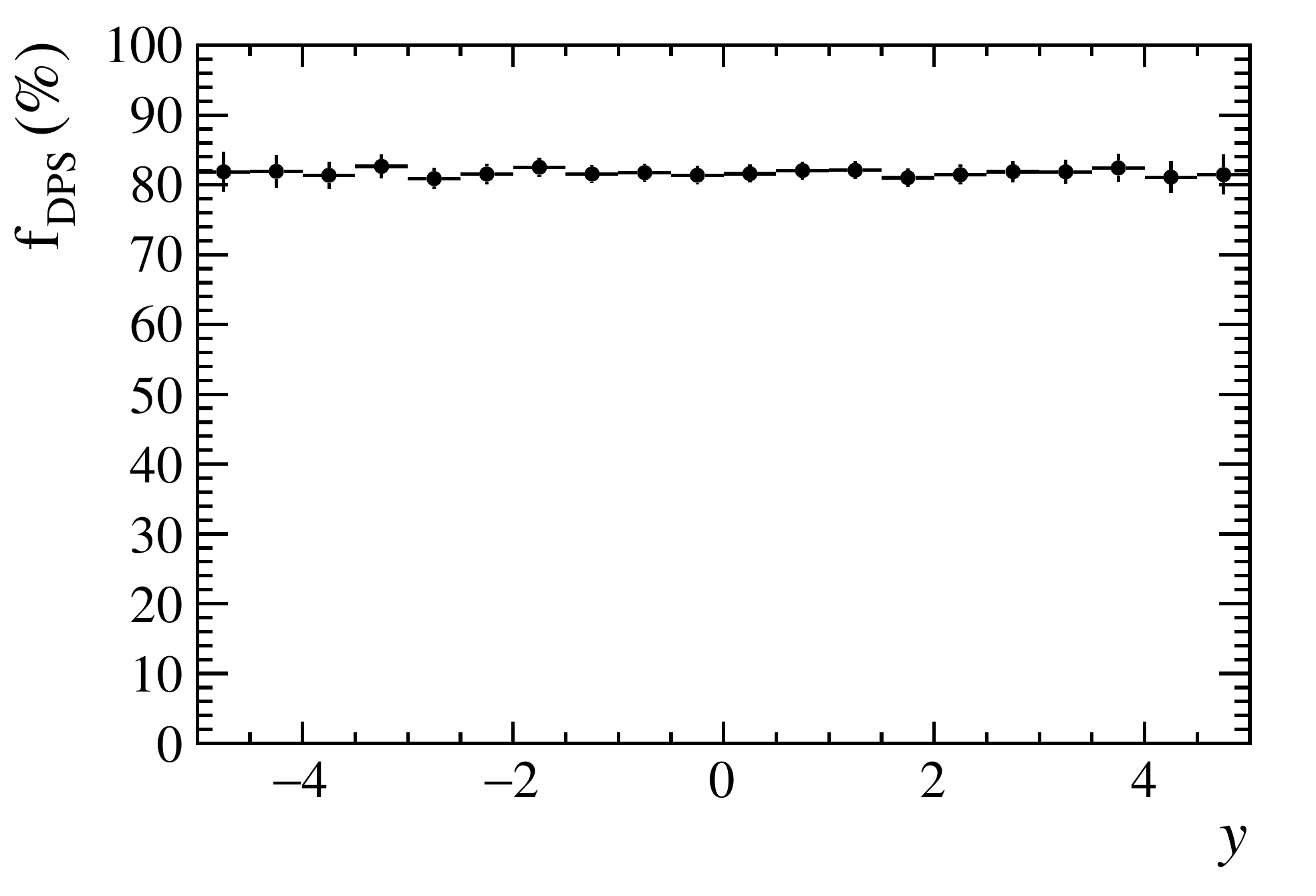}
    \caption{Fraction of \Bcp decays predicted to be produced by DPS processes as a function of (left) \pt and (right) rapidity in simulations samples produced by \pythia.}
    \label{fig:Bc_f_DPS}
\end{figure}

The predicted fraction, $f_{\rm DPS}\equiv \sigma(\Bcp)_{\rm DPS}/[\sigma(\Bcp)_{\rm SPS}+\sigma(\Bcp)_{\rm DPS}]$, of \Bcp mesons produced in DPS processes varies as a function of \pt (Fig.~\ref{fig:Bc_f_DPS}), implying the effects would be most pronounced at low-\pt. This would motivate measuring the relative cross sections as a function of the number of tracks in different \pt regions. The DPS fraction is not found to vary as a function of rapidity. 

The contributions from DPS production mechanisms can also be studied in events with one quarkonium and two singly-heavy hadrons, as discussed in Section~\ref{sec:associated_production}. These final states have the advantage that quarkonia can be efficiently reconstructed using leptonic final states. The singly heavy hadrons could be reconstructed exclusively, using a range of different final states, or inclusively, by exploiting the typical topologies of heavy-meson decays.

\subsection{Exclusive reconstruction of associated singly-heavy hadrons}
For singly-heavy hadrons containing $c$-quarks the branching fractions of the main experimentally-efficient decay channels constitute a reasonable fraction of the total decay width, as listed in Table~\ref{tab:charm_branching_fractions}. Therefore it may be feasible to exclusively reconstruct a sufficient fraction of the charm hadrons to make the proposed measurements. Indeed, the cross-section of quarkonia plus one singly-heavy charm hadron have already been performed~\cite{LHCb:2012aiv,LHCb:2015wvu}.  
\begin{table}[h]
    \centering
    \begin{tabular}{ l c}
    \hline \hline
          Decay mode & $\mathcal{B} (\%)$   \\
               \hline 
         $\decay{\Dz}{\Km\pip}$         & $3.95\pm0.03$\\
          $\decay{\Dz}{\Km\pip\pim\pip}$ &$8.22\pm0.14$ \\
        \hline
         $\decay{\Dp}{\Km\pip\pip}$     & $9.36\pm0.16$\\
        \hline
         $\decay{\Dsp}{\Kp\Km\pip}$     & $5.39\pm0.15$\\
          $\decay{\Dsp}{\pip\pim\pip}$   & $1.08\pm0.04$\\
        \hline
         $\decay{\Lc}{\proton\Km\pip}$& $6.28\pm0.32$\\
        \hline \hline 
    \end{tabular}
    \caption{Branching fractions of experimentally efficient decay channels for different charm hadron species from Ref~\cite{ParticleDataGroup:2020ssz}.}
    \label{tab:charm_branching_fractions}
\end{table}

However, for hadrons contain $b$-quarks the typical decay channels have much smaller branching fractions and may decay further to charm hadrons, therefore the fraction of events containing a quarkonia in which it is possible to  exclusively reconstruct two $X_{b}$ hadrons is small.

\subsection{Inclusive reconstruction of associated singly-heavy hadrons}

It is possible to inclusively reconstruct singly-heavy hadrons by taking advantage of the common features of many decay channels, namely a secondary decay vertex that is significantly displaced from the primary interaction. 
In order to measure the kinematic relationships between the singly- and doubly-heavy hadrons the position of displaced vertices must be measured, along with a flavour tag labelling the vertex as a \bquark- or \cquark-hadron.

Alternatively, the singly-heavy hadrons could be reconstructed as charm and beauty jets. In this case the reconstruction of the full jet may allow a better approximation of the parton-level kinematics to be determined rather than the heavy hadron kinematics. Jet flavour-tagging algorithms use similar displaced vertex tagging algorithms as already discussed to determine if a jet is the result of a heavy flavour quark. Recent measurements of heavy dijets at LHCb claim a reconstruction efficiency for $b\bar{b}$ and $c\bar{c}$ dijets of about 15\% and 1.5\%, respectively~\cite{LHCb:2020frr}. Therefore in a significant fraction of events containing quarkonia it may be feasible to additionally reconstruct these jets, assuming that the rate of fake jets can be sufficiently controlled.

\section{Conclusions}

Studies have been performed using the \pythia Monte Carlo event generator to investigate the contribution from DPS to the production of doubly-heavy hadrons. New \texttt{UserHooks} have been developed for \pythia to produce events with one or more heavy quarks more efficiently, enabling the samples to be produced in a realistic time frame. Comparisons have been made to SPS predictions from \bcvegpy{} and \genxicc. The studies  show significant and measurable differences in the ratio of cross sections between doubly- and singly- heavy hadrons as a function of the collision multiplicity that can discriminate between SPS and DPS production mechanisms. Further, the DPS mechanism is sensitive to the assumed scenario for colour reconnections. A range of measurements have been proposed, including cross-section ratios for \Bcp and \Xiccpp decays as well as a combination of cross-section ratios and relative transverse directions in events containing a quarkonium and two singly-heavy hadrons. 

\section*{Acknowledgements}
This work is supported by the Monash Warwick Alliance as part of the Monash Warwick Alliance in Particle Physics (UE, MS, PS, MV), and by the Australian Research Council via Discovery Projects DP170100708 (PS) and DP210102707 (UE, TH). MV is supported by the grant
ERC-CoG-865469 SPEAR. 

\bibliographystyle{LHCb}
\bibliography{main}
\end{document}

%% file: lhcb-symbols-def.tex
\usepackage{xspace} 
\usepackage{upgreek}

\newcommand{\offsetoverline}[2][0.1em]{\kern #1\overline{\kern -#1 #2}}%

\def\MagUp {\mbox{\em Mag\kern -0.05em Up}\xspace}

\ifthenelse{\boolean{uprightparticles}}%
{

 \def\Ppi         {\ensuremath{\uppi}\xspace}

 \def\Ppsi        {\ensuremath{\uppsi}\xspace}

 \def\PDelta      {\ensuremath{\Delta}\xspace}                 
 \def\PXi         {\ensuremath{\Xi}\xspace}                 
 \def\PLambda     {\ensuremath{\Lambda}\xspace}                 
 \def\PSigma      {\ensuremath{\Sigma}\xspace}                 
 \def\POmega      {\ensuremath{\Omega}\xspace}                 
 \def\PUpsilon    {\ensuremath{\Upsilon}\xspace}

 \def\PB      {\ensuremath{\mathrm{B}}\xspace}                 
                  
 \def\PD      {\ensuremath{\mathrm{D}}\xspace}

 \def\PJ      {\ensuremath{\mathrm{J}}\xspace}                 
 \def\PK      {\ensuremath{\mathrm{K}}\xspace}

 \def\Pb      {\ensuremath{\mathrm{b}}\xspace}                 
 \def\Pc      {\ensuremath{\mathrm{c}}\xspace}

 \def\Pi      {\ensuremath{\mathrm{i}}\xspace}

 \def\Pp      {\ensuremath{\mathrm{p}}\xspace}

 \def\Ps      {\ensuremath{\mathrm{s}}\xspace}

}
{

 \def\Ppi         {\ensuremath{\pi}\xspace}

 \def\Ppsi        {\ensuremath{\psi}\xspace}                 
                  
 \mathchardef\PDelta="7101
 \mathchardef\PXi="7104
 \mathchardef\PLambda="7103
 \mathchardef\PSigma="7106
 \mathchardef\POmega="710A
 \mathchardef\PUpsilon="7107
                  
 \def\PB      {\ensuremath{B}\xspace}                 
                  
 \def\PD      {\ensuremath{D}\xspace}

 \def\PJ      {\ensuremath{J}\xspace}                 
 \def\PK      {\ensuremath{K}\xspace}

 \def\Pb      {\ensuremath{b}\xspace}                 
 \def\Pc      {\ensuremath{c}\xspace}

 \def\Pi      {\ensuremath{i}\xspace}

 \def\Pp      {\ensuremath{p}\xspace}

 \def\Ps      {\ensuremath{s}\xspace}

}

\makeatletter
\ifcase \@ptsize \relax% 10pt
  \newcommand{\miniscule}{\@setfontsize\miniscule{4}{5}}% \tiny: 5/6
\or% 11pt
  \newcommand{\miniscule}{\@setfontsize\miniscule{5}{6}}% \tiny: 6/7
\or% 12pt
  \newcommand{\miniscule}{\@setfontsize\miniscule{5}{6}}% \tiny: 6/7
\fi
\makeatother

\DeclareRobustCommand{\optbar}[1]{\shortstack{{\miniscule (\rule[.5ex]{1.25em}{.18mm})}
  \\ [-.7ex] $#1$}}

\def\squark    {{\ensuremath{\Ps}}\xspace}

\def\cquark    {{\ensuremath{\Pc}}\xspace}
\def\cquarkbar {{\ensuremath{\overline \cquark}}\xspace}

\def\bquark    {{\ensuremath{\Pb}}\xspace}
\def\bquarkbar {{\ensuremath{\overline \bquark}}\xspace}

\def\pion   {{\ensuremath{\Ppi}}\xspace}

\def\pip    {{\ensuremath{\pion^+}}\xspace}
\def\pim    {{\ensuremath{\pion^-}}\xspace}

\def\kaon    {{\ensuremath{\PK}}\xspace}
  \def\Kbar    {{\kern 0.2em\overline{\kern -0.2em \PK}{}}\xspace}

\def\KorKbar {\kern 0.18em\optbar{\kern -0.18em K}{}\xspace}

\def\Kp      {{\ensuremath{\kaon^+}}\xspace}
\def\Km      {{\ensuremath{\kaon^-}}\xspace}

  \def\Dbar    {{\kern 0.2em\overline{\kern -0.2em \PD}{}}\xspace}
\def\D       {{\ensuremath{\PD}}\xspace}

\def\DorDbar {\kern 0.18em\optbar{\kern -0.18em D}{}\xspace}
\def\Dz      {{\ensuremath{\D^0}}\xspace}

\def\Dp      {{\ensuremath{\D^+}}\xspace}

\def\Dsp     {{\ensuremath{\D^+_\squark}}\xspace}

\def\B       {{\ensuremath{\PB}}\xspace}
\def\Bbar    {{\ensuremath{\kern 0.18em\overline{\kern -0.18em \PB}{}}}\xspace}

\def\BorBbar    {\kern 0.18em\optbar{\kern -0.18em B}{}\xspace}
\def\Bz      {{\ensuremath{\B^0}}\xspace}

\def\Bu      {{\ensuremath{\B^+}}\xspace}

\def\Bp      {{\ensuremath{\Bu}}\xspace}

\def\Bs      {{\ensuremath{\B^0_\squark}}\xspace}

\def\Bc      {{\ensuremath{\B_\cquark^+}}\xspace}
\def\Bcp     {{\ensuremath{\B_\cquark^+}}\xspace}

\def\jpsi     {{\ensuremath{{\PJ\mskip -3mu/\mskip -2mu\Ppsi\mskip 2mu}}}\xspace}

\def\Y#1S{\ensuremath{\PUpsilon{(#1S)}}\xspace}
\def\OneS  {{\Y1S}}

\def\proton      {{\ensuremath{\Pp}}\xspace}

\def\Lz          {{\ensuremath{\PLambda}}\xspace}

\def\LorLbar     {\kern 0.18em\optbar{\kern -0.18em \PLambda}{}\xspace}

\def\Xires       {{\ensuremath{\PXi}}\xspace}

\def\Lc          {{\ensuremath{\Lz^+_\cquark}}\xspace}

\def\Xicp        {{\ensuremath{\Xires^+_\cquark}}\xspace}

\def\Xiccp       {{\ensuremath{\Xires^+_{\cquark\cquark}}}\xspace}
\def\Xiccpp      {{\ensuremath{\Xires^{++}_{\cquark\cquark}}}\xspace}

\newcommand{\decay}[2]{\mbox{\ensuremath{#1\!\to #2}}\xspace}         % {\Pa}{\Pb \Pc}

\def\to                 {\ensuremath{\rightarrow}\xspace}

\def\AT#1     {\ensuremath{A_{\mathrm{T}}^{#1}}\xspace}           % 2

\def\C#1      {\ensuremath{\mathcal{C}_{#1}}\xspace}                       % 9
\def\Cp#1     {\ensuremath{\mathcal{C}_{#1}^{'}}\xspace}                    % 7
\def\Ceff#1   {\ensuremath{\mathcal{C}_{#1}^{\mathrm{(eff)}}}\xspace}        % 9  
\def\Cpeff#1  {\ensuremath{\mathcal{C}_{#1}^{'\mathrm{(eff)}}}\xspace}       % 7
\def\Ope#1    {\ensuremath{\mathcal{O}_{#1}}\xspace}                       % 2
\def\Opep#1   {\ensuremath{\mathcal{O}_{#1}^{'}}\xspace}                    % 7

\newcommand{\tev}{\ifthenelse{\boolean{inbibliography}}{\ensuremath{~T\kern -0.05em eV}}{\ensuremath{\mathrm{\,Te\kern -0.1em V}}}\xspace}
\newcommand{\gev}{\ensuremath{\mathrm{\,Ge\kern -0.1em V}}\xspace}
\newcommand{\mev}{\ensuremath{\mathrm{\,Me\kern -0.1em V}}\xspace}
\newcommand{\kev}{\ensuremath{\mathrm{\,ke\kern -0.1em V}}\xspace}
\newcommand{\ev}{\ensuremath{\mathrm{\,e\kern -0.1em V}}\xspace}
\newcommand{\mevc}{\ensuremath{{\mathrm{\,Me\kern -0.1em V\!/}c}}\xspace}
\newcommand{\gevc}{\ensuremath{{\mathrm{\,Ge\kern -0.1em V\!/}c}}\xspace}
\newcommand{\mevcc}{\ensuremath{{\mathrm{\,Me\kern -0.1em V\!/}c^2}}\xspace}
\newcommand{\gevcc}{\ensuremath{{\mathrm{\,Ge\kern -0.1em V\!/}c^2}}\xspace}
\newcommand{\gevgevcc}{\ensuremath{{\mathrm{\,Ge\kern -0.1em V^2\!/}c^2}}\xspace} % for \pt^2 in CEP
\newcommand{\gevgevcccc}{\ensuremath{{\mathrm{\,Ge\kern -0.1em V^2\!/}c^4}}\xspace} % for q^2

\def\ms   {\ensuremath{{\mathrm{ \,ms}}}\xspace}

\def\gsim{{~\raise.15em\hbox{$>$}\kern-.85em
          \lower.35em\hbox{$\sim$}~}\xspace}
\def\lsim{{~\raise.15em\hbox{$<$}\kern-.85em
          \lower.35em\hbox{$\sim$}~}\xspace}

\def\sqs   {\ensuremath{\protect\sqrt{s}}\xspace}

\def\pt         {\ensuremath{p_{\mathrm{T}}}\xspace}

\def\bcvegpy    {\mbox{\textsc{Bcvegpy}}\xspace}

\def\pythia     {\mbox{\textsc{Pythia}}\xspace}

\def\tell1  {TELL1\xspace}
\def\ukl1   {UKL1\xspace}

\newcommand{\eg}{\mbox{\itshape e.g.}\xspace}
\newcommand{\ie}{\mbox{\itshape i.e.}\xspace}